\documentclass[journal ]{new-aiaa}

\usepackage{graphicx}
\usepackage{amsmath}
\usepackage[version=4]{mhchem}
\usepackage{siunitx}
\usepackage{longtable,tabularx}
\usepackage{float}
\usepackage{graphicx}
\usepackage{xcolor}
\usepackage{multirow}
\usepackage{array, booktabs, makecell}
\usepackage{subcaption}
\usepackage{booktabs}
\usepackage{adjustbox}
\usepackage[tableposition=top]{caption} 
\usepackage{bm}
 \usepackage{bmpsize}

\usepackage{booktabs}
\newcolumntype{L}[1]{>{\raggedright\arraybackslash}p{#1}}
\usepackage{siunitx} 

\def\frac#1#2{{\textstyle{#1\over#2}}}

\def\diag{{\rm diag}}

\def\sign{{\rm sign}}

\def\obs{{\rm obs}}

\def\red#1{{\color{red} #1}}%

\newcommand{\D}[1]{\mathrm{d}{#1}}

\def\Real{\mathbb{R}}

\def\bi{\begin{itemize}}
\def\ei{\end{itemize}}
\def\bd{\begin{description}}
\def\ed{\end{description}}
\def\ben{\begin{enumerate}}
\def\een{\end{enumerate}}
\def\bv{\begin{verbatim}}
\def\ev\end{verbatim}

\def\calN{{\mathcal N}}
\def\N{{\calN}}

\def\calI{{\mathcal{I}}}

\def\s{{\sigma}}

\def\v{{\varepsilon}}

\def\T{{ \mathrm{\scriptscriptstyle T} }}

\def\pr{{\rm Pr}}

\DeclareMathOperator*{\E}{E}

\def\2to{{\ {\buildrel 2\over \longrightarrow}\ }}

\def\I1ton{{$I_1,\ldots,I_n$}}
\def\X1ton{{$X_1,\ldots,X_n$}}
\def\Y1ton{{$Y_1,\ldots,Y_n$}}
\def\Z1ton{{$Z_1,\ldots,Z_n$}}
\def\R1ton{{$R_1,\ldots,R_n$}}
\def\e1ton{{$e_1,\ldots,e_n$}}
\def\t1ton{{$t_1,\ldots,t_n$}}
\def\x1ton{{$x_1,\ldots,x_n$}}
\def\y1ton{{$y_1,\ldots,y_n$}}
\def\z1ton{{$z_1,\ldots,z_n$}}

\def\Pvalue{{P-value}}

\def\calI{{\mathcal{I}}}

\def\calG{{\mathcal{G}}}
\def\o{{\mathrm{o}}}
\def\hat{\widehat}
\def\red#1{{\color{black} #1}}%
\title{\red{Space Oddity? A Statistical Formulation of Conjunction Assessment}}

\author{S.~Elkantassi  \footnote{Ph.D candidate,  Institute of Mathematics, Ecole Polytechnique F\'ed\'erale de Lausanne (EPFL), Station 8, 1015 Lausanne, Switzerland} 
and A.~C.~Davison\footnote{Professor, Institute of Mathematics, Ecole Polytechnique F\'ed\'erale de Lausanne (EPFL), Station 8, 1015 Lausanne, Switzerland}
}
\affil{Institute of Mathematics, Ecole Polytechnique F\'ed\'erale de Lausanne (EPFL), Lausanne, 1015,   Switzerland}

\begin{document}

\maketitle

\begin{abstract}
Satellite conjunctions involving `near misses' of space objects are becoming increasingly likely.  One approach to risk analysis for them involves the computation of the collision probability, but this has been regarded as having some counter-intuitive properties \red{and its interpretation has been debated.  This paper formulates an approach to satellite conjunction based on a simple statistical model and discusses inference on the miss distance between the two objects, both when the relative velocity can be taken as known and when its uncertainty must be taken into account. It is pointed out that the usual collision probability estimate can be badly biased, but that highly accurate inference on the miss distance is possible.  The ideas are illustrated with case studies and Monte Carlo results that show its excellent performance.}
\end{abstract}

\red{\section*{Nomenclature}}
{\small 
\noindent(Nomenclature entries should have the units identified)

\red{{\renewcommand\arraystretch{1.0}
\noindent\begin{longtable*}{@{}l @{\quad=\quad} l@{}}
$f(y;\vartheta)$& probability density function for data $y$, depends on unknown parameter $\vartheta$\\
$f(\vartheta)$& prior density for $\vartheta$ in Bayesian formulation of satellite conjunction problem\\
$y$& generic value of data vector\\
$y^\o$& observed value of data vector\\
$\Omega$& dispersion matrix of $y$, considered to be known\\
$\vartheta$ & column vector of parameters of statistical model\\
$\psi$	& scalar interest parameter for statistical model, usually the unknown miss distance (m or km)\\
$\psi_{\rm min}$	& combined hard-body radius HBR(m or km)\\
$\psi_0$& safety threshold (m or km), with $\psi_0>\psi_{\rm min}$\\
$\lambda$& column vector of nuisance parameters for statistical model\\
$\pr(\cdot)$& probability of event `$\cdot$'\\
$\E(\cdot)$&expectation with respect to distribution $\pr(\cdot)$\\
$\pr_0(\cdot)$& probability of event `$\cdot$' computed under null hypothesis $\psi=\psi_0$\\
$p_c$ & collision probability\\
$\hat p_c$ & estimated collision probability\\
$r(\psi)$& likelihood root, an approximate pivot\\
$q(\psi)$& correction to likelihood root\\
$r^*(\psi)$& modified likelihood root $r(\psi)+ \log\{q(\psi)/r(\psi)\}/r(\psi)$, an approximate pivot\\
$r^*_B(\psi)$& Bayesian version of $r^*(\psi)$\\
$w(\psi)$& Wald statistic, an approximate pivot\\
$I_{1-2\alpha}$&Two-sided $(1-2\alpha)$  confidence interval for unknown parameter $\psi$\\
$L_\alpha,U_\alpha$& lower and upper limits to an equi-tailed  $(1-2\alpha)$ confidence interval\\
$\Phi(\cdot)$& cumulative probability function for standard normal, $\N(0,1)$,  distribution\\
$z_p$& $p$ quantile of the standard normal distribution, satisfying $\Phi(z_p)=p$ for $0<p<1$\\
$\mu$&true current position vector for secondary space object relative to first (m or km)\\
$\nu$&true current velocity vector for secondary space object relative to first (m/s or km/s)\\
$\|\cdot\|$ & Euclidean norm of the vector `$\cdot$'\\
$\beta$& angle between $\mu$ and $\nu$ (radians)\\
$\theta_1,\theta_2,\phi_1,\phi_2$& angles of spherical polar coordinates (radians)\\
$\xi$&two-dimensional projection of $\mu$ in the direction of $\nu$ onto the encounter plane (m or km)\\
$x$&two-dimensional projection of $y$ in the direction of $\nu$ onto the encounter plane (m or km)\\
$D$&$\diag(d_1^2,d_2^2)$, variance matrix of $x$ (m$^2$ or km$^2$)\\ 
$I(\cdot)$& indicator function for event `$\cdot$'\\
$L(\vartheta)$&likelihood function, i.e., $f(y;\vartheta)$ regarded as a function of unknown parameter $\vartheta$\\
$\ell(\vartheta)$&log-likelihood function, i.e., $\log f(y;\vartheta)$ regarded as a function of unknown parameter $\vartheta$\\
$\jmath(\vartheta)$& observed information matrix, i.e., $-\partial^2\ell(\vartheta)/\partial\vartheta\partial\vartheta^\T$\\
$\jmath_{\lambda\lambda}(\vartheta)$& $(\lambda,\lambda)$ block of observed information matrix, i.e., $-\partial^2\ell(\vartheta)/\partial\lambda\partial\lambda^\T$\\
$\jmath_p(\psi)$& profile information, $|\jmath(\hat\vartheta_\psi)|/|\jmath_{\lambda\lambda}(\hat\vartheta_\psi)|$\\
$\varphi(\vartheta)$& constructed parameter used for higher-order approximation\\
\multicolumn{2}{@{}l}{Sub- and super-scripts}\\
$\o$ &indicates quantity evaluated for observed data $y=y^\o$\\
$\hat\vartheta$& indicates maximum likelihood estimator of parameter $\vartheta$\\
$\hat\lambda_\psi$& indicates maximum likelihood estimator of parameter $\lambda$ when parameter $\psi$ is fixed\\
$\hat\vartheta_\psi$& indicates maximum likelihood estimator of parameter $\vartheta$  when parameter $\psi$ is fixed, i.e., $\hat\vartheta_\psi=(\psi,\hat\lambda_\psi)$\\
$x^{\T}$& transpose of $x$
\end{longtable*}}}
}

\newpage
\section{Introduction}
The expansion of the aerospace industry and the increasing number of space objects, especially in Low Earth Orbit where most spacecraft operate, means that risk assessment and collision avoidance manoeuvres are vital  to ensure their safety. A great deal of effort has been put into conjunction assessment for orbiting objects, generally by attempting to estimate the  probability of collision between them, as in \citep[][Chapter~5]{Chen.etal:2017} or \citep[][Section~11.7]{Vallado:2013}.  This is calculated at the time of the closest approach using the estimated position and velocity vectors for the two objects and the associated error covariances and, in a short-term conjunction, can be expressed as an integral of a two-dimensional Gaussian probability density function over the collision cross-sectional area. Although unavailable in explicit form, it can readily be evaluated numerically \citep{Foster:1992,Chan:1997,Alfano:2005,Alfano:2006,Patera:2005,Garcia-Pelayo.Hernando-Ayuso:2016}. However both more precise and less precise measurement reduce the collision probability, a `dilution' property that has been seen as paradoxical, and its interpretation is not seen as clear-cut \citep{Balch:2019}.   Another criterion for risk assessment is the closest approach, or miss, distance, as a miss distance that is likely to be lower than a specified safety threshold indicates a situation that requires action. \red{This distance is estimated repeatedly over the approach to a conjunction  and the collision probability is updated accordingly.}  There are many algorithms to compute the miss distance, often as the root of a polynomial equation \citep{Alfano:1993,Gronchi:2005,Armellin:2010} \red{that depends on the relative path of the objects, which may be highly nonlinear when they are far apart.}

 In this work, we formulate a statistical model for conjunction assessment that resolves the apparent difficulties with the collision probability and suggest that the miss distance is a more appropriate focus of interest.  We discuss inference for this distance based on standard likelihood theory \citep[chapter 9]{Cox.Hinkley:1974}, and also describe an improved theory that is both highly accurate and should give results similar to a Bayesian formulation \citep{Brazzale.Davison.Reid:2007}. Our approach is based on significance functions \citep{Hjort:2020} and provides both point and interval estimates for the miss distance, with the intervals containing the true miss distance with a specified probability under the model.  We can also test whether the true distance is higher than the safety threshold, in order to inform decisions about avoidance manoeuvres \citep{Neyman:1937}.
 
The paper is organized as follows. In Section~\ref{conjunction.sec} we formulate the conjunction assessment problem in statistical terms and discuss the relationship between the conjunction probability and miss distance.  \red{We point out that the conjunction probability tends to be biased downwards} and elucidate the so-called dilution paradox of the former.  In Section~\ref{inference.sect}, we introduce significance functions and discuss how they provide calibrated frequentist inference, present elements of modern likelihood inference and link them to the Bayesian approach.  In Section~\ref{likelihood.sec} we apply these ideas to the satellite conjunction problem and in Section~\ref{numerical.sect} study their properties in Monte Carlo studies and case studies.  Section~\ref{further.sect} \red{comments on possible extensions and limitations of the approach}.

\section{Statistical modeling of conjunction assessment}\label{conjunction.sec}

\subsection{Problem formulation}

Modelling and analysis of the relative motion of two space objects has been successfully applied to many space missions \citep{Klinkrad:2006}. Algebraic models for relative errors were proposed by \citet{Hill:1878} and \citet{Clohessy:1960}  and developed in \citet{Chen.etal:2017}.   Many authors \citep{Foster:1992,Chan:1997,Alfriend:1999aa,Alfano:2005,Patera:2005} have considered the two objects as ellipsoids and attempted to estimate the probability that they will collide.  As mentioned above, this probability has been considered by some to have paradoxical properties \citep{Balch:2019}, but these evaporate when the problem is formulated using a statistical model, as we shall see.  Recall that a parametric statistical model treats the available data $y$ as the realisation of a random variable whose probability density function $f(y; \vartheta)$ is determined by unknown parameters $\vartheta$, and the \red{usual goal is to use $y$ for inference about the value of a scalar parameter $\psi=\psi(\vartheta)$.}  Below we take $\psi$ to be the miss distance and suppose that a collision occurs if the two objects pass within a minimum distance of each other, namely the combined hard-body radius, $\psi_{\rm min}$.  \red{We assume that the conjunction is sufficiently close that the relative motion can be taken to be linear; \citet{Hall:2021} summarises the literature on probability calculation for nonlinear and repeated conjunctions.}

Suppose initially that the current positions $\mu_{s1}$ and $\mu_{s2}$ and velocities  $\nu_{s1}$ and $\nu_{s2}$ of the two objects are known; all these are $3\times 1$ vectors.  Define  $\mu=\mu_{s2}-\mu_{s1}$ and $\nu=\nu_{s2}-\nu_{s1}$, so the second object is considered relative to an origin at the first.  In this frame of reference and under linear  relative motion, the second object traverses the line $\mu + t\nu$, where $t\in\Real$.  Its distance from the origin, $(\mu + t \nu)^\T(\mu + t \nu)$,  is minimised by choosing $t = -\nu^\T\mu/\nu^\T\nu$, at which point the minimum squared distance is 
$
\psi^2=\mu^\T \mu-\left(\mu^\T \nu\right)^{2} / \nu^\T\nu  
$.  
In terms of spherical polar coordinates we have
\begin{eqnarray}
\label{mu.eq} \mu&=&\| \mu \| \left(\sin\theta_1\cos\phi_1, \sin\theta_1\sin\phi_1,\cos\theta_1\right)^\T,\\
\label{nu.eq} \nu &= &\|\nu\|\left(\sin\theta_2\cos\phi_2, \sin\theta_2\sin\phi_2,\cos\theta_2\right)^\T,
\end{eqnarray}
where $\|\cdot\|$ denotes the Euclidean norm and $0\leq \theta_1,\theta_2\leq \pi$ and $0\leq \phi_1,\phi_2<2\pi$ are  the polar and azimuthal angles for $\mu,\nu$. 
The minimum distance between the two objects, the miss distance, is 
\begin{equation}
\psi = \|\mu\| (1-\cos^2\beta)^{1/2}  = \|\mu\|  |\sin\beta|,
\label{min_distance}
\end{equation}
where  $\beta$, the angle between the location and velocity vectors $\mu$ and $\nu$, satisfies 
\begin{equation}
\cos\beta = \sin\theta_1\cos\phi_1\sin\theta_2\cos\phi_2+ \sin\theta_1\sin\phi_1\sin\theta_2\sin\phi_2 + \cos\theta_1\cos\theta_2. \label{cosalpha}
\end{equation}
When $\beta=0$ we distinguish two cases: if $\mu^\T\nu<0$ the second object will pass through the origin, leading to a collision, whereas if $\mu^\T\nu>0$ the second object is heading away from the origin, so its current position is the closest it will come to the first object. 
More generally, $\psi<\|\mu\|$ only if $\cos\beta <0$, i.e., $\pi/2< \beta < 3\pi/2$.  For $\mu$ and $\nu$ to be collinear but pointing in opposite directions we need $\phi_2 = \pi + \phi_1$ and $\theta_2=\pi-\theta_1$, and then $\cos\beta = \cos(\theta_1+\theta_2) = -1$, so $\beta=\pi$ and hence $\psi=0$, as expected.  To lighten the notation below we write $\lambda=( \theta_1,\phi_1,\|\nu\|,\theta_2,\phi_2)$, and $\vartheta=(\psi,\lambda)$. 

In the above deterministic setting the collision probability $p\equiv p(\vartheta)$ takes two possible values,
\begin{equation}\label{pc.eq}
p(\vartheta) = \begin{cases} 0, & \psi>\psi_{\rm min},\\ 1, &0\leq  \psi\leq \psi_{\rm min}, \end{cases}
\end{equation}
and a decision-maker can make an ideal decision.  In reality, of course, both $\mu$ and $\nu$ are observed with error, and we follow the literature and assume that the available observations on the positions and velocities of the two objects have a multivariate normal distribution with known covariance  matrix.  If so, then  the vector $y$ containing the observed position and velocity of the second object relative to the first has a six-dimensional normal distribution, and we suppose that this has mean vector $\eta(\vartheta)=\left\{\mu(\psi, \lambda)^\T,\nu(\lambda)^\T \right\}^{\T}$ given by equations~\eqref{mu.eq}--\eqref{cosalpha} and known $6\times 6$ covariance matrix $\Omega^{-1}$, whose inverse  $\Omega$ is known as the dispersion matrix.   \red{In reparametrizing the six-dimensional model to $(\psi, \lambda),$ it is important that the new parameters are variation independent. If the allowable values of $\psi$ were to depend on $\lambda$, this would introduce irregularities into the statistical model and the usual asymptotic theory would not apply. Such problems arise when the miss distance $\psi$ is expressed in Cartesian coordinates.}

\subsection{\red{Two-dimensional case}}\label{2D.section}

\red{A simplified version of the problem treats the relative velocity vector $\nu$ as known.  In this case the last three components of $y$ and of $\eta(\vartheta)$ corresponding to the relative velocity can be dropped, and only the $3\times 3$ corner of  $\Omega^{-1}$ pertaining to the relative position need be retained.  The conjunction can then be visualised in the encounter plane, also called the conjunction plane, which is normal to $\nu$ and passes through the origin.  The observed position $y$ of the second object, its true position $\mu$, and the density of $y$, can be projected orthogonally in the direction of $\nu$ into this plane, leading to respective projected observed and true positions $x$ and $\xi$ and bivariate normal density $f(x;\xi)$ for $x$ with unknown mean vector $\xi$ and known diagonal covariance matrix $D=\diag(d_1^2,d_2^2)$; see Appendix~A.  As shown in the left-hand panel of Figure~\ref{fig:fig1},  $\xi$ and $x$ are two-dimensional vectors: the true point at which the second object will pass through the encounter plane, $\xi$, is at a distance $\psi=\|\xi\|$ from the origin, and the collision probability is zero unless $\psi\leq \psi_{\rm min}$.  The right-hand panel of Figure~\ref{fig:fig1} shows how the collision probability is computed: the density is assumed to be centered at $x$ and the estimator $\hat p_c$ is the integral of this density over the disk of radius $\psi_{\rm min}$ around the origin.  Thus we can write $\hat p_c = p_c(x)$, where 
\begin{equation}\label{pc-defn}
p_c(\xi) = \int_{\{x': \|x'\|\leq \psi_{\rm min}\}}  f(x';\xi)\, \D{x'}.
\end{equation}
One might regard $p_c(x)$ as an estimate of $p_c(\xi)$, since the unknown $\xi$ is replaced with the known $x$ in computing the integral.  Seen through the lens of the statistical model, however, it is not obvious why $p_c(\xi)$ is of interest, as it is the probability that the noisy observation $x$ will appear to pass inside the hard-body radius, rather than the probability~\eqref{pc-defn} that the satellite itself will do so.  Moreover the symmetry of $f(x;\xi)$ around $\xi$ shown in the left-hand panel of Figure~\ref{fig:fig1} implies that there is more probability outside the circle of radius $\psi$ than inside, so $x$ will tend to be further away from the origin than $\xi$. Another way to see this is to note that the Euclidean norm of $x$, $\|x\|$, satisfies 
$$
\E( \|x\|^2) = \E(x_1^2+x_2^2)= \xi_1^2+\xi_2^2+d_1^2+d_2^2 = \psi^2\left\{1+(d_1^2+d_2^2)/\psi^2\right\}, 
$$ 
and hence $\E\{\|x\|\}\approx\psi\left\{1+(d_1^2+d_2^2)/\psi^2\right\}^{1/2}$ for large $\psi$ , i.e., the mean length of $x$ exceeds $\psi$ by an amount that depends on $\xi$ and $D$.  Hence $\hat p_c= p_c(x)$ will tend to be smaller than $p_c(\xi)$.  Figure~\ref{fig:fig2} shows the effect of this in a test case described more fully in Section~\ref{CARA.sec}.  Each panel takes $\xi=(11.84, -1.36)^\T$m, so $\psi= 11.92$m, and compares the values of $p_c(\xi)$ with  boxplots showing the values of $p_c(x)$ for $20,000$ values of $x$ generated as bivariate normal, $\N_2\{\xi, c^2\diag (d^2_1,d^2_2)\}$, with different values of $c^2$. The probability of collision is computed using a contour integral transformation \cite{Patera:2005}, which is then approximated numerically using the trapezoidal rule. The left-hand panel for $c^2=10^{-2},$ shows that although $p_c(\xi) \approx10^{-3}$, around 25\% of the values of $p_c(x)$ are less than $10^{-4}$.  Overall the values of $p_c(x)$ can vary over many orders of magnitude and can be much lower than $p_c(\xi)$.  Figure~\ref{fig:fig3}, which shows the values of $\xi$ and $10^3$ simulated values of $x$ for three values of $c^2=10^{-2},1,2$,  illustrates how $p_c(x)$ varies greatly when $\psi_{\rm min}=5$m and when $\psi_{\rm min}=20$m. Clearly the bias is also present when the velocity is not treated as known.}


\begin{figure}[t]
\centering
\includegraphics[width=\textwidth]{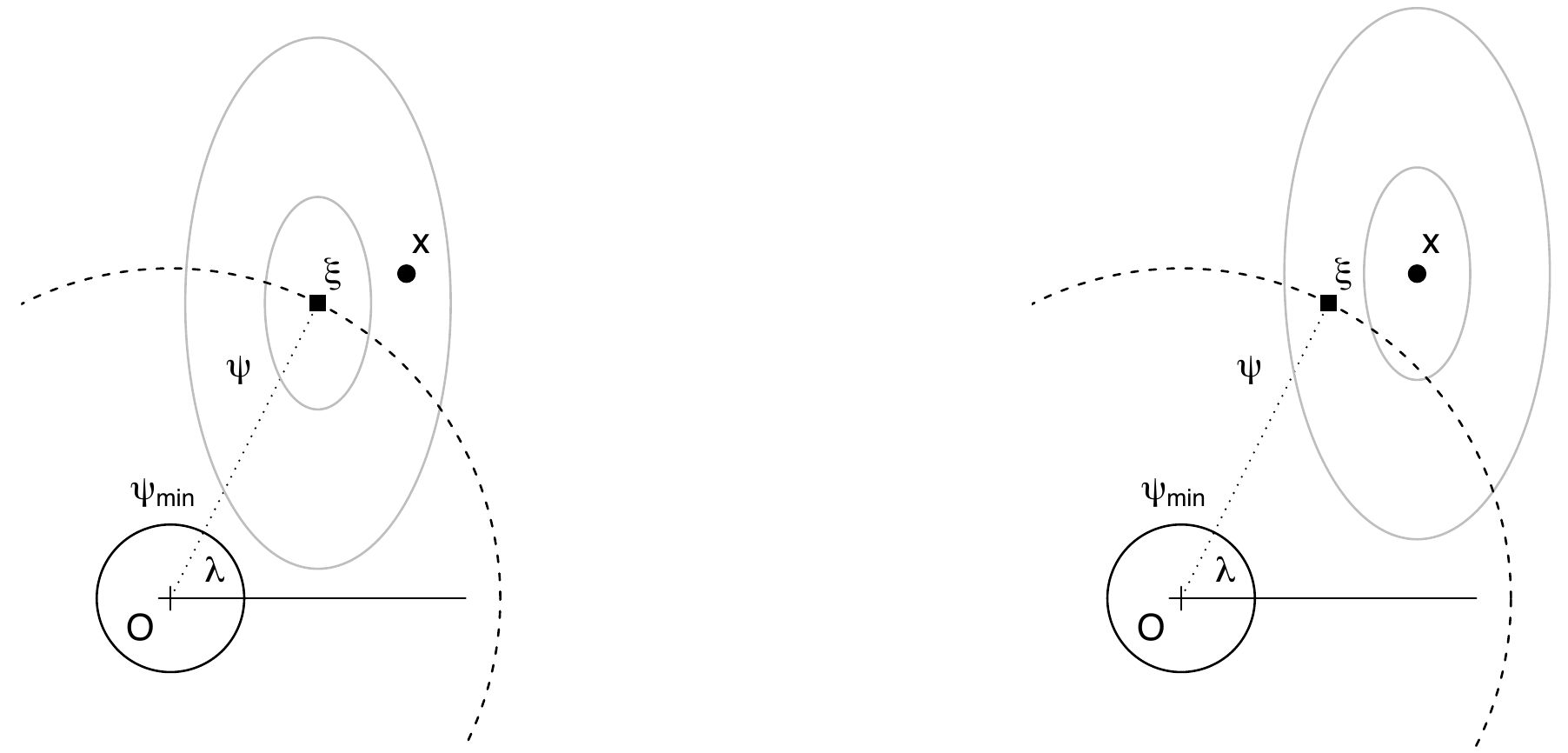}
 \caption{Statistical formulation of satellite conjunction in the encounter plane.  The primary object is at the origin $O$ and the solid circle around it has radius $\psi_{\rm min}$, the combined hard-body radius.  The true position at which the second object will cross the encounter plane, $\xi$, can be expressed in terms of the angular coordinates $\psi$ and $\lambda$.  The noisy observation of the second object will cross the encounter plane at $x$.  Left panel: the ellipses indicate the true density of $x$, with mean at $\xi$.  Right panel: the ellipses indicate the assumed density of $x$ when computing the conjunction probability estimate $\hat p_c$.}
 \label{fig:fig1}
\end{figure}

\begin{figure}[t]
\centering
\includegraphics[width=0.49\textwidth]{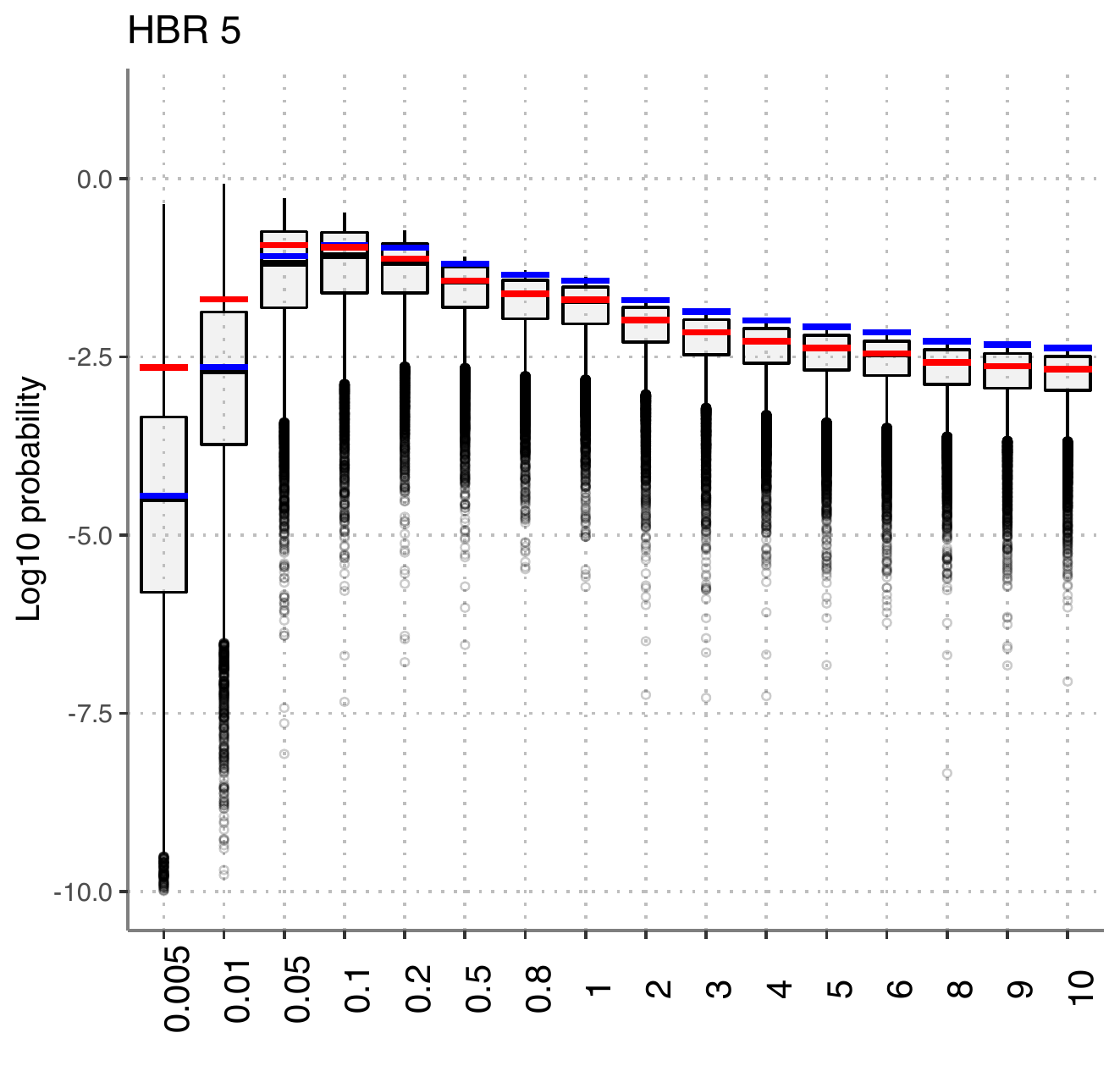}
\includegraphics[width=0.49\textwidth]{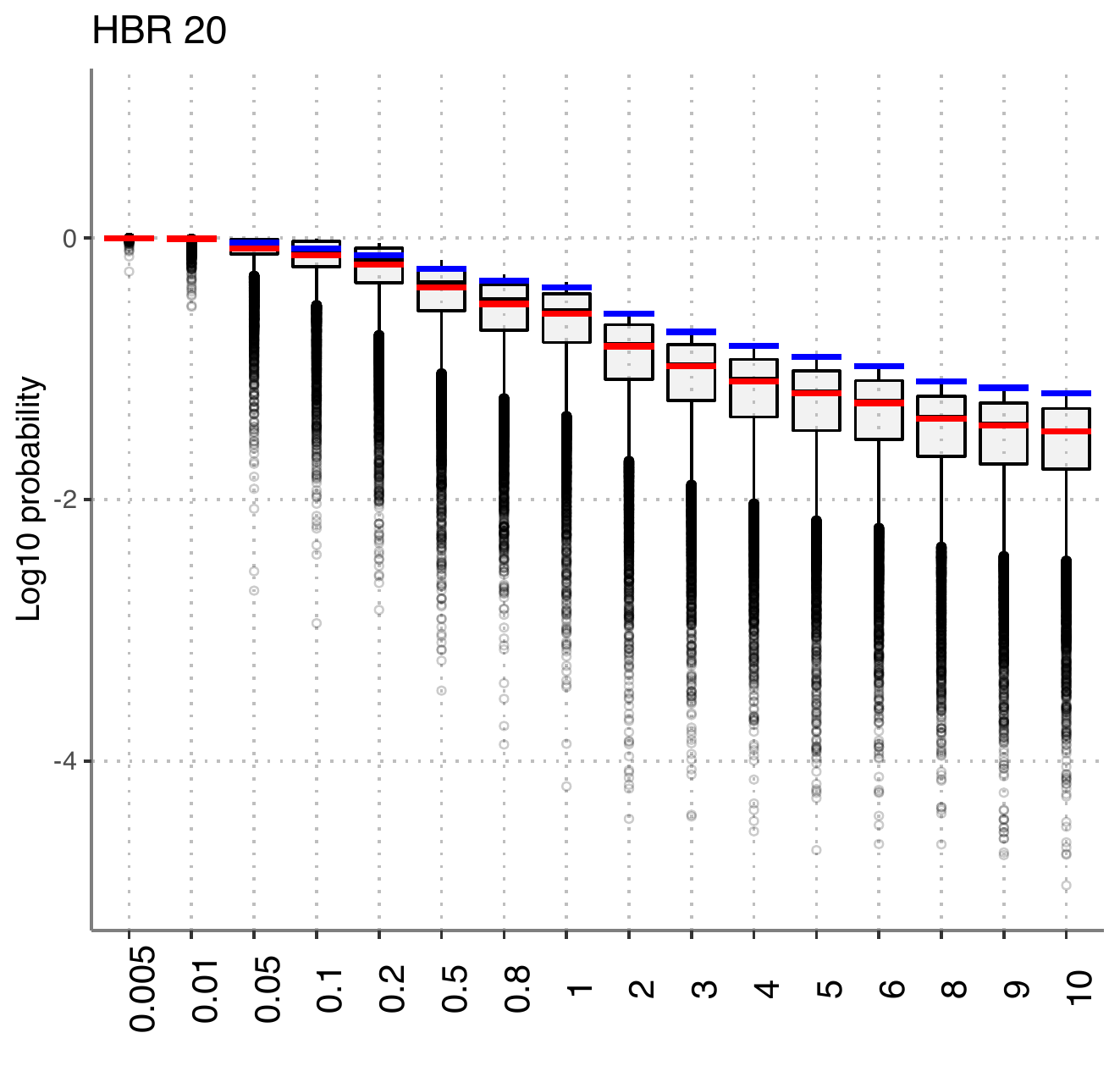}
 \caption{Behaviour of $\hat p_c = p_c(x)$ for hard-body radius $\psi_{\rm min}=5$m (left) and 20m (right) in Case Study C, with $\psi\approx 11.92$m, as a function of the variance of  $x\sim \N_2\{\xi, c^2\diag (d^2_1,d^2_2)\}$.  For each $c^2$ in $0.005, 0.01, \ldots, 10$ we computed $p_c(\xi)$ (blue segments) and $p_c(x)$ (boxplots) for 20,000 simulated values of $x$.  The average values of $p_c(x)$ are shown by the red segments.  In the left-hand panel $\psi>\psi_{\min}$, so $p_c(\xi)\to 0$ as $c\to 0$, whereas in the right-hand panel $\psi<\psi_{\min}$, so $p_c(\xi)\to 1$ as $c\to 0$.}
 \label{fig:fig2}
\end{figure}

\begin{figure}
     \centering
     \begin{subfigure}[b]{0.45\textwidth}
         \centering
         \includegraphics[width=\textwidth]{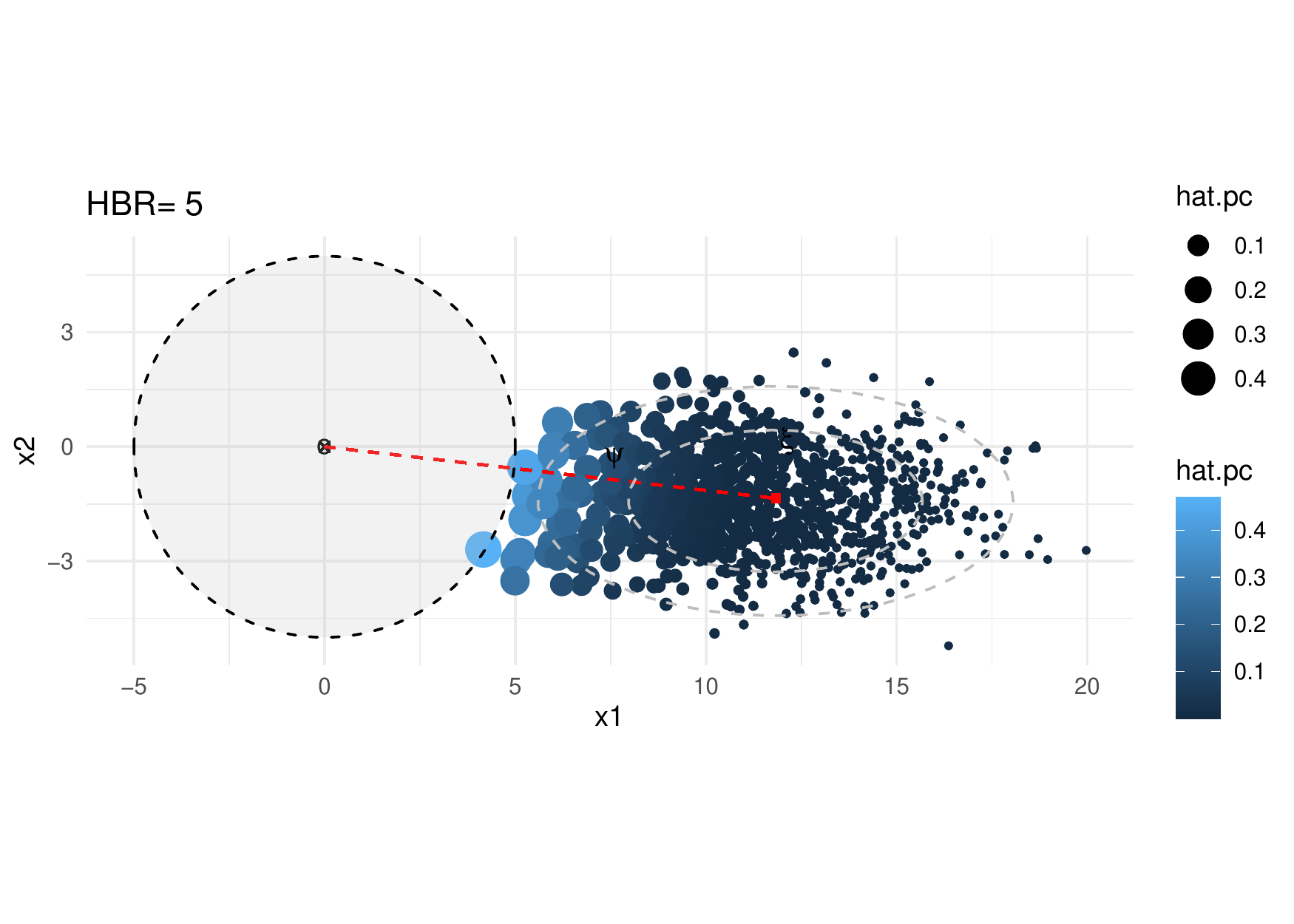}
     \end{subfigure}
     \begin{subfigure}[b]{0.45\textwidth}
         \centering
  \includegraphics[width=\textwidth]{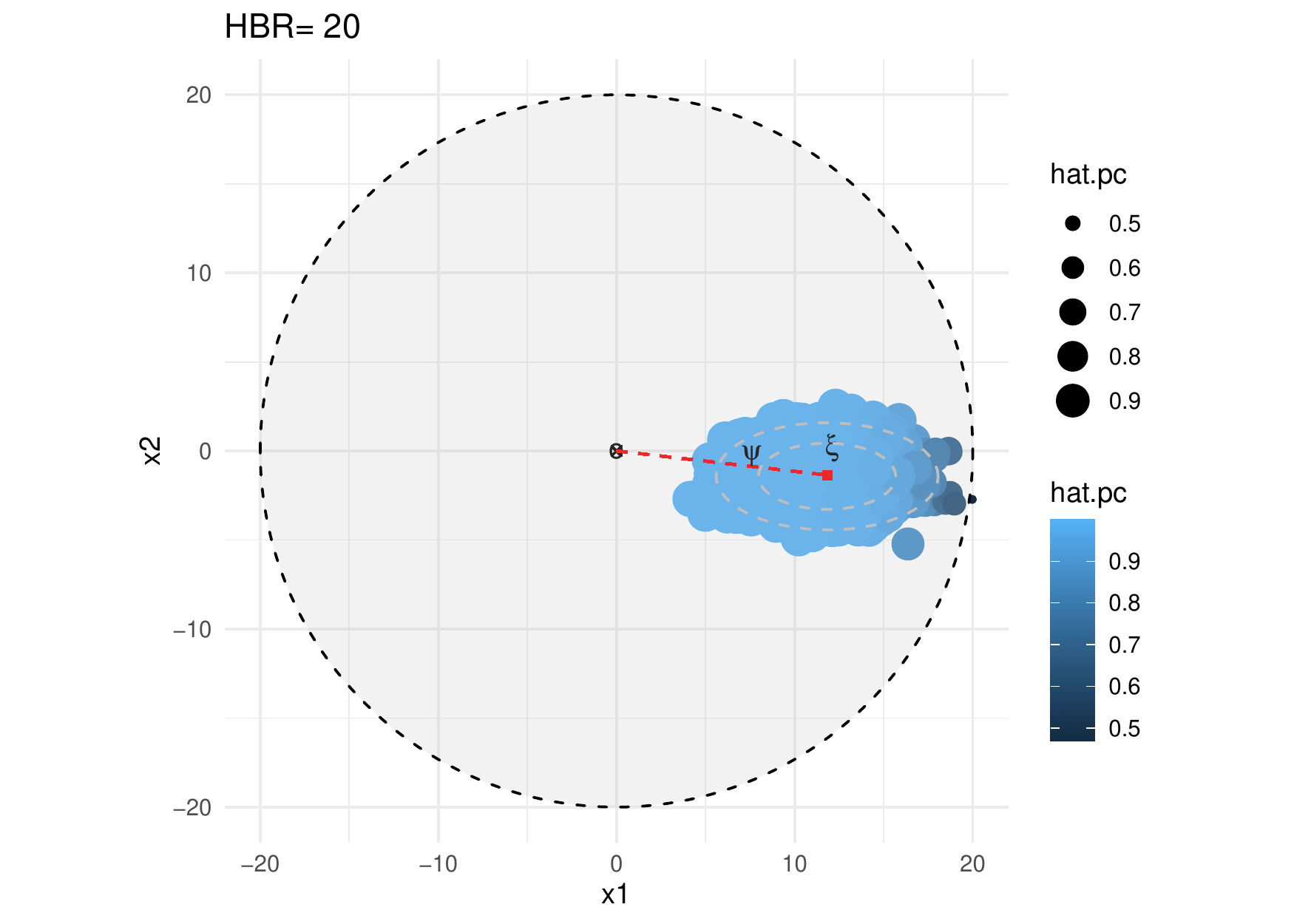}
     \end{subfigure}\\
    
     \begin{subfigure}[b]{0.45\textwidth}
         \centering
         \includegraphics[width=\textwidth]{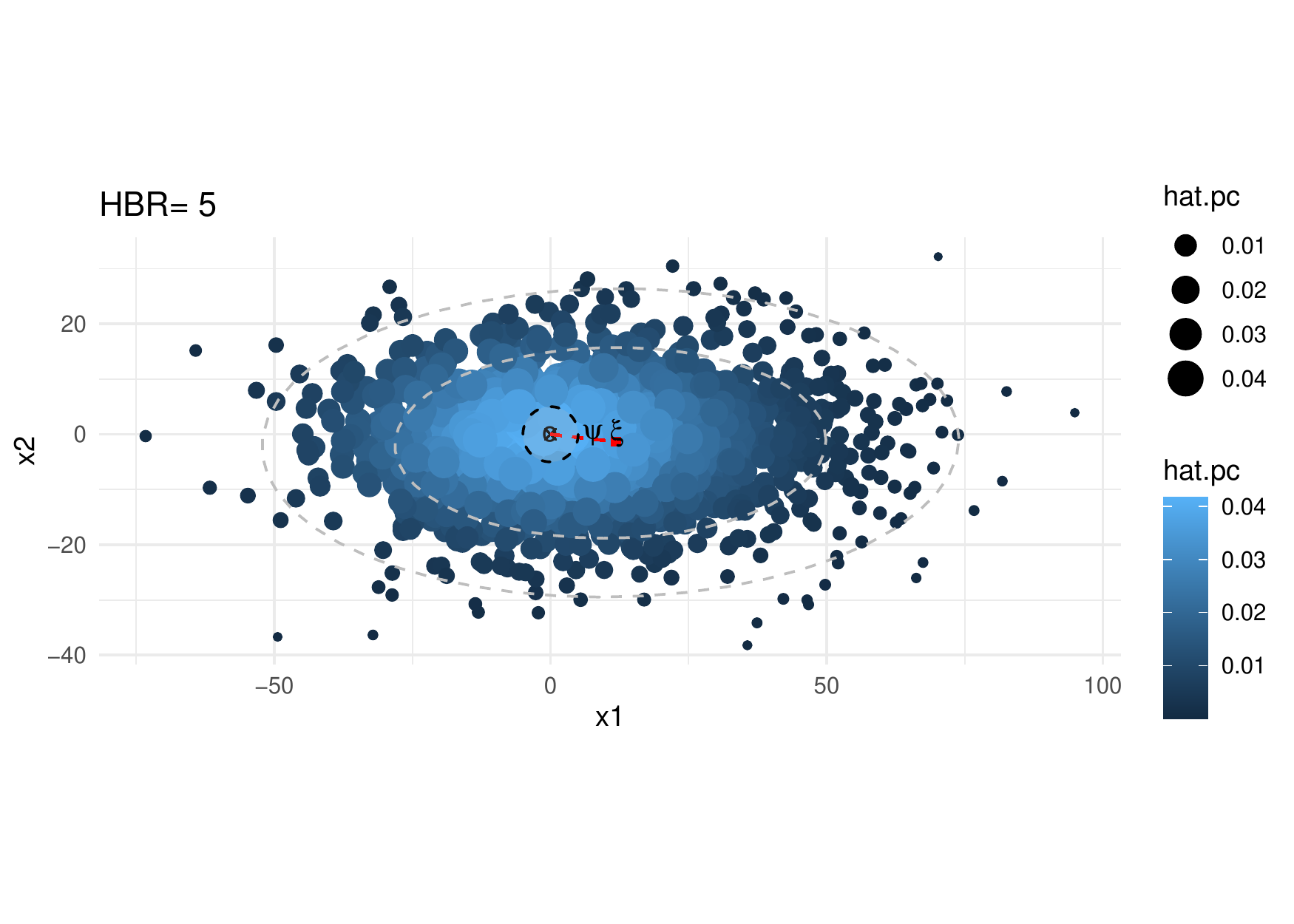}
         \label{fig:y equals x}
     \end{subfigure}
     \begin{subfigure}[b]{0.45\textwidth}
         \centering
  \includegraphics[width=\textwidth]{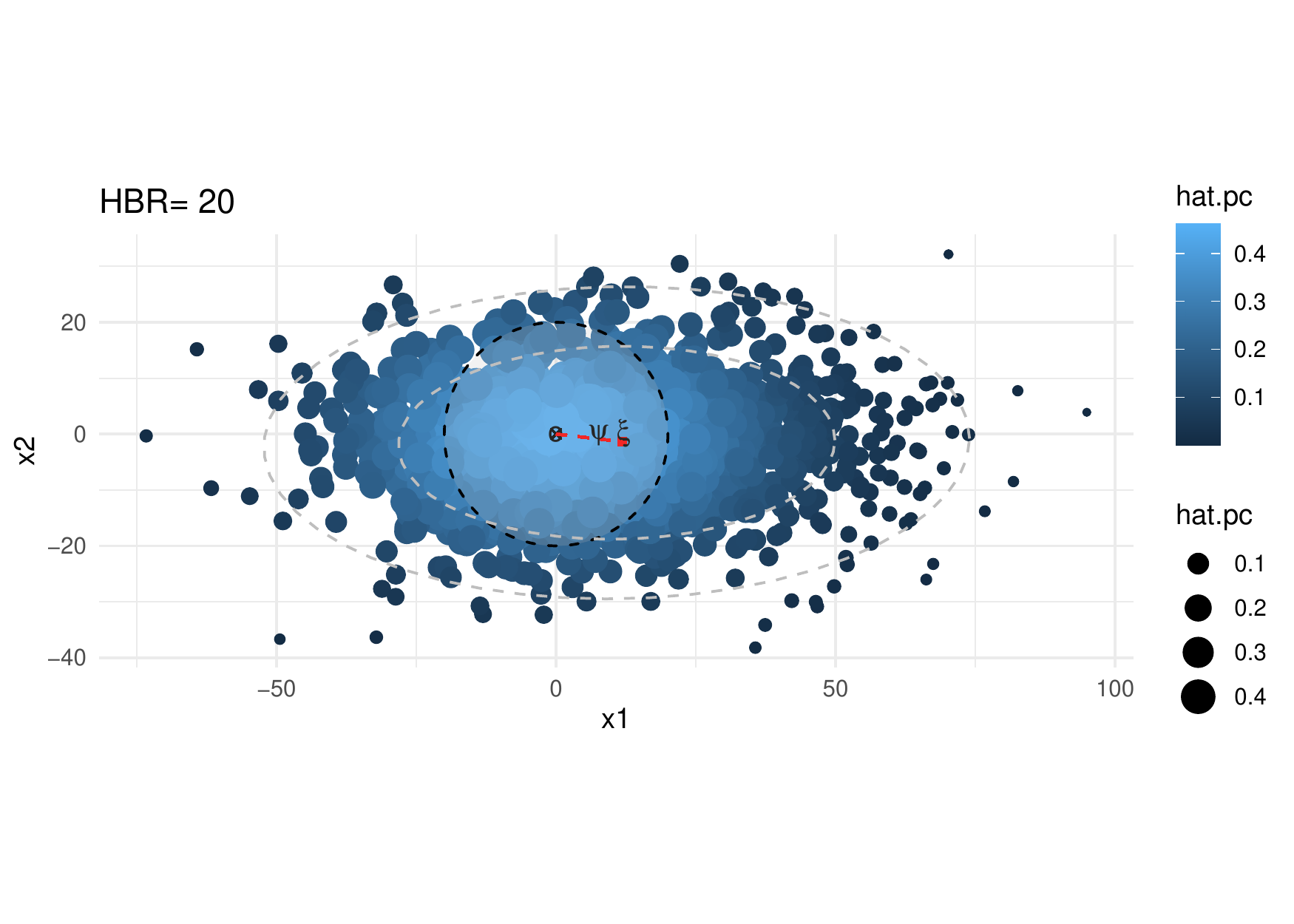}

     \end{subfigure}
          \begin{subfigure}[b]{0.45\textwidth}
         \centering
         \includegraphics[width=\textwidth]{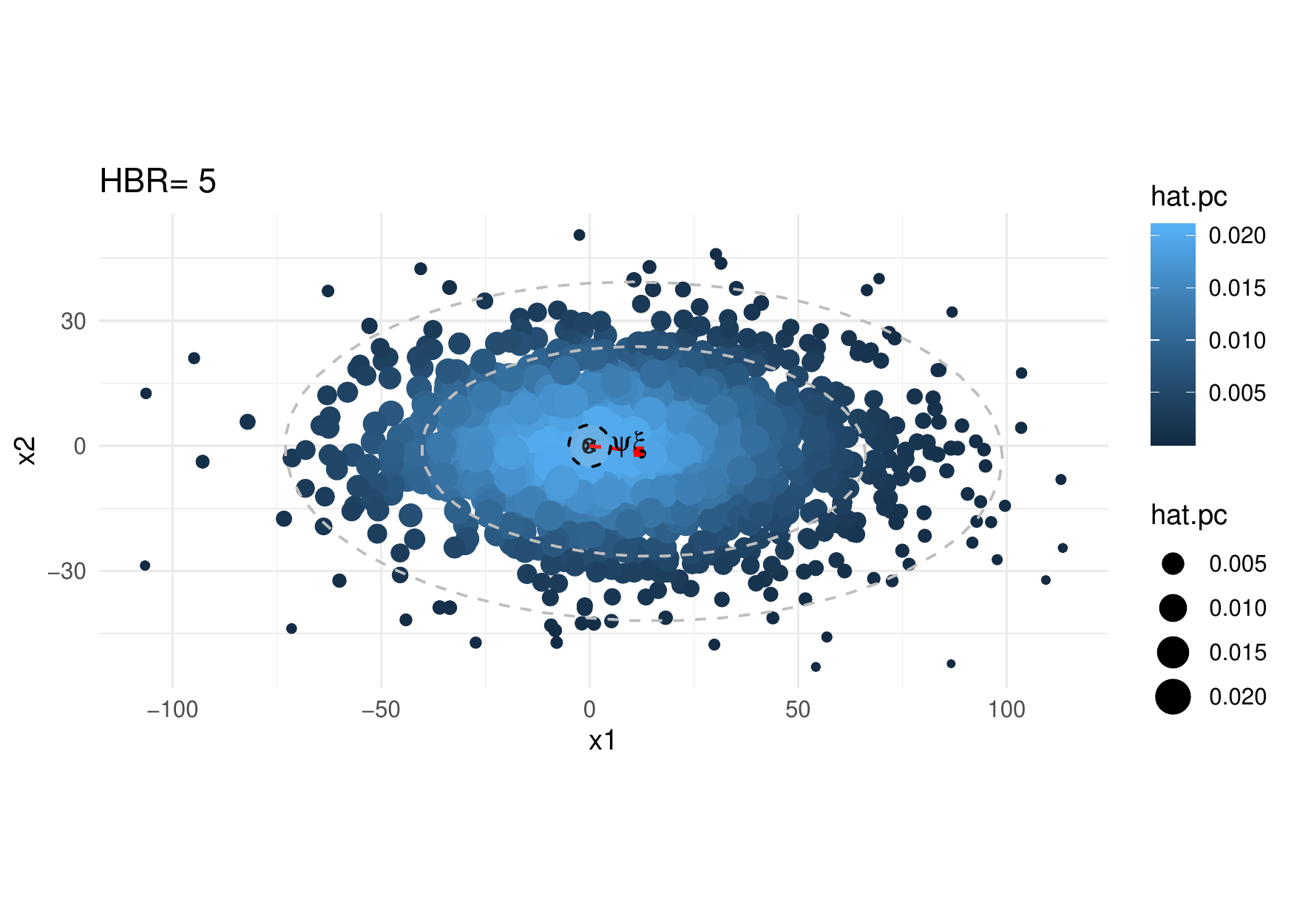}
     \end{subfigure}
     \begin{subfigure}[b]{0.45\textwidth}
         \centering
  \includegraphics[width=\textwidth]{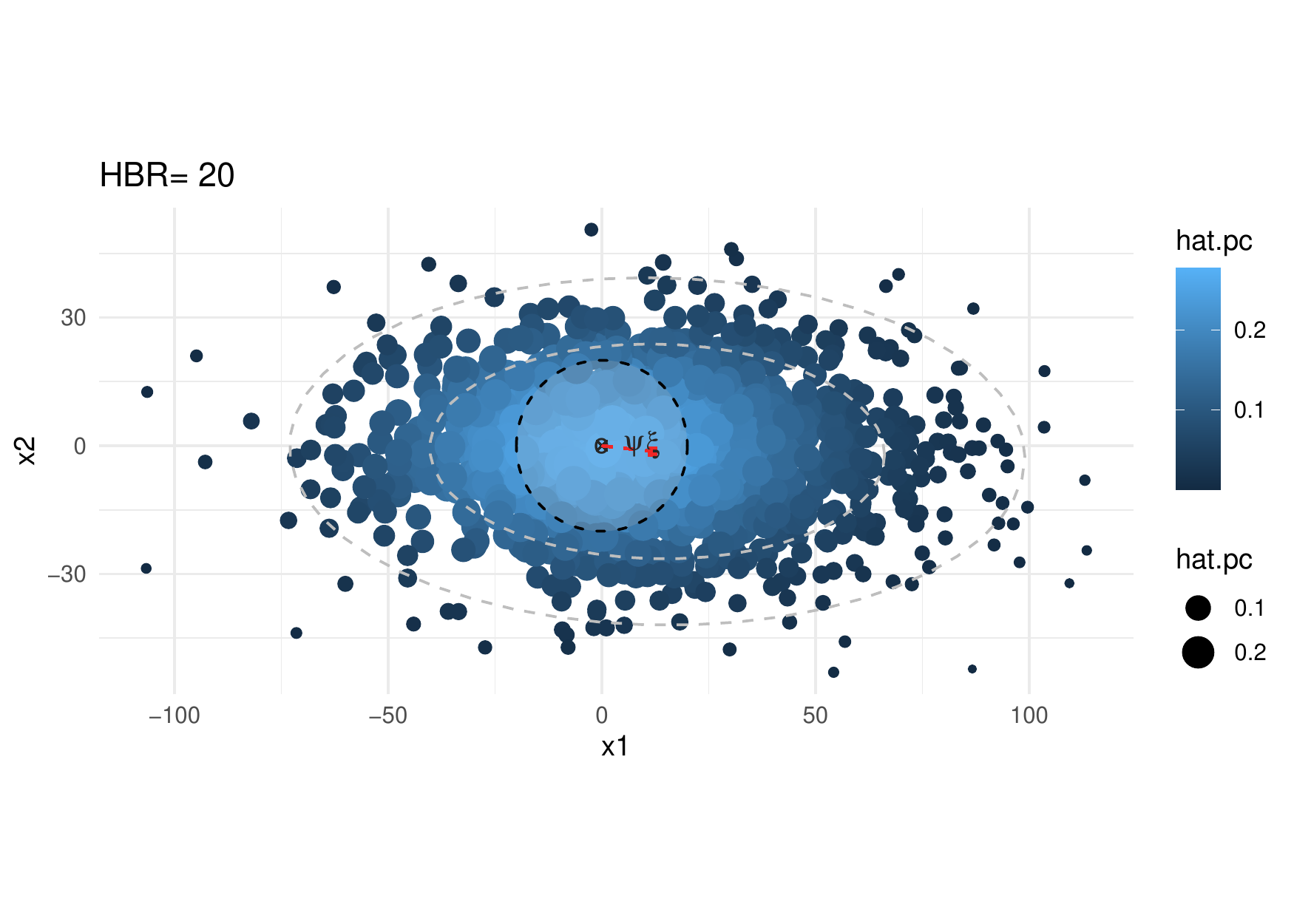}

     \end{subfigure}
 \caption{Scatter plots showing 1000 observed locations in the conjunction plane sampled as $x\sim\mathcal{N}_2\left\{\xi,c^2\text{diag}(d_1^2,d_2^2)\right\}$ with $c^2=10^{-2},1$, and $2$ (top to bottom). The true position of the second object, $\xi$, is shown by the red point and the observed positions, $x$, are shown by blue points whose sizes and shades depend on the value of $p_c(x)$.  The shaded grey disks show the hard-body radius, $\psi_{\rm min}=5$~m on the left and $\psi_{\rm min}=20$~m on the right.}
        \label{fig:fig3}
\end{figure}

\red{Despite the comments above, one reason to use $\hat p_c$ is that it is a Bayesian estimator.  If a prior density $f(\xi)$ for $\xi$ is placed on the encounter plane, then the posterior probability that $\xi$ lies within the hard-body radius is given in terms of the posterior density of $\xi$, i.e., 
$$
 f(\xi\mid x) = {f(x;\xi)f(\xi)\over \int f(x;\xi)f(\xi) \, \D{\xi}}  , 
 $$
 by
\begin{equation}\label{Bayes.pc}
\pr(\psi\leq \psi_{\rm min}\mid x) = \pr(\|\xi\|\leq \psi_{\rm min}\mid x) = \int_{\{\xi: \|\xi\|\leq \psi_{\rm min}\}} f(\xi\mid x)\, \D{\xi}, 
\end{equation}
and if $f(\xi)$ is constant and $f(x;\xi)$ is bivariate normal, then  $f(\xi\mid x) = f(x; \xi)$ and~\eqref{Bayes.pc} equals $\hat p_c = p_c(x)$. This explains why  $\hat p_c$ is a plausible estimator of~\eqref{pc.eq}, but does not alter its downward bias.  Moreover a constant prior for $\xi$ is improper, with the undesirable property that the ratio of the probability inside any disk around the origin is zero relative to the probability outside that disk, thus expressing a prior belief that $\xi$ is infinitely far from the origin, i.e., the second object will traverse the encounter plane  infinitely far from the first.  This gives an alternative explanation of the behaviour seen in Figure~\ref{fig:fig2}.}

\subsection{Conjunction probability or miss distance?}

Two further points are immediately clear from the above very simple statistical model for satellite conjunction.

First, the conventional target of inference, the true collision probability~\eqref{pc.eq}, \red{depends on the unknown relative position and velocity of the two objects, but not on the covariance matrix $\Omega^{-1}$ or the data $y$.  The collision probability is generally estimated by $\hat p_c$, which depends on both $\Omega^{-1}$ and $y$}.   The `paradoxical' behaviour whereby $\hat p_c$ is very tiny when the data have a very large variance and then increases when that variance decreases is the behaviour of an estimator, not of a parameter of the model.  The estimator $\hat p_c$ depends on $\Omega^{-1}$, and as the variance decreases we expect that either $\hat p_c\to 1$, if a collision will occur, or $\hat p_c\to 0$, otherwise; in both cases $\hat p_c\to p(\vartheta)$, as we should expect when the data become noiseless.  Thus probability dilution is the natural behaviour of an estimator in response to changes in the variability of the underlying data, not a probability paradox.  

Second, the fact that $p(\vartheta)$ takes just two values whereas the miss distance $\psi$ takes values in a continuum suggests that $\psi$ is a better overall target of inference.  For example, the maximum likelihood estimator of $p(\vartheta)$ is $I(\hat\psi\leq \psi_{\rm min})$, where $I(\cdot)$ and $\hat\psi$ denote the indicator function and the maximum likelihood estimator of $\psi$, and clearly $\hat\psi$ is more informative.  Moreover in a Bayesian framework the unknown parameters are regarded as random variables and  the posterior probability of collision given the data is obtained by integrating over the posterior density of $\psi$ given $y$; c.f.~\eqref{Bayes.pc} when $\nu$ is known.  In this setting also the collision probability is based on the available knowledge about the miss distance $\psi$, which is the more fundamental quantity.  
In Section~\ref{bayes.sec} we discuss Bayesian inference in more detail, and   in Section~\ref{likelihood.sec} we show how inference on $\psi$ provides a significance probability with an interpretation akin to that of $\hat p_c$. 

Motivated by these considerations, we turn to inference on $\psi$ based on the observed value of $y$.  If the data suggest that $\psi$ is lower than a safety threshold $\psi_0$, possibly with $\psi_0>\psi_{\rm min}$ for a safety margin, then action to avert a collision should be considered.  
Our goal below is therefore inference on the unknown miss distance $\psi$,  allowing for the fact that $\lambda$ is also unknown and must be estimated; in the six-dimensional case we can write the relative distance vector using~\eqref{mu.eq} but with $\|\mu\|$ replaced by $\psi/|\sin\beta|$, for $\beta\neq \pi$.    \red{In statistical terms the scalar $\psi$ is the primary object of inference, the so-called interest parameter, whereas the $5\times 1$ vector $\lambda$ of nuisance parameters, while essential for realistic modelling, is of only secondary concern.  In the special case with known velocity vector $\nu$ and considering only the encounter plane, the miss distance $\psi$ remains the parameter of interest and its interpretation is unchanged, but the nuisance parameter $\lambda$ is scalar.  This simplifies the problem but the statistical issue remains the same.}

We return to this model in Section~\ref{likelihood.sec} after outlining elements of the theory of inference.  

\section{Inference}\label{inference.sect}

\subsection{Calibration, \red{decisions} and significance functions}\label{sig.sect}

Statistical inference involves statements about the properties of a probability distribution that is assumed to have given rise to observed data $y^\o$.  \red{In the simplest setting the distribution depends only on a scalar parameter $\psi$, which is the focus of interest,}  and the likelihood function $L(\psi)=f(y^\o;\psi)$ is used to compare the plausibility of different values of $\psi$ as explanations for $y^\o$.  The best-fitting model is provided by the maximum likelihood estimate based on $y^o$, $\hat\psi^\o$, and the relative likelihood function $L(\psi)/L(\hat\psi^o)$, which has maximum value 1, allows values of $\psi$ to be compared.  A `pure likelihood' approach \citep[e.g.,][Chapter 3]{Edwards:1972} treats any $\psi$ for which $L(\psi)\geq u\,L(\hat\psi^o)$ as plausible, but with $u$ chosen essentially arbitrarily.  In practice further information is typically \red{required in order to choose $c$}.  

Bayesian inference treats $\psi$ as a random variable and \red{is based on} a density $f(\psi)$ that weights values of $\psi$ according to their plausibility prior to seeing the data. This is updated in light of the observed data $y^\o$ using Bayes' formula, resulting in the posterior distribution function
\begin{equation}\label{bayes.eq}
\pr(\psi\leq \psi_0\mid y^\o) = {\int_{-\infty}^{\psi_0} L(\psi) f(\psi)\D{\psi}\over \int_{-\infty}^{\infty} L(\psi) f(\psi)\D{\psi}}.
\end{equation}
Clearly this calculation depends on the prior density $f(\psi)$; if this is badly chosen then~\eqref{bayes.eq} may have poor properties when used repeatedly.  \red{The most obvious choice of prior in the satellite conjunction setting is uniform on the position of the secondary space object, but this has the undesirable properties mentioned in Section~\ref{2D.section}.}

\red{When the losses due to possible evasive actions can be specified, the data can be used to choose the action that minimises the expected posterior loss.  For the simplest possible formulation in the collision avoidance context, suppose that $\psi$ represents the unknown miss distance, that the two actions  $a=0$ and $a=1$ correspond to `do nothing' and `take evasive action', and that the loss $l_{ae}$ when action $a$ is taken and event $e$ occurs is as given in Table~\ref{costs.tab}; the loss in doing nothing in the case of no collision is zero. If $\psi\leq \psi_0$ results in a collision, then the posterior expected loss due to taking action $a\in\{0,1\}$ is
$$
l_{a0}\,\pr(\psi>\psi_0\mid y^\o) + l_{a1} \,\pr(\psi\leq \psi_0\mid y^\o) = l_{a0}(1-p) + l_{a1}p, 
$$
say, which is minimised by doing nothing if $l_{01}p<l_{10}(1-p)+l_{11}p$.  Equivalently, evasive action should be taken if $\pr(\psi\leq \psi_0\mid y^\o) \geq  l_{10}/(l_{10}+l_{01}-l_{11})$.  Thus if the losses are known explicitly, they provide a threshold for action, and the use of a decision rule such as `take evasive action if the posterior probability of collision exceeds $10^{-4}$' corresponds to an implicit ratio of losses.  This decision setup could be made more realistic, and in any case would only be regarded as guidance in a practical setting;  our point is that it provides a rational basis for considering action when~\eqref{bayes.eq} exceeds a threshold, and explicitly links that threshold to potential losses.}

\begin{table}
\red{\caption{Basic decision analysis for satellite conjunction, with losses $l_{ae}$ corresponding to action $a$ and event $e$.}
\label{costs.tab}
\begin{center}
\begin{tabular}{lcc}
\hline
Action&\multicolumn{2}{c}{Event}\\
\cline{2-3}
&No collision&Collision\\
\hline
Do nothing, $a=0$&$l_{00}=0$&$l_{01}$\\
Evasive action, $a=1$&$l_{10}$&$l_{11}$\\
\hline
\end{tabular}
\end{center}}
\end{table}

\red{Bayesian inference has certain advantages, but other approaches to calibrating the likelihood are often preferred.  Repeated sampling inference invokes hypothetical repetition of the random experiment that is presumed to have led to the observed data \citep[pp.~33--38]{Fisher:1973}.}  In the simplest case $\hat\psi^\o$ is regarded as a realization of a random variable $\hat\psi$ that has a normal distribution, $\N(\psi,\lambda^2)$, under repeated sampling, with $\lambda$ known.  This implies that 
\begin{equation} \label{conf.eq}
\pr(\hat\psi\leq \hat\psi^o;\psi) = \Phi\left\{(\hat\psi^\o-\psi)/\lambda\right\}, 
\end{equation}
where $\Phi$ denotes the standard normal cumulative distribution function.  The significance function~\eqref{conf.eq}, also called the confidence distribution or \Pvalue\ function \citep{Fraser:2017,Fraser:2019,Schweder.Hjort:2016}, is then used for inference on $\psi$.  For example, the null hypothesis that $\psi=\psi_0$ can be tested against the alternative that $\psi>\psi_0$ by computing the significance probability 
\begin{equation}\label{pval.eq}
p_{\rm obs} = \pr\left(\hat\psi\geq \hat\psi^\o;\psi_0\right) , 
\end{equation}
small values of which are regarded as evidence against the null hypothesis in favour of the alternative; here $p_{\rm obs} =1-\Phi\left\{(\hat\psi^\o-\psi_0)/\lambda\right\}$.  Likewise a two-sided $(1-2\alpha)\times 100\%$ confidence interval $\calI_{1-2\alpha}$ for the value of $\psi$ underlying the data, the so-called `true value', has as its upper and lower limits $U_\alpha$ and $L_\alpha$ the solutions to the equations
$$
\pr(\hat\psi\leq \hat\psi^o;U_{\alpha}) =\alpha, \quad \pr(\hat\psi\leq \hat\psi^o;L_\alpha) =1-\alpha,
$$
and this yields $\calI_{1-2\alpha}=(L_\alpha,U_\alpha) = (\hat\psi^\o -\lambda z_{1-\alpha}, \hat\psi^\o -\lambda z_{\alpha})$, where $z_p$, the $p$ quantile of the standard normal distribution, satisfies $\Phi(z_p)=p$ for $0<p<1$.   The limits of $\calI_{1-2\alpha}$ simplify to the familiar $\hat\psi^\o\pm\lambda z_{1-\alpha}$ on recalling that $z_{1-\alpha}=-z_{\alpha}>0$.   In this ideal case the inferences are perfectly calibrated: under repeated sampling with $\psi=\psi_0$ the significance probability $p_{\rm obs} $ has an exact uniform distribution and $\calI_{1-2\alpha}$ contains $\psi_0$ with probability exactly $1-2\alpha$, for any $\alpha\in(0,0.5)$.  When $\psi=\psi_0$, therefore, there is a probability $p_\obs$ that a decision to reject this hypothesis in favor of the alternative based on a significance probability $p_\obs$ will be incorrect. 

Significance functions are decreasing in the parameter and have the properties of survivor functions, so the confidence density based on differentiating~\eqref{conf.eq}, 
\begin{equation}\label{conf-dens.eq}
- {\partial\pr(\hat\psi\leq \hat\psi^o;\psi)\over \partial \psi} , 
\end{equation}
has the formal properties of a probability density function for $\psi$.  The confidence density and its integral, the confidence distribution, can be regarded as frequentist summaries of the information about $\psi$ based on the observed data; see \citet{Schweder.Hjort:2016}.  Unlike the posterior density obtained from~\eqref{bayes.eq}, no prior information is involved, and despite the use of the word `density' for~\eqref{conf-dens.eq}, $\psi$ is regarded as an unknown constant. 

\subsection{Approximate inference}

In Section~\ref{sig.sect}, $\hat\psi$ was assumed to have an exact normal distribution, but the argument there often applies approximately when some measure of precision, typically the sample size $n$, becomes large.  Under mild regularity conditions  the maximum likelihood estimator $\hat\psi$ has an approximate normal distribution centered at the true parameter $\psi$ with variance $1/\jmath(\hat\psi)$, where $\jmath(\psi) = - \partial^2 \log L(\psi)/\partial\psi^2$ is the observed information.  This allows calibration of $\psi$ using the Wald statistic
$$ 
w(\psi) = \jmath(\hat\psi)^{1/2}(\hat\psi-\psi),
$$ 
corresponding to a quadratic approximation to the log likelihood function, or using the likelihood root 
$$ 
r(\psi)=\sign(\hat\psi-\psi)\{2\log L(\hat\psi)-2\log L(\psi)\}^{1/2}.
$$ 
Both $w(\psi)$ and $r(\psi)$ are approximate pivots: they are functions of the data and parameter and have approximate standard normal distributions \red{under repeated sampling} if $\psi$ equals its true value.  The approximations introduce so-called first-order error, of size $O(n^{-1/2})$, in~\eqref{conf.eq}.  For more details, see, for example, Chapter~9 of \citet{Cox.Hinkley:1974}.

\begin{figure}[t]
\centering
  \includegraphics[width=0.47\textwidth]{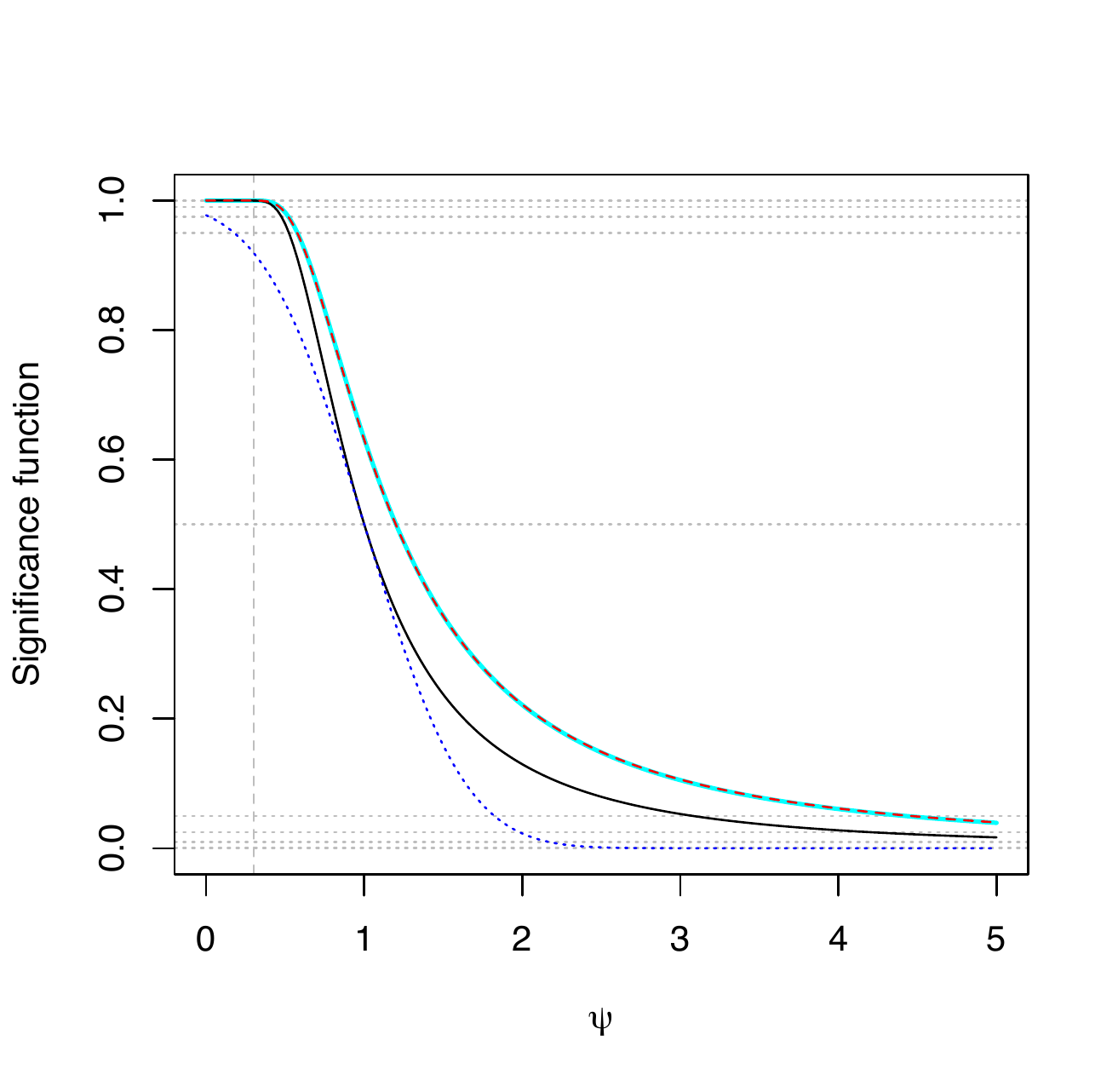}
  \includegraphics[width=0.47\textwidth]{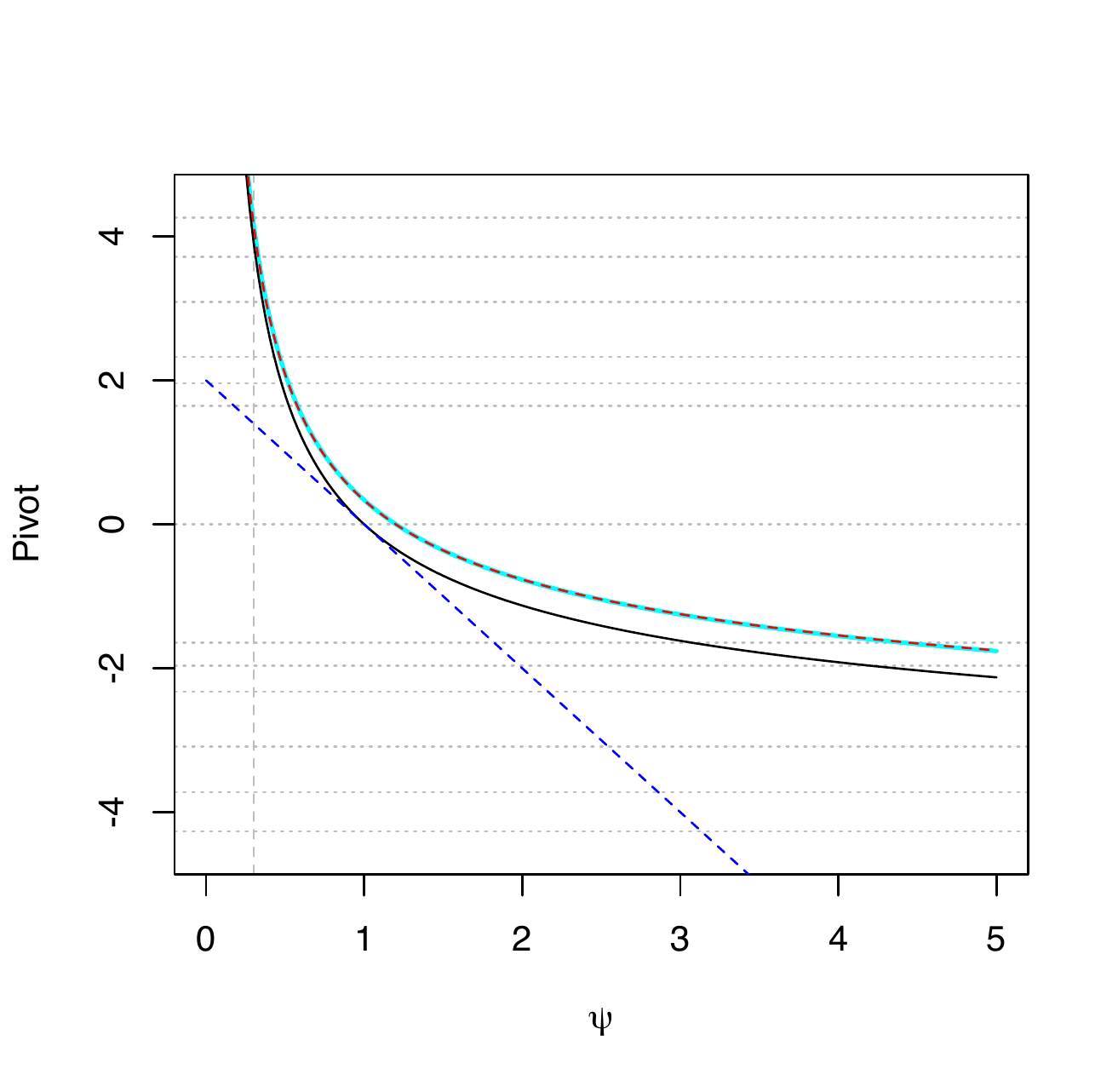}
 \caption{Left: Significance functions based on likelihood root $r^\o(\psi)$ (solid black),Wald statistic $w(\psi)$ (dotted blue), exact (wide cyan) and modified likelihood root $r^{*\o}(\psi)$ (red dashes).  The dashed vertical line shows $\psi_0=0.3$, and the horizontal lines correspond to probabilities $10^{-6}, 10^{-5}$, $10^{-4}$, $10^{-3}$, $10^{-2}$, 0.025, 0.05, 0.5, 0.95, 0.975, 0.999, 0.9999, 0.99999 and 0.999999.  Right: functions from (a) transformed to the standard normal scale.}
\label{fig:fig4}
\end{figure}

To illustrate these approximations, consider a single observation from the Rayleigh distribution, which arises as the length $y$ of a bivariate normal vector whose components are independent $N(0,\psi^2)$ variables.  The probability density function for $y$ is
$$
f(y;\psi) = {y\over \psi^2} e^{-y^2/(2\psi^2)}, \quad y>0, \quad \psi>0, 
$$
and one can readily check that $\hat\psi=y/\surd{2}$, $\jmath(\hat\psi) = 4/\hat\psi^2$ and 
$$
w(\psi) = 2(1-\psi/\hat\psi), \quad r(\psi) =\sign(\hat\psi-\psi)[2\{2\log(\psi/\hat\psi) + (\hat\psi/\psi)^2 -1\}]^{1/2}.
$$
The left-hand panel of Fig~\ref{fig:fig4} shows the significance functions $\Phi\left\{w^\o(\psi)\right\}$ and $ \Phi\left\{r^\o(\psi)\right\}$ when $y^\o=\sqrt{2}$, so $\hat\psi$ has observed value $\hat\psi^\o = 1$.  The functions are decreasing in $\psi$ and the maximum likelihood estimate and the limits of the 90\% confidence interval are the values of $\psi$ for which the functions equal $0.5$ and $0.05, 0.95$, respectively. Evidence against  the null hypothesis $\psi=\psi_0=0.3$ in favor of the alternative $\psi>\psi_0$ is found from the \Pvalue~\eqref{pval.eq} given by the intersections of the significance functions with the vertical line $\psi=0.3$; for the Wald statistic we obtain $1-\Phi\left\{w^\o(\psi_0)\right\}=0.0808$ and for the likelihood root we obtain $1- \Phi\left\{r^\o(\psi_0)\right\}=4.33\times 10^{-5}$. \red{Thus these pivots give quite different evidence about $\psi$.  
The right-hand panel of Fig~\ref{fig:fig4}, which shows how pivots themselves depend on $\psi$, makes it easier to read off confidence intervals;  using $r^\o(\psi)$ to obtain one-sided intervals $(L_\alpha,\infty)$, for example, we see that $\psi_0$ lies below the intervals for $\alpha=10^{-2}$, $10^{-3}$ and $10^{-4}$, and just inside that for $\alpha=10^{-5}$.  A nested set of confidence intervals such as this is more informative than a single significance probability based on $r^\o(\psi_0)$.}

In the Rayleigh example $\hat\psi/\psi$ has a known distribution, so the left-hand side of~\eqref{conf.eq} provides an exact significance function.  The sample size here is $n=1$, so it is no surprise that the exact function differs greatly from the large-sample approximations based on $w^\o(\psi)$ or $r^\o(\psi)$; the exact significance probability  for testing $\psi_0=3$ is $1.49\times 10^{-5}$.  Remarkably, however, an approximation that treats the modified likelihood root 
\begin{equation}\label{rstar.eq}
r^*(\psi) = r(\psi) + {1\over r(\psi)}\log\left\{{q(\psi)\over r(\psi)}\right\}
\end{equation}
as standard normal is essentially exact; its significance probability is $1.55\times 10^{-5}$, giving a relative error of $3.9\%$.   The approximation based on $r^*(\psi)$ is said to be third-order accurate, i.e., its error is $O(n^{-3/2})$, and moreover this error is relative, so high accuracy can be expected even when the true probabilities are small.   Here $n=1$, so the approximation yields quasi-perfect calibration of different values of $\psi$.  In this particular example $q(\psi) = 1-\hat\psi^2/\psi^2$, but in general it is uglier; see Appendix~B.

\subsection{Several parameters}\label{several.sect}

Models with a single scalar parameter are unusual: other parameters are usually present and must be estimated.   Despite this, near-exact inferences remain possible in many cases, as we now outline.  

Suppose now that the density function $f(y;\vartheta)$ depends on an unknown $d$-dimensional parameter $\vartheta$, which comprises a scalar interest parameter $\psi$ and a nuisance parameter $\lambda$; $\psi$ and $\lambda$ are supposed to be variation independent.  As $\lambda$ is unknown, it must be replaced by an estimate, and we see from~\eqref{conf.eq} that this would introduce errors.  

In this situation the maximum likelihood estimate $\hat\vartheta^\o=(\hat\psi^\o,\hat\lambda^\o)$ maximises the log-likelihood $\ell(\vartheta) = \log f(y^\o;\vartheta)$ with respect to $\vartheta$, and the partial maximum likelihood estimate $\hat\vartheta^\o_\psi=(\psi,\hat\lambda_\psi^\o)$ maximizes $\ell(\vartheta)$  with respect to $\lambda$ for fixed $\psi$.   The large-sample properties of the maximum likelihood estimator $\hat\vartheta$ under repeated sampling are well-established  \citep[Chapter 9]{Cox.Hinkley:1974}: as the sample size $n\to\infty$ and under mild regularity conditions, $\hat\vartheta$ has an approximate $d$-dimensional normal distribution with mean the true parameter $\vartheta$ and covariance matrix $\jmath(\hat\vartheta)^{-1}$, where $\jmath(\vartheta) = -\partial^{2} \ell(\vartheta) / \partial \vartheta \partial \vartheta^\T$ is the $d\times d$ observed information matrix and $\vartheta^\T$ is the transpose of the $d\times 1$ vector $\vartheta$.  Under these conditions the error committed by replacing parameters in~\eqref{conf.eq} by their estimates is $O(n^{-1/2})$, giving  first-order approximations, and the same error is committed by treating the likelihood root
\begin{equation}\label{r.eq}
r(\psi) = \sign(\hat\psi-\psi)\left[ 2\left\{ \ell(\hat\vartheta) - \ell(\hat\vartheta_\psi)\right\}\right]^{1/2}
\end{equation}
or the Wald statistic
\begin{equation}\label{wald.eq}
w(\psi)=\jmath_{\mathrm{p}}(\hat{\psi})^{1 / 2}(\hat{\psi}-\psi), 
\end{equation}
as standard normal; here 
$$
\jmath_{\mathrm{p}}({\psi})=\dfrac{\left| \jmath\left(\hat{\vartheta}_{\psi}\right)\right|}{\left|\jmath_{\lambda \lambda}\left(\hat{\vartheta}_{\psi}\right)\right|}, 
$$
$|\cdot |$ indicates the determinant, and $\displaystyle \jmath_{\lambda \lambda}(\vartheta)$ is the $\displaystyle(\lambda,\lambda)$ corner of  $\jmath(\vartheta)$.  The improvement to third-order accuracy seen in the scalar case is again given by~\eqref{rstar.eq}, now using~\eqref{r.eq} and 
\begin{equation}\label{q.eq}
q(\psi)=\dfrac{\left | \varphi(\hat{\vartheta})-\varphi(\hat{\vartheta_{\psi}})~~\varphi_{\lambda}(\hat{\vartheta_{\psi}})  \right|}{\left |\varphi_{\vartheta} (\hat{\vartheta})\right |}  
\dfrac{\left |\jmath(\hat{\vartheta}) \right |^{1/2}}{\left | \jmath_{\lambda\lambda}(\hat{\vartheta}_{\psi})\right|^{1/2} } .
\end{equation}
Here $\varphi(\vartheta)$ is a $d\times 1$ constructed parameter that can be viewed as a directional derivative of the log-likelihood, and $\varphi_\vartheta(\vartheta) =\partial\varphi(\vartheta)/\partial\vartheta^\T$ and $\varphi_\lambda(\vartheta) =\partial\varphi(\vartheta)/\partial\lambda^\T$ are respectively $d\times d$ and $d\times (d-1)$ matrices; see Appendix~B.

In applications, expressions~\eqref{r.eq}--\eqref{q.eq} are evaluated at the observed data $y^\o$ and the corresponding estimates $\hat\vartheta^\o$ and $\hat\vartheta^o_{\psi}$, leading to the approximate significance function
\begin{equation}\label{r-approx.eq}
\pr(\hat\psi\leq \hat\psi^o;\psi) =\Phi\left\{r^{*\o}(\psi)\right\}
\end{equation}
which can be used in the same way as~\eqref{conf.eq}.  The relative error committed in using this expression, $O(n^{-3/2})$, is appreciably lower than that using a normal approximation to the likelihood root~\eqref{r.eq}.

%
\subsection{Bayesian approximation}\label{bayes.sec}

In the setting with several parameters, expression~\eqref{bayes.eq} generalizes to the marginal posterior distribution of the scalar interest parameter $\psi$, i.e., 
\begin{equation}\label{bayes2.eq}
\pr(\psi\leq \psi_0\mid y^\o) = {\int_{-\infty}^{\psi_0} \int L(\psi,\lambda ) f(\psi,\lambda)\,\D{\lambda}\,\D{\psi}\over \int_{-\infty}^{\infty} \int L(\psi,\lambda ) f(\psi,\lambda)\,\D{\lambda}\,\D{\psi}},
\end{equation}
where $f(\psi,\lambda)$ is the prior density.  On approximating both integrals using Laplace's method, it turns out that \citep[Section~8.7]{Brazzale.Davison.Reid:2007}
\begin{equation}\label{B-approx.eq}
\pr(\psi\leq \psi_0\mid y^\o) = \Phi\left\{ r_B^{*\o}(\psi_0)\right\}\left\{1+ O(n^{-1})\right\}, 
\end{equation}
where $ r_B^{*\o}(\psi)$ is defined by equation~\eqref{rstar.eq} but with $q^\o(\psi)$ replaced by 
\begin{equation}\label{qB.eq}
q_B(\psi) =  \left. {\D{\ell(\hat\vartheta_\psi)}\over \D{ \psi}}\right. 
\times 
{|\jmath_{\lambda\lambda}(\hat\vartheta_{\psi})|^{1/2}\over  |\jmath(\hat\vartheta)|^{1/2}} \times
{f(\hat\vartheta)\over f(\hat\vartheta_\psi)}
\end{equation}
evaluated at the observed data $y=y^\o$ and the corresponding estimates $\hat\vartheta=\hat\vartheta^\o$ and $\hat\vartheta_\psi= \hat\vartheta^o_{\psi_0}$.  Thus there is a close parallel between frequentist and the Bayesian approximations and, with an appropriate choice of the prior, the probabilities that the corresponding confidence intervals contain $\psi_0$ differ by only $O(n^{-1})$.  
The Jeffreys prior gives inferences invariant to 1--1 transformation of $\vartheta$ and \red{thus is often regarded as a natural choice, though it is criticised by \citet{Fraser.etal:2016}.  In the conjunction problem this prior has the undesirable property mentioned at the end of Section~\ref{2D.section} of attributing zero probability to any sphere around the origin, relative to outside that region.}   

In a similar context \citet{Davison.Sartori:2008} compare the performances of $ r_B^{*\o}(\psi)$ and $ r^{*\o}(\psi)$ and find that the former performs rather worse.  The Bayesian confidence intervals are slightly shorter and tend to contain the true parameter less often, \red{and we would expect this in the present setting also, owing to the downward bias of $\hat p_c$}.  Thus we do not implement Bayesian approximations; rather we find it reassuring that Bayesian and frequentist inferences can be approached in the same way. If reliable prior information was available on each likely conjunction, then a Bayesian approach would be indicated. 

Further details of the Bayesian approximation can be found in Appendix~C.
\section{Inference for conjunction assessment}\label{likelihood.sec}

\subsection{Likelihood}

In this section we apply the standard theory described in Section~\ref{several.sect} to the statistical model described in Section~\ref{conjunction.sec}.  After dropping irrelevant additive constants, the log-likelihood function is 
 \begin{equation}\label{logL.eq}
 \ell(\vartheta)= -\frac{1}{2}\{y-\eta(\vartheta)\}^{\T}\Omega \{y-\eta(\vartheta)\},
 \end{equation} 
where $\vartheta=(\psi,\lambda)$ contains the miss distance $\psi$ for which inference is required and the five-dimensional nuisance parameter vector $\lambda=( \theta_1,\phi_1,\|\nu\|,\theta_2,\phi_2)$.  
The maximum likelihood estimator $\hat{\vartheta}$ based on the observed relative distance and velocity contained in the $6\times 1$ vector $y$ satisfies $\eta(\hat{\vartheta})=y$ and the observed information matrix is 
 \begin{equation}\label{jhat.eq}
\jmath(\hat{\vartheta})=\left. {\partial\eta^\T(\vartheta)\over \partial\vartheta}  \Omega {\partial\eta(\vartheta)\over \partial\vartheta^\T}\right\vert_{\vartheta=\hat{\vartheta}},
 \end{equation} 
where ${\partial\eta^\T(\vartheta)/ \partial\vartheta}$ is a $6\times 6$ matrix. Hence $\ell(\hat\vartheta)=0$, $\hat\vartheta$ is simply a transformation of $y$, and $\hat\vartheta_\psi=(\psi,\hat\lambda_\psi)$ minimises the weighted sum of squares in~\eqref{logL.eq} for fixed $\psi$. These quantities allow inference on $\psi$ based on the likelihood root~\eqref{r.eq} and the Wald statistic~\eqref{wald.eq}, while the more accurate modified likelihood root~\eqref{rstar.eq} also requires~\eqref{q.eq}.  The normal model is a curved exponential family \citep[Section~5.2]{Davison:2003} in terms of $\vartheta$, so we can take $\varphi(\vartheta)=\eta(\vartheta)$,  the computation of $r^*(\psi)$ only involves $\eta(\vartheta)$ and its derivatives, and expression~\eqref{q.eq} simplifies to 
 \begin{equation}\label{useful.eq}
q(\psi) = | y-\eta(\hat\vartheta_{\psi})~~\eta_{\lambda}(\hat\vartheta_{\psi})  |\times  |\Omega|^{1/2} \times |\jmath_{\lambda\lambda}(\hat\vartheta_{\psi})|^{-1/2}.
\end{equation} 
Appendices~B ,~C and~D contain the detailed calculations for this, for the Bayesian approximation with the Jeffreys prior $ |\eta_\vartheta(\vartheta)|$, and implementation details.

Here there are six parameters and a single six-dimensional observation $y$, so the sample size is $n=1$, and it appears that we cannot expect large-sample approximations to apply.  However the covariance matrix for an average of $n$ independent observations would be $(n\Omega)^{-1}$, so a large sample size $n$ is equivalent to a small variance for the observations or equivalently large $\Omega$, which is the correct gauge of  accuracy.

\red{As we saw in  Section~\ref{2D.section}, the model simplifies greatly when the velocity vector $\nu$ is known.  In this case $y$ and $\eta(\vartheta)$ are replaced by the projections of the observed position vector and the true position of the second object into the encounter plane, i.e., $x=(x_1,x_2)$ and $\xi=(\xi_1,\xi_2) = (\psi\cos\lambda, \psi\sin\lambda)$, and as the variance matrix of $x$ can be diagonalised, the log likelihood becomes (see Figure~\ref{fig:fig1})
\begin{equation}\label{radial.lik}
\ell(\psi, \lambda) = -{1\over 2} \left\{{(x_1-\psi\cos\lambda)^2\over d_1^2} +  {(x_2-\psi\sin\lambda)^2\over d_2^2} \right\} , \quad \psi>0, 0\leq\lambda<2\pi,
\end{equation}
}

\red{The approach advocated by \citet{Carpenter:2019} is related to the discussion above, but the confidence statements therein are based on the marginal quantiles of the miss distance distribution rather than on likelihood theory.  That approach does not allow for the uncertainty about the nuisance parameter and appears to be equivalent to basing inference on the Wald statistic, which can perform very poorly in nonlinear settings.}

\subsection{\red{Evidence and decisions}}

\red{In Section~\ref{sig.sect} we argued that significance functions such as~\eqref{pval.eq} and~\eqref{r-approx.eq} allow inference on the true miss distance $\psi$, either by constructing confidence intervals or as an assessment of the evidence that $\psi$ equals some particular value $\psi_0$.  In the present context $\psi_0$ might be a safety threshold, and then one approach to inference on $\psi$ is to test the null hypothesis $H_0:\psi=\psi_0$ against the alternative hypothesis $H_+:\psi>\psi_0$, with evasive action to be considered if~\eqref{r-approx.eq}  exceeds some threshold $\v$, i.e., $H_0$ cannot be rejected at level $\v$.  In practice $\v$ is often taken to be $10^{-4}$.   If the significance probability is correctly calibrated and the true miss distance is $\psi_0$, then the false positive probability, that of considering action unnecessarily, would be $1-\v$, whatever $\v$ is chosen.  The choice of  $H_+$ as alternative hypothesis ensures that $p_\obs$ is small when the estimated conjunction probability $\hat p_c$ and the Bayesian posterior probability $\pr(\psi\leq \psi_0\mid y)$ would also be small, despite their different interpretations and properties.  \citet{Hejduk:2019} argue that the null hypothesis $\psi\leq \psi_0$ is unnatural, since it implies that the `null' situation is to anticipate a collision, but it appears more important to us that small values of $p_\obs$ correspond to small conjunction probabilities, which is ensured by the above  setup.  Our approach is supported by regarding hypothesis testing as attempting  `proof by stochastic contradiction': the null hypothesis represents an assumption that is regarded as absurd (disproved) when the corresponding significance level is sufficiently small.  If so, it makes sense to take $\psi=\psi_0$ as the null hypothesis, as we hope that this will be contradicted by the data and no evasive action need be considered.}

\red{The interpretation of the threshold $\v$ in terms of an elementary decision analysis was described in Section~\ref{sig.sect}, and since the purpose of conjunction analysis is to assist decision-making, consideration of losses seems an appropriate basis for choosing $\v$. Although our earlier discussion suggested considering evasive action when the posterior probability $\pr(\psi\leq \psi_0\mid y)$ exceeds $\v$, it seems better to replace $\pr(\psi\leq \psi_0\mid y)$ by a significance probability $p_\obs $ computed as $ \Phi\left\{r^{\o}(\psi_0)\right\}$ or $ \Phi\left\{r^{*\o}(\psi_0)\right\}$, which are approximately uniformly distributed under $H_0$.   }

\red{The abuse of hypothesis tests has been much discussed \citep[e.g.,][]{Carpenter:2017}, and it is often suggested that they be systematically replaced by confidence intervals. Plotting $\Phi\left\{r^{*\o}(\psi)\right\}$ as a function of $\psi$, as in Figure~\ref{fig:fig4}, allows the construction of confidence intervals  for $\psi$, the form of which expresses the uncertainty and suggests what power is available: a narrow interval corresponds to more precise estimation and hence higher power for rejecting hypotheses such as $H_0$ above.  Unlike a two-sided $(1-2\alpha)$ confidence interval $[L_{\alpha},U_{\alpha}]$, one-sided  intervals such as $[L_{\alpha},+\infty)$ or  $[0,U_{\alpha})$ give no information about the accuracy of the estimate, so a two-sided interval provides a better basis for risk assessment.  A one-sided confidence interval $(L_\v,\infty)$ that does not contain $\psi_0$ leads  to the same decision as observing $p_\obs<\v$, and it seems partly a matter of taste which is preferred: computing $p_\obs$ alone is quicker but is less informative than a plot of $\Phi\left\{r^{*\o}(\psi)\right\}$. In navigating the literature it is enlightening to recognize that hypothesis testing has a variety of distinct uses \citep{Cox:2020}, one of which is to flag situations that merit more detailed scrutiny.   In conjunction analysis one might therefore plot significance functions only when $p_\obs>\v$, thereby focusing on those conjunctions requiring careful consideration.  }

\section{Numerical results}\label{numerical.sect}

\subsection{General}\label{numerical.general}

Below we investigate the accuracy of the normal approximations to the Wald statistic, the likelihood root and the modified likelihood root in two scenarios.  We do so in terms of one-sided error rates for confidence intervals $(L_\alpha,U_\alpha)$ for the true miss distance $\psi_0$ and use $\pr_0$ to indicate probability computed when $\psi=\psi_0$.  An ideal two-sided equi-tailed confidence interval with coverage probability $1-2\alpha$ should satisfy $\pr_0(L_\alpha\leq\psi_0\leq U_\alpha) = 1-2\alpha$ and have one-sided left-tail and right-tail error rates $\pr_0(\psi_0<L_\alpha) $ and $ \pr_0(U_\alpha<\psi_0)$ both equal to $\alpha$, for any $\alpha\in(0,0.5)$.  Departures from this will indicate deficiencies of the confidence intervals and the corresponding tests, whereas close agreement will indicate that the inferences are well-calibrated.  \red{As the pivots are decreasing in $\psi$ and should ideally have standard normal distributions, inaccurate left-tail error rates correspond to the pivot departing from normality in its upper tail.  In the present setting, accurate left-tail error rates are most important, since they correspond to well-calibrated significance probabilities and confidence intervals of form $(L_\alpha,\infty)$. }

The form of the covariance matrix used in practice depends on the type of the conjunction \citep[Chapter 5]{Chen.etal:2017}. In a short-term conjunction, uncertainty on the velocity is negligible compared to uncertainty on the position.  In a long-term conjunction, the motion is nonlinear and the computations are more involved. In both cases, the quality of risk assessment depends heavily on the covariance matrix.  Below we suppose that the error covariance matrix for the relative distance and velocity of the second satellite relative to the first is given by
$$
\Omega^{-1}=\left[
\begin{array}{cc}
{P}_{1} & {P}_{12}  \\
{P}_{12}  & {P}_{2}
\end{array}
\right],
$$
where $P_{1}$, $P_{2}$, and $P_{12}$ are the position,  the velocity  and the cross-correlation covariance matrices, \red{of units $\mathrm{m}^{2}$, $\mathrm{m}^{2} \mathrm{~s}^{-2}$ and $\mathrm{m}^{2} \mathrm{~s}^{-1}$ respectively.  The six eigenvalues of $\Omega^{-1}$ are difficult to interpret physically, and can vary greatly.}

\red{In our first two case studies,  we assume that $P_{12}= 0_{3 \times 3}$, and choose $P_{1}  = \tau \sigma^2 I_3$ and $P_{2}  =\sigma^{2}I_3$.  This choice implies that the standard deviation of position errors along each axis direction is $\sqrt{\tau} \sigma$  (km) and the standard deviation of velocity errors is $\sigma$ (km/s).   Uncertainty on the position is typically larger than that on the velocity, and then  $\tau>1$.}

\subsection{Case study A}

The relative quantities and spherical coordinates of two satellites are as given in Table~\ref{tab:tab2}.  The relative distance is around 102 km and the relative speed around 11.7 km/s, the value of $\s^2$ varies from $10^{-3}$ km$^2$ to $2$ km$^2$, and that of $\tau$ varies from 1 to 3.

\begin{table}[t!]
\center
\caption{Conjunction geometry of case studies: A: a simulated example; B,  the U.S. and
Russian satellite collision event;  and C an event with high probability of conjunction.}
\label{tab:tab2}
\begin{tabular}{c rrr}
\hline 
Variable & \multicolumn{3}{c}{Case study}\\
\cline{2-4}
&A&B&C\\
\hline 
Miss distance (m) 		    &$ 10^3 \times35.267$             & 698.011    				        & 11.917 \\
$\Delta X$ (m) 					&$-10^3 \times100$   				& $-$258.909 		            & $-$7.678  \\
$\Delta Y$ (m) 					&$-10^3 \times20  $                 & $-$635.813 			        & $-$9.152 \\
$ \Delta Z $ (m) 				&0                                                 & 126.229 				         & 0.564 \\
$\Delta V_x$ (km/s) 		&10 												& 10.580					 & 9.926 \\  
$\Delta V_y$ (km/s) 		& 6 													&  $-$3.733              &  $-$9.653 \\
$\Delta V_z$ (km/s)          & 1												& 3.126        		     &$-$4.110\\
$\theta_1$ 		& 1.570  						& 1.389        			         & 1.618 \\ 
$\theta_2$ 		& 1.485 						& 1.299					          & 1.860 \\   
$\phi_1$ 			&$-$2.944				&1.957 					         & $-$2.269 \\
$\phi_2$				 &$-$2.944 				& $-$0.339                    & $-$ 0.772 \\
\hline 
\end{tabular}
\end{table}



Table~\ref{tab:tab3} shows the error rates for the Wald statistic, the likelihood root $r$ and the modified likelihood root $r^*$ based on $10^4$ datasets simulated for various combinations of values of $\sigma$ and $\tau$.  For very small $\sigma^2$ all three sets of error rates are close to the nominal values, but problems with the Wald statistic and to a lesser extent the likelihood root start to appear when $\sigma^2\geq 10^{-1}$, with the left-tail error systematically too high and the right-tail error systematically too low.  The modified root behaves much better overall, though its right-tail error also rises as  $\sigma^2$ increases.  As mentioned in Section~\ref{numerical.general},  tests of $H_+$ require accuracy in the left tail, so right-tail error is less important.

\begin{figure}[h]
\begin{center}

\includegraphics[width=.3\textwidth]{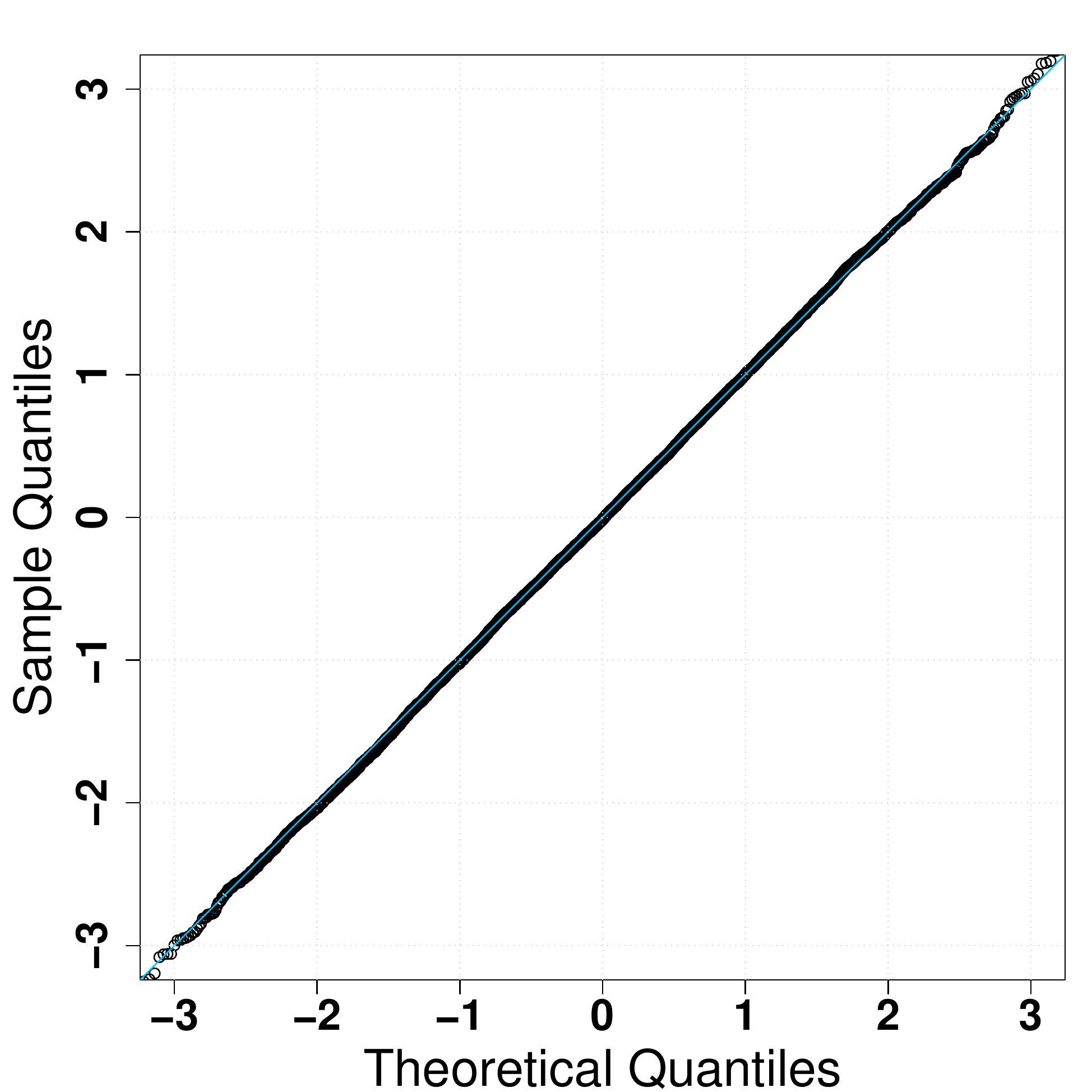}
\includegraphics[width=.3\textwidth]{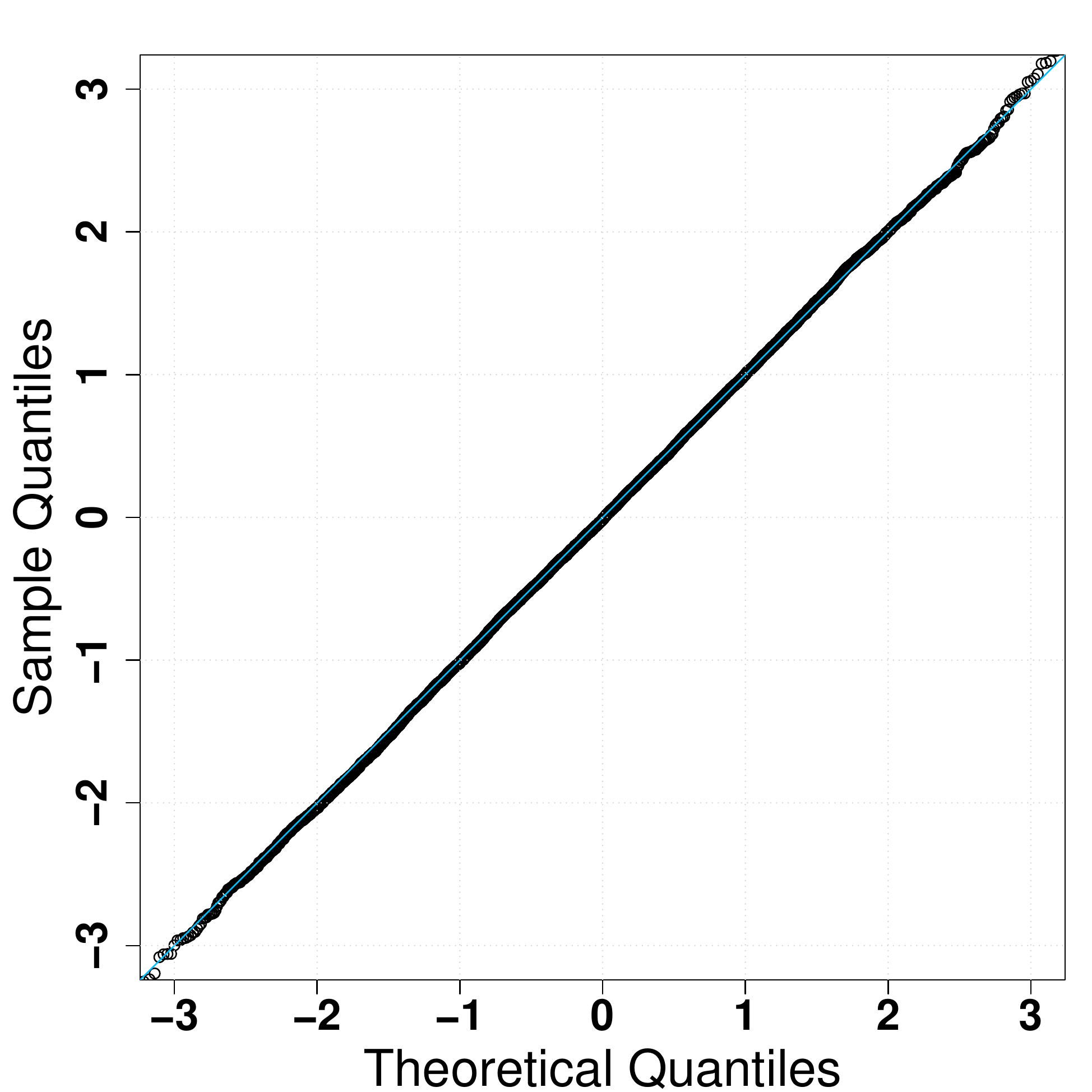}
\includegraphics[width=.3\textwidth]{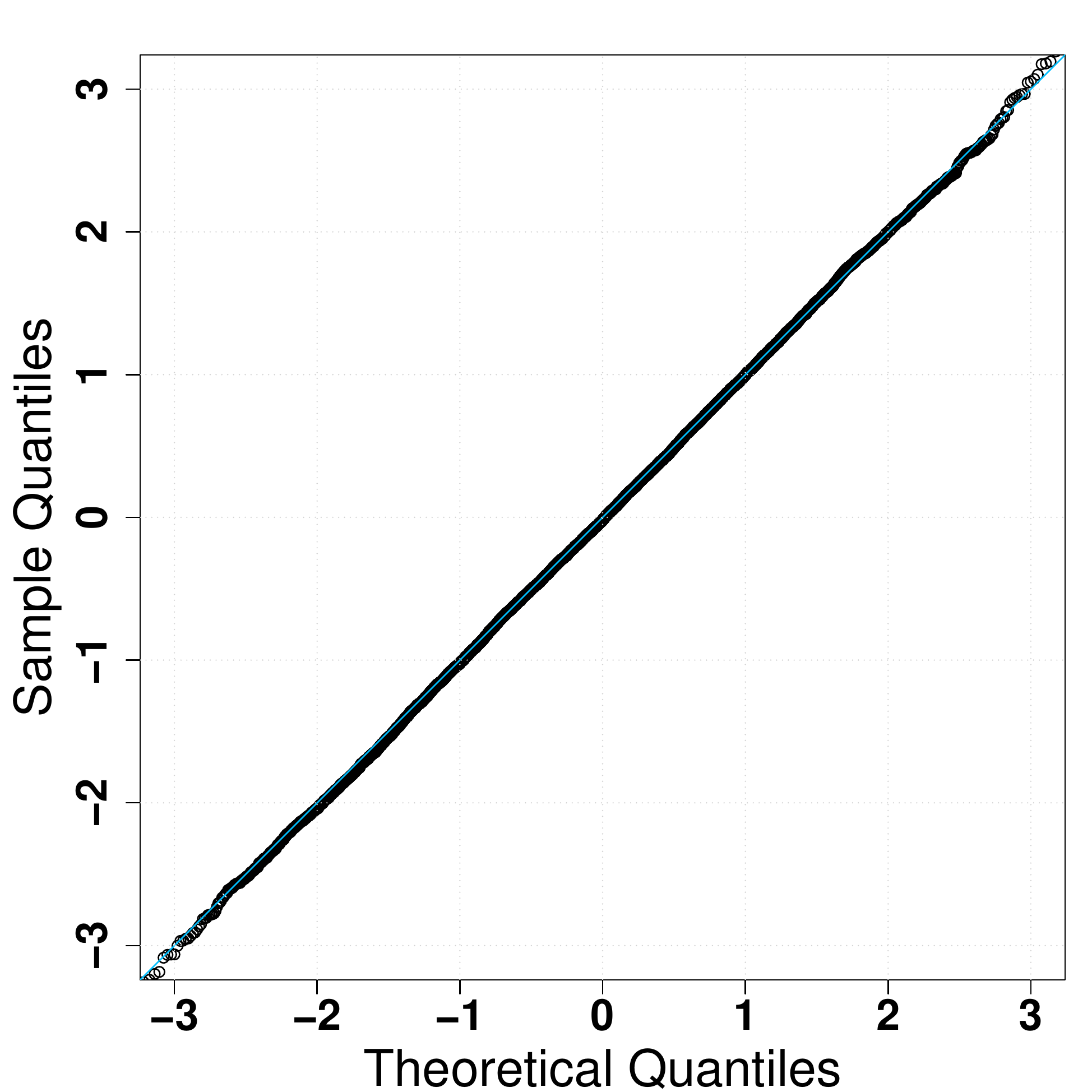}\\
\includegraphics[width=.3\textwidth]{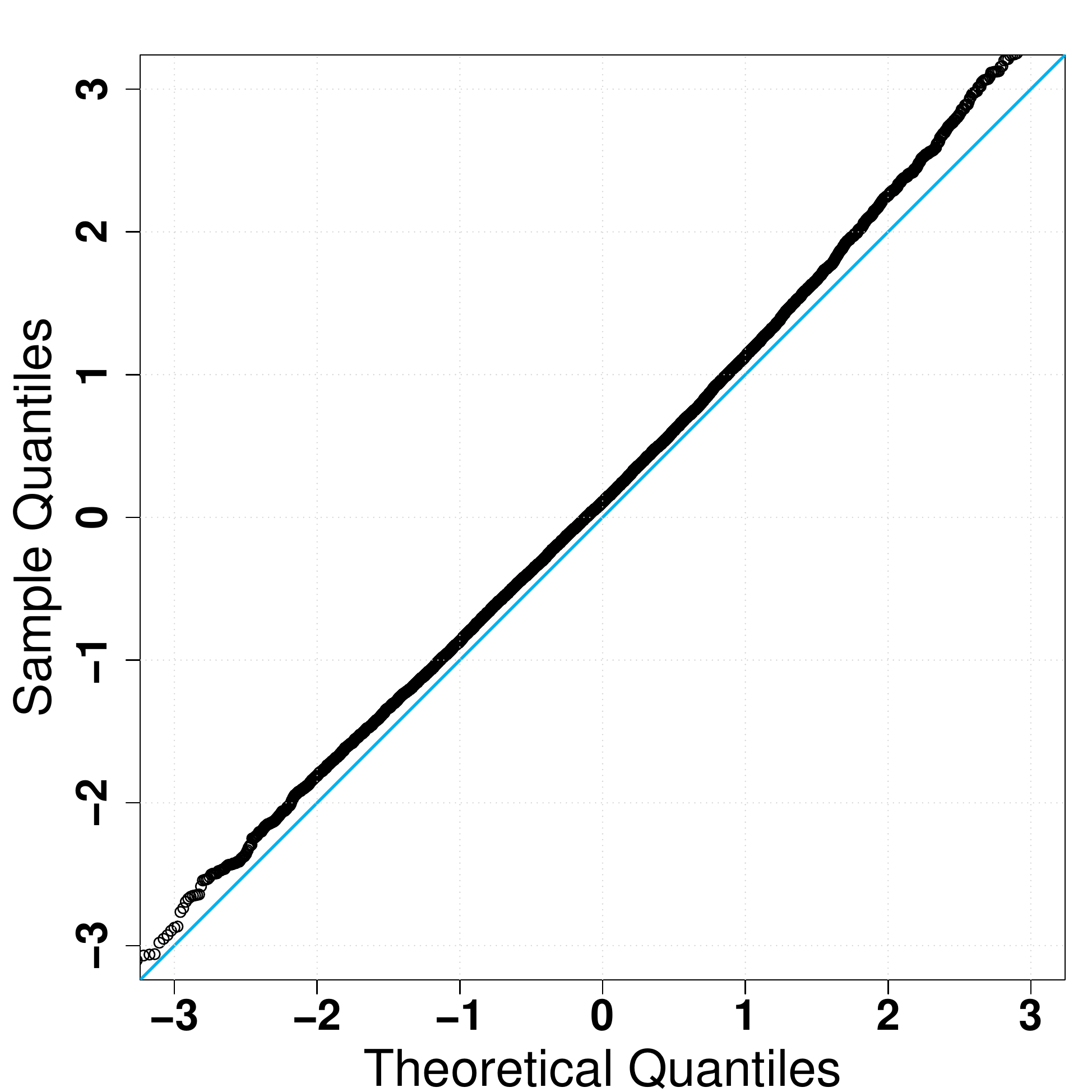}
\includegraphics[width=.3\textwidth]{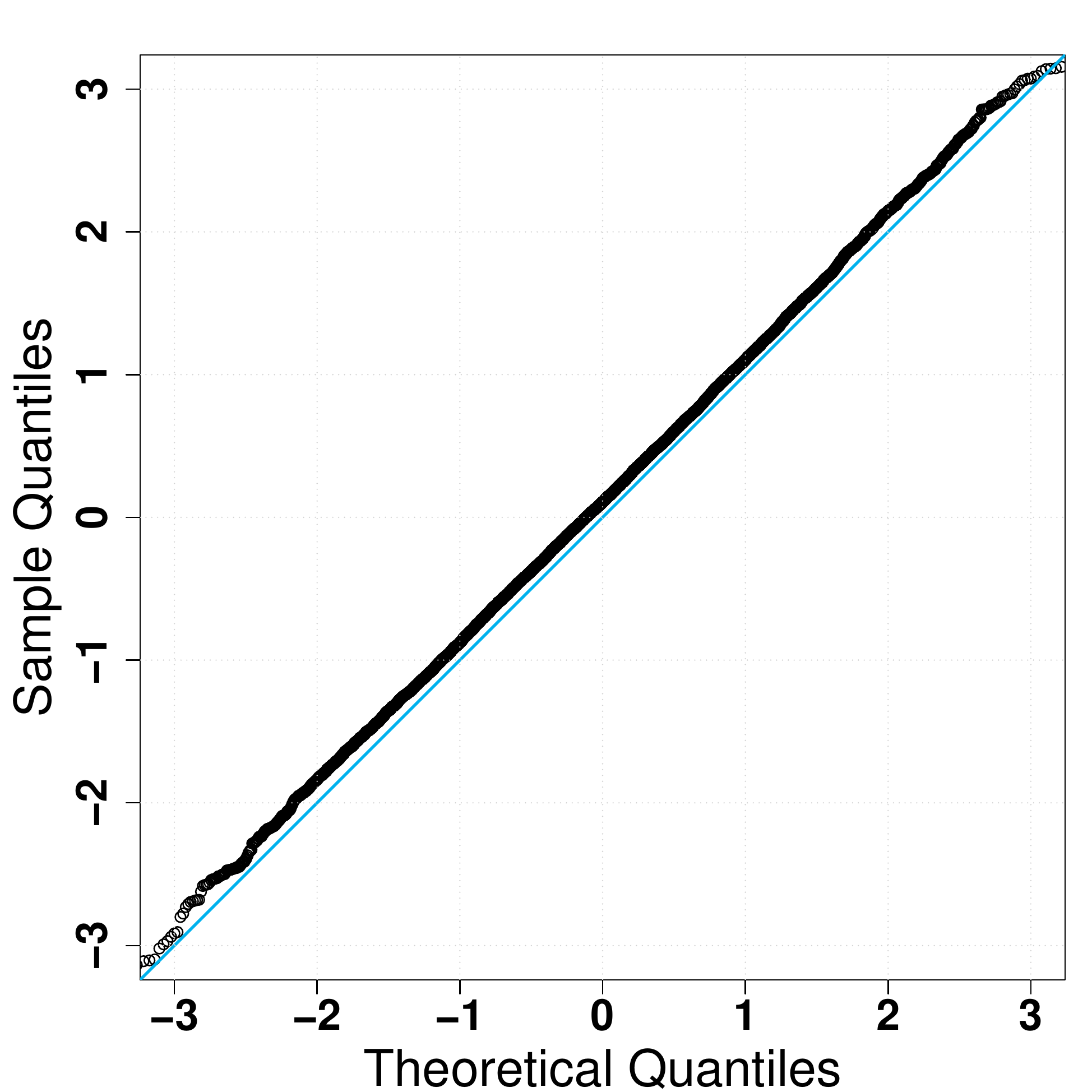}
\includegraphics[width=.3\textwidth]{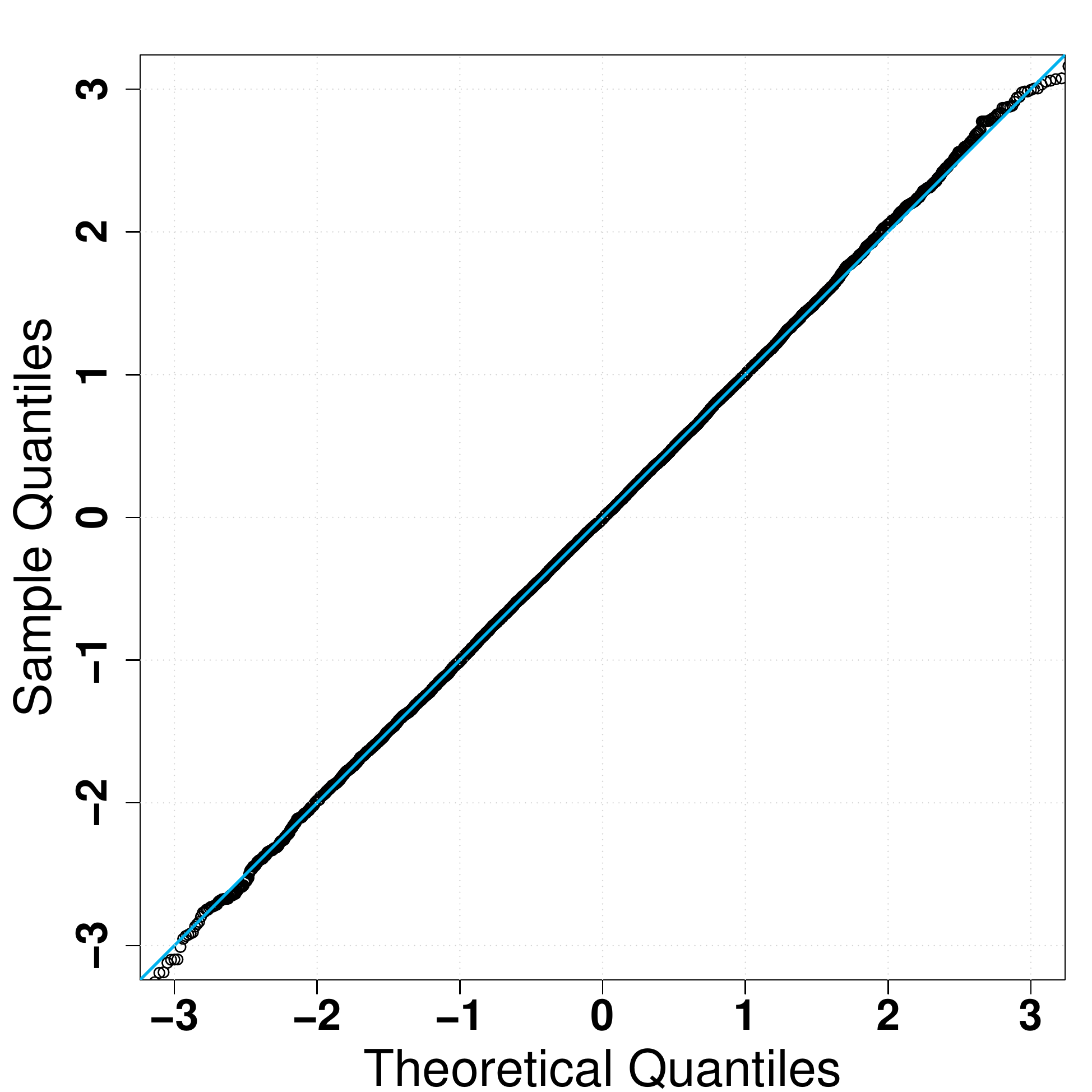}\\
\includegraphics[width=.3\textwidth]{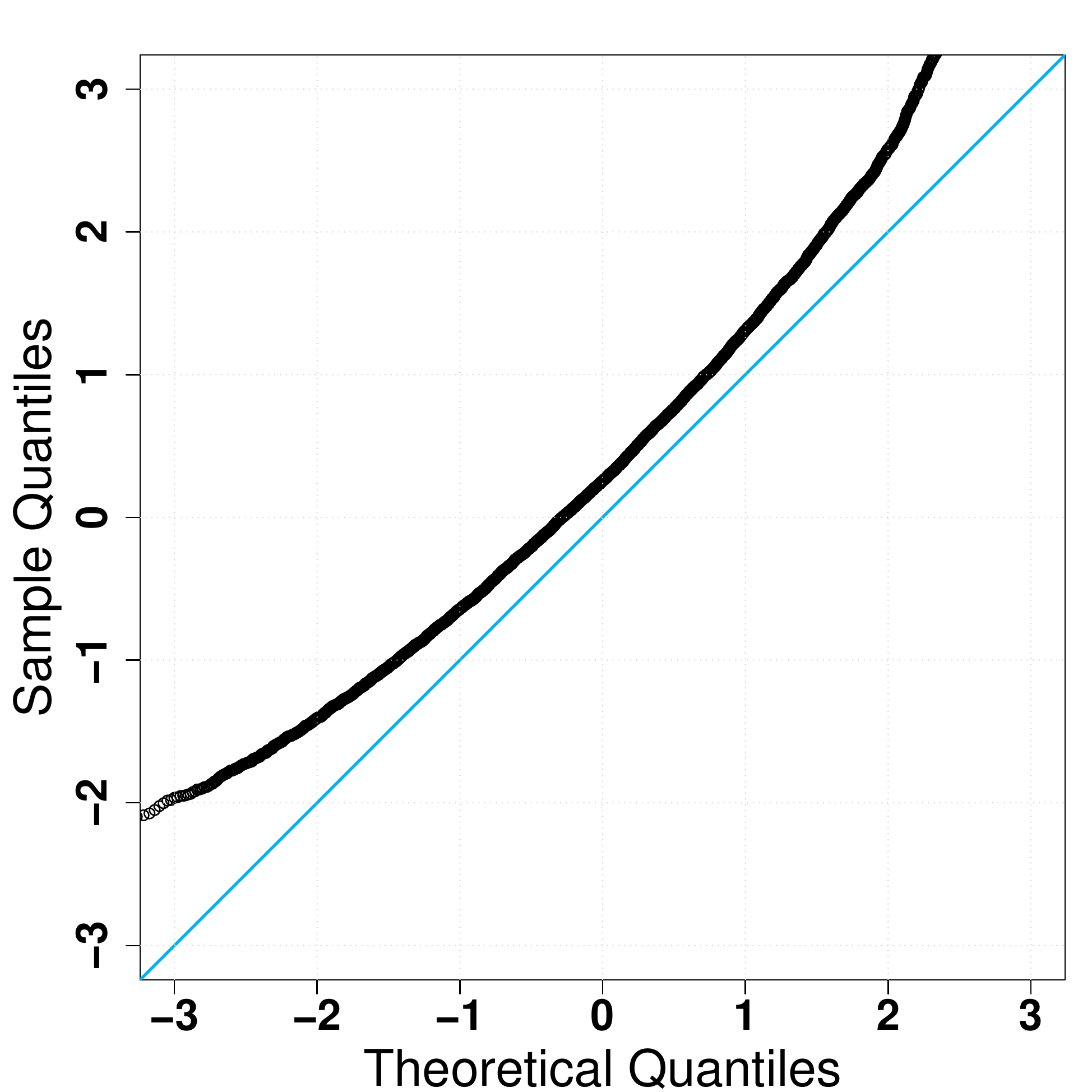}
\includegraphics[width=.3\textwidth]{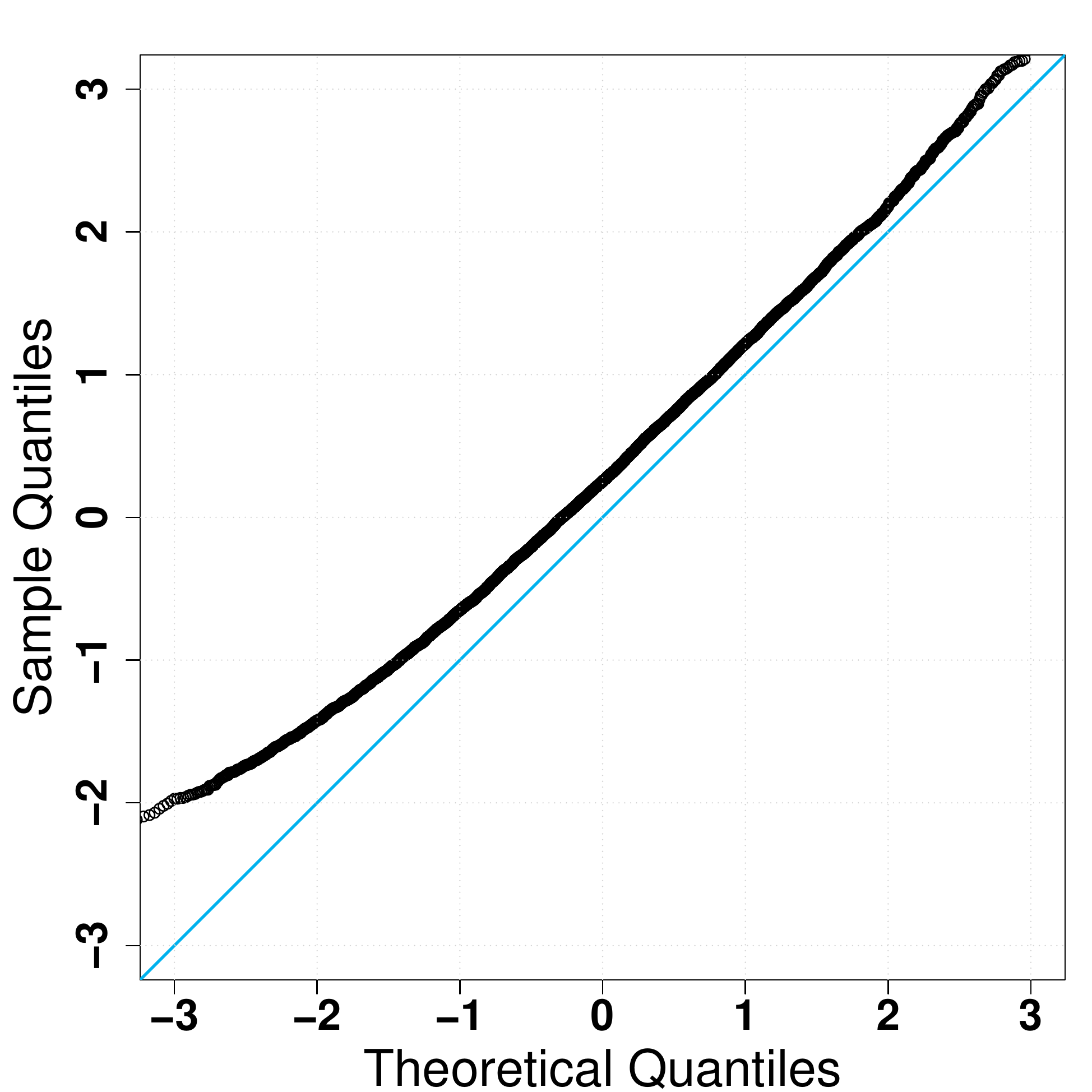}
\includegraphics[width=.3\textwidth]{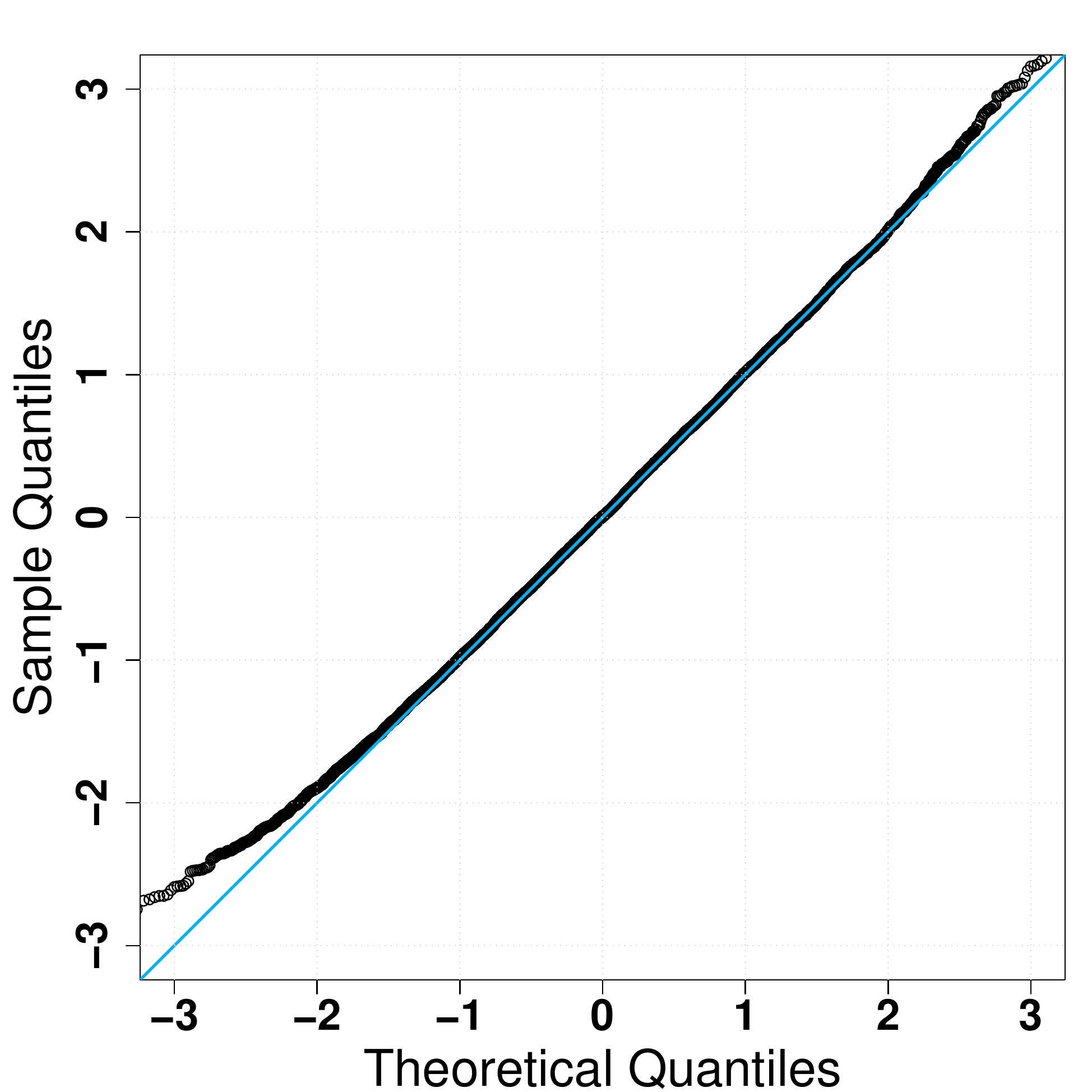}
\includegraphics[width=.3\textwidth]{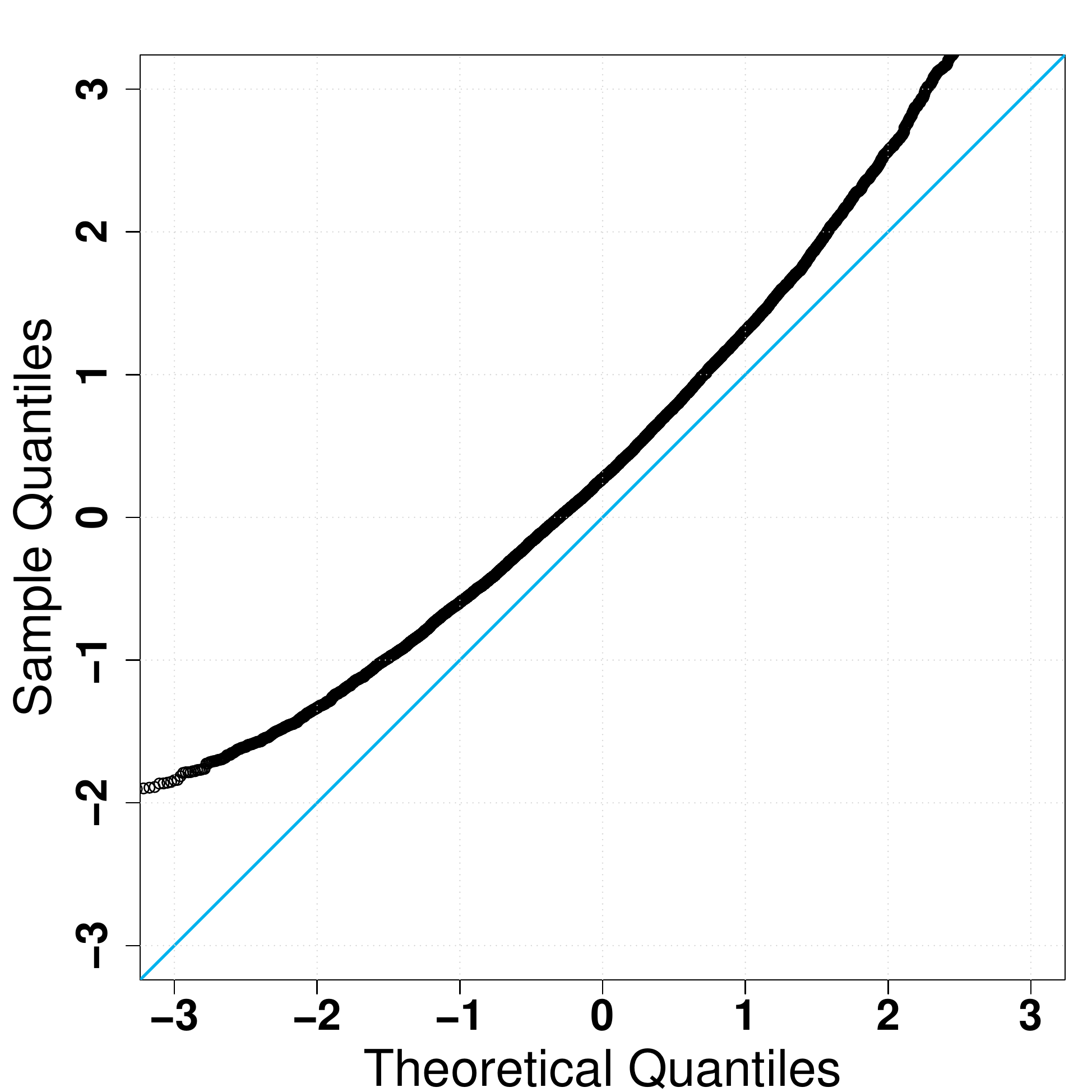}
\includegraphics[width=.3\textwidth]{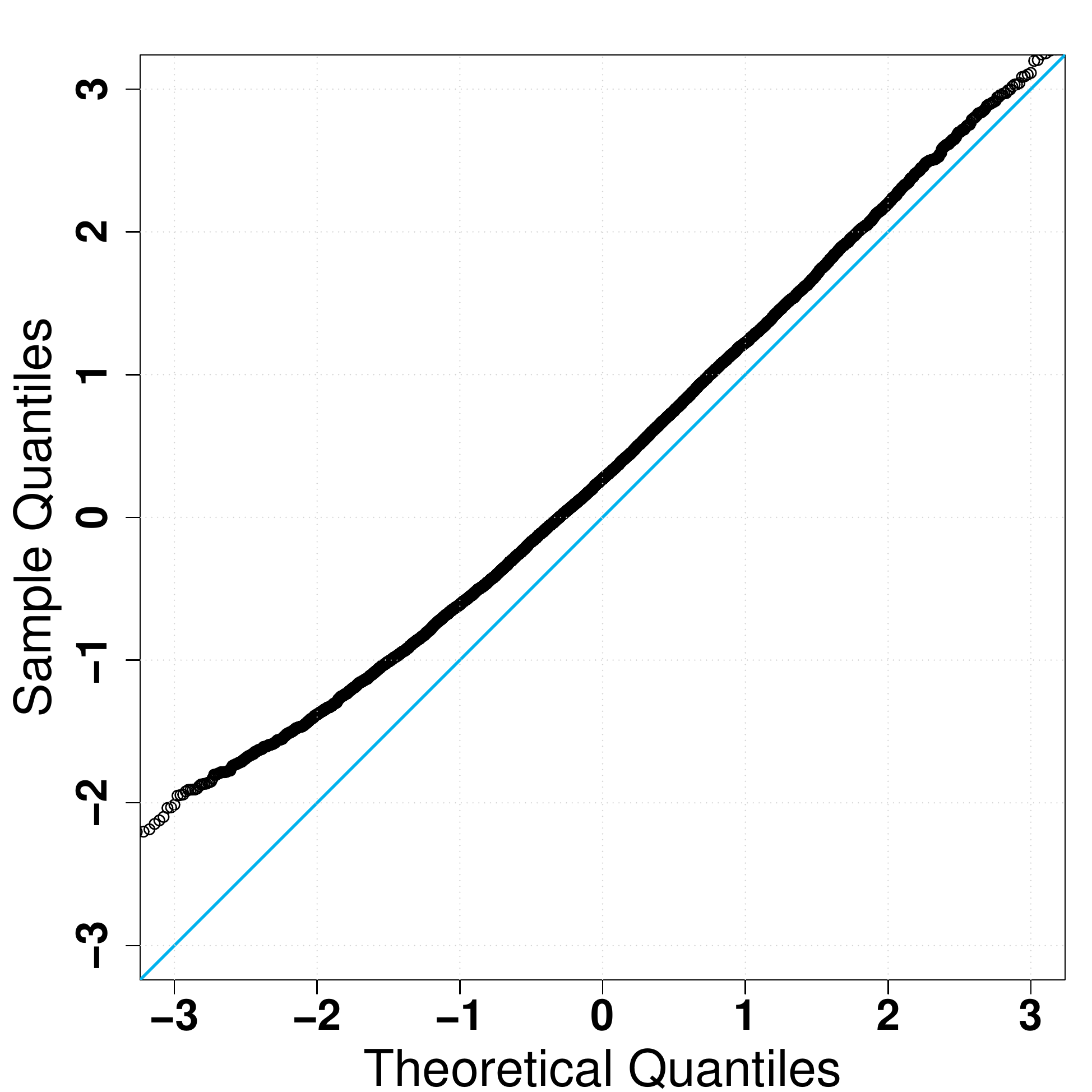}
\includegraphics[width=.3\textwidth]{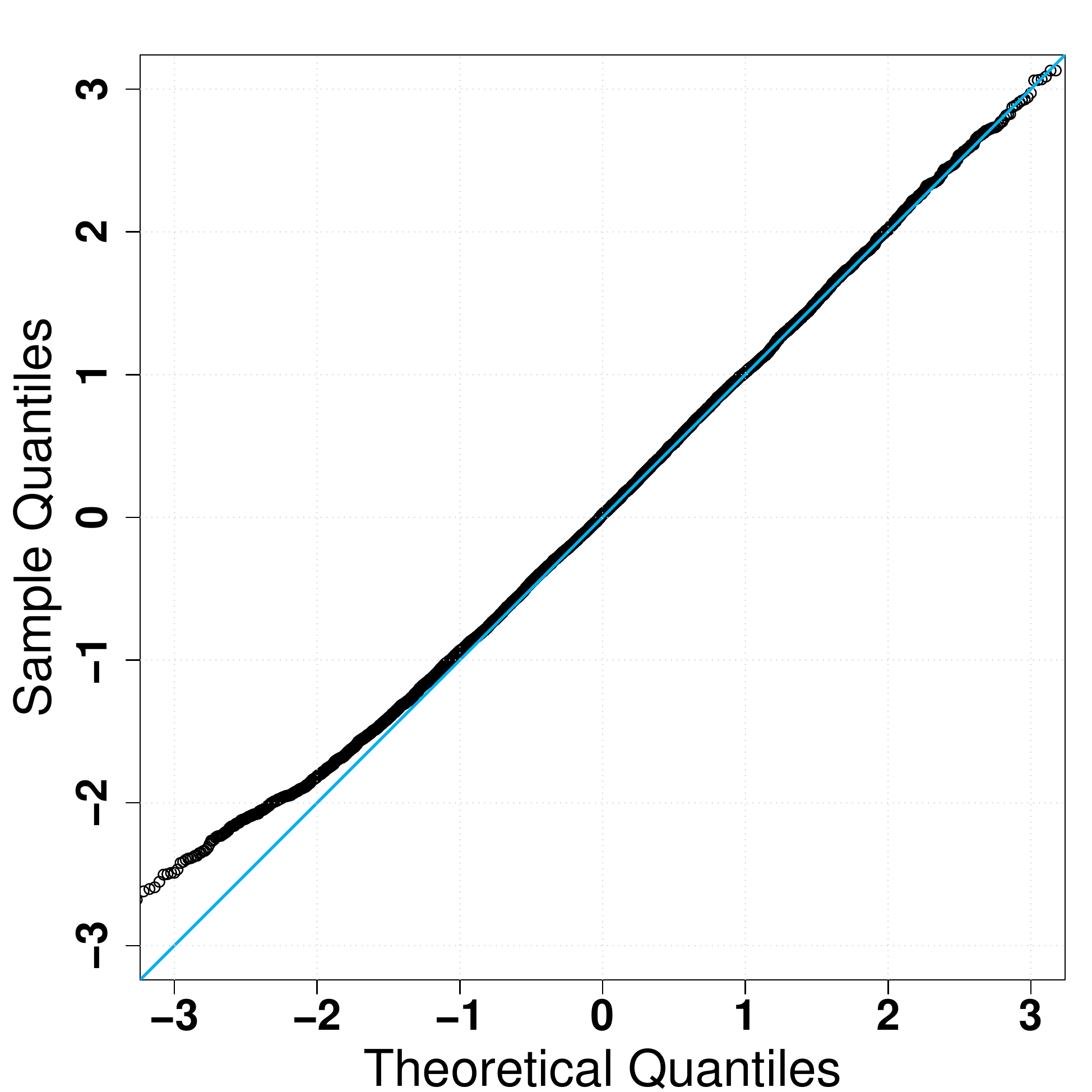}

\end{center}

\caption{Normal QQ-plots of $w(\psi^o)$ (left), $r(\psi^o)$ (middle) and $r^*(\psi^o)$ (right) based on $10^4$ Monte Carlo sample quantiles, with $(\sigma^2,\tau)$ equal to $(10^{-3},1), (1,1), (5,2), (5,5)$ (top to bottom).}

\label{fig:fig5}
\end{figure}

These remarks are confirmed by the Gaussian QQ-plots of simulated values of the three quantities in Figure~\ref{fig:fig5}.  If the distribution is exactly Gaussian, then the confidence intervals are exactly calibrated, so a departure from the line of unit slope through the origin implies a lack of calibration.  For  small $\sigma^2$, all three statistics have standard normal distributions and give comparable results, but for larger $\sigma^2$, the Wald statistic and the likelihood root are shifted to the right and right-skewed, more strikingly for larger values of $\tau$.  \red{This reflects the upward bias of the estimated distance, discussed in Section~\ref{2D.section}, and explains the asymmetric error rates in Table~\ref{tab:tab3}, with lower probabilities for the right than for the left.} The asymmetry increases with larger uncertainties on the relative distance and velocity and with smaller nominal error rates. 

The third column of Figure~\ref{fig:fig5} shows that the modified likelihood root $r^*$ corrects the departure from normality in the upper tail even for $\sigma^2=5$, and its error rates are closer to the nominal rates in all cases considered. For $\sigma^2>2$ and for $1\%$ nominal levels, the Wald statistic and the likelihood root show extreme overcoverage on the right and undercoverage on the  left; although the modified likelihood root provides a considerable improvement, its right-tail error is somewhat smaller than the nominal value.

\begin{table}[p]
\center
\caption{ Left and right error rates ($\%$) for two-sided nominal $10 \%, 5 \%,$ and $1 \%$ confidence intervals for the true parameter $\psi_0$ of Case Study A, estimated from $10^4$ Monte Carlo samples. The standard errors (SE) appear in the last line.}
\label{tab:tab3}
\begin{adjustbox}{width=\textwidth,totalheight=0.9\textheight,keepaspectratio}
\begin{tabular}{ccccccccc}
\hline \hline 
&&\multicolumn{3}{c}{Left tail $(\%)$}&&\multicolumn{3}{c}{Right tail $(\%)$} \\
\cline{3-5} \cline{7-9}
Uncertainty & Statistic&5 & 2.5 & 0.5&& 5 & 2.5 & 0.5 \\
\hline
$(\sigma^2,\tau)=(10^{-3},1)$& $w$ & $5.22$ & $2.58$ & $0.54$ && $5.03$ & $2.49$ & $0.49$ \\ 
 &$r$ &$5.20$ & $2.56$ & $0.53$ && $5.04$ & $2.51$ & $0.50$ \\ 
 &$r^*$ &$5.15$ & $2.54$ & $0.53$ && $5.13$ & $2.54$ & $0.51$ \\ 
\hline
$(\sigma^2,\tau)=(10^{-3},2)$&  $w$&  $5.43$ & $2.61$ & $0.43$ && $4.96$ & $2.61$ & $0.59$ \\ 
 &$r$ & $5.43$ & $2.61$ & $0.43$ && $4.98$ & $2.61$ & $0.61$ \\ 
 &$r^*$ & $5.40$ & $2.58$ & $0.43$ && $5.02$ & $2.62$ & $0.62$ \\ 
\hline
$(\sigma^2,\tau)=(10^{-3},3)$&  $w$ &  $4.93$ & $2.48$ & $0.53$ && $4.83$ & $2.46$ & $0.54$ \\ 
 &$r$ &$4.91$ & $2.47$ & $0.52$ && $4.88$ & $2.46$ & $0.55$ \\ 
 &$r^*$ & $4.85$ & $2.40$ & $0.50$ && $4.90$ & $2.49$ & $0.55$ \\ 
\hline
$(\sigma^2,\tau)=(10^{-1},1)$&  $w$ &  5.45 & 2.88 & 0.64 && 4.15 & 2.09&  0.37\\
 &$r$ &  5.26 & 2.78 & 0.59 &&  4.27 & 2.15 & 0.37  \\
&$r^*$ &  4.89&  2.53&  0.54  && 4.65& 2.37& 0.42 \\

\hline
$(\sigma^2,\tau)=(10^{-1},2)$&  $w$ & $5.62$ &$2.98$ &$0.67$&& $4.60$ & $2.28$ & $0.50$ \\
&$r$ & $5.48$ & $2.75$ & $0.61$ && $4.70$ & $2.32$& $0.54$\\
&$r^*$& $5.06$ & $2.48$ &  $0.54$ && $5.06$ & $2.56$ &  $0.64$\\
\hline
$(\sigma^2,\tau)=(10^{-1},3)$&  $w$ &  $5.66$ & $2.86$ & $0.65$ && $4.49$ & $2.26$ & $0.49$ \\ 
&$r$ & $5.54$ & $2.70$ & $0.63$ && $4.58$ & $2.35$ & $0.50$ \\ 
&$r^*$ &$5.06$ & $2.50$ & $0.59$ && $4.92$ & $2.57$ & $0.53$ \\ 
\hline
$(\sigma^2,\tau)=(1,1)$&  $w$ & $7.21$ & $4.10$ & $1.04$ && $3.18$ & $1.37$ & $0.16$ \\ 
 &$r$ &  $6.45$ & $3.36$ & $0.68$ && $3.35$ & $1.53$ & $0.21$ \\ 
 &$r^*$& $5.39$ & $2.64$ & $0.55$ && $4.56$ & $2.37$ & $0.31$ \\ 
\hline
$(\sigma^2,\tau)=(1,2)$ &  $w$ &  $6.58$ & $3.72$ & $0.92$ && $3.64$ & $1.50$ & $0.24$ \\ 
 &$r$ &$5.92$ & $3.16$ & $0.57$ && $3.80$ & $1.62$ & $0.29$ \\ 
 &$r^*$ & $4.81$ & $2.40$ & $0.41$ && $5.32$ & $2.59$ & $0.48$ \\ 
\hline 
$(\sigma^2,\tau)=(1,3)$  &  $w$  &$6.40$ & $3.50$ & $0.96$ && $3.73$ & $1.64$ & $0.25$ \\ 
&$r$&$5.74$ & $3.03$ & $0.64$ && $3.94$ & $1.76$ & $0.25$ \\ 
&$r^*$ &$4.74$ & $2.49$ & $0.54$ && $5.19$ & $2.72$ & $0.46$ \\ 
\hline
$(\sigma^2,\tau)=(2,1)$ &  $w$ &  $7.78$ & $4.57$ & $1.41$ && $2.12$ & $0.69$ & $0.03$ \\ 
&$r$  &  $6.52$ & $3.56$ & $0.78$ && $2.27$ & $0.75$ & $0.03$ \\ 
&$r^*$ &$5.19$ & $2.64$ & $0.58$ && $4.46$ & $1.85$ & $0.10$ \\  
\hline
$(\sigma^2,\tau)=(2,2)$&  $w$  &  $7.94$ & $4.59$ & $1.33$ && $2.30$ & $0.71$ & $0.05$ \\ 
 &$r$  & $6.77$ & $3.49$ & $0.71$ && $2.43$ & $0.88$ & $0.05$ \\ 
&$r^*$ &  $5.18$ & $2.60$ & $0.54$ && $4.62$ & $2.09$ & $0.18$ \\ 
 \hline
$(\sigma^2,\tau)=(2,3)$&  $w$  & $7.67$ & $4.07$ & $1.15$ && $2.19$ & $0.73$ & $0$ \\ 
 &$r$  &$6.49$ & $3.24$ & $0.65$ && $2.39$ & $0.86$ & $0$ \\ 
&$r^*$ & $4.71$ & $2.34$ & $0.49$ && $4.48$ & $1.98$ & $0.11$ \\  
 \hline
 SE && 0.22 &0.16 & 0.07 &&  0.22 &0.16& 0.07 \\
 \hline \hline
\end{tabular}
\end{adjustbox}
\end{table}

\subsection{Case study B}\label{USUSSR.sect}

\begin{figure}[t]
  \includegraphics[width=0.5\textwidth]{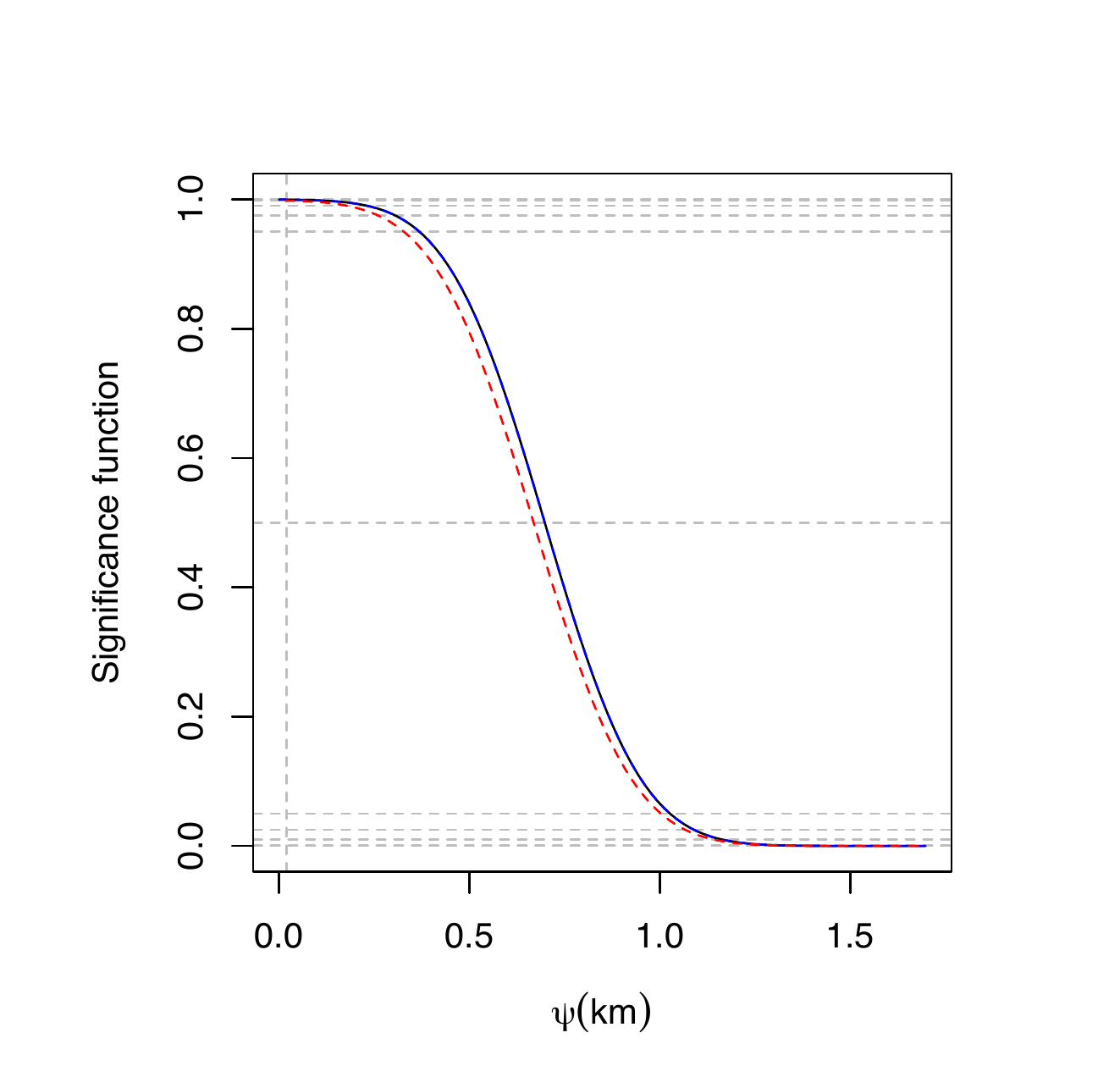}
  \includegraphics[width=0.5\textwidth]{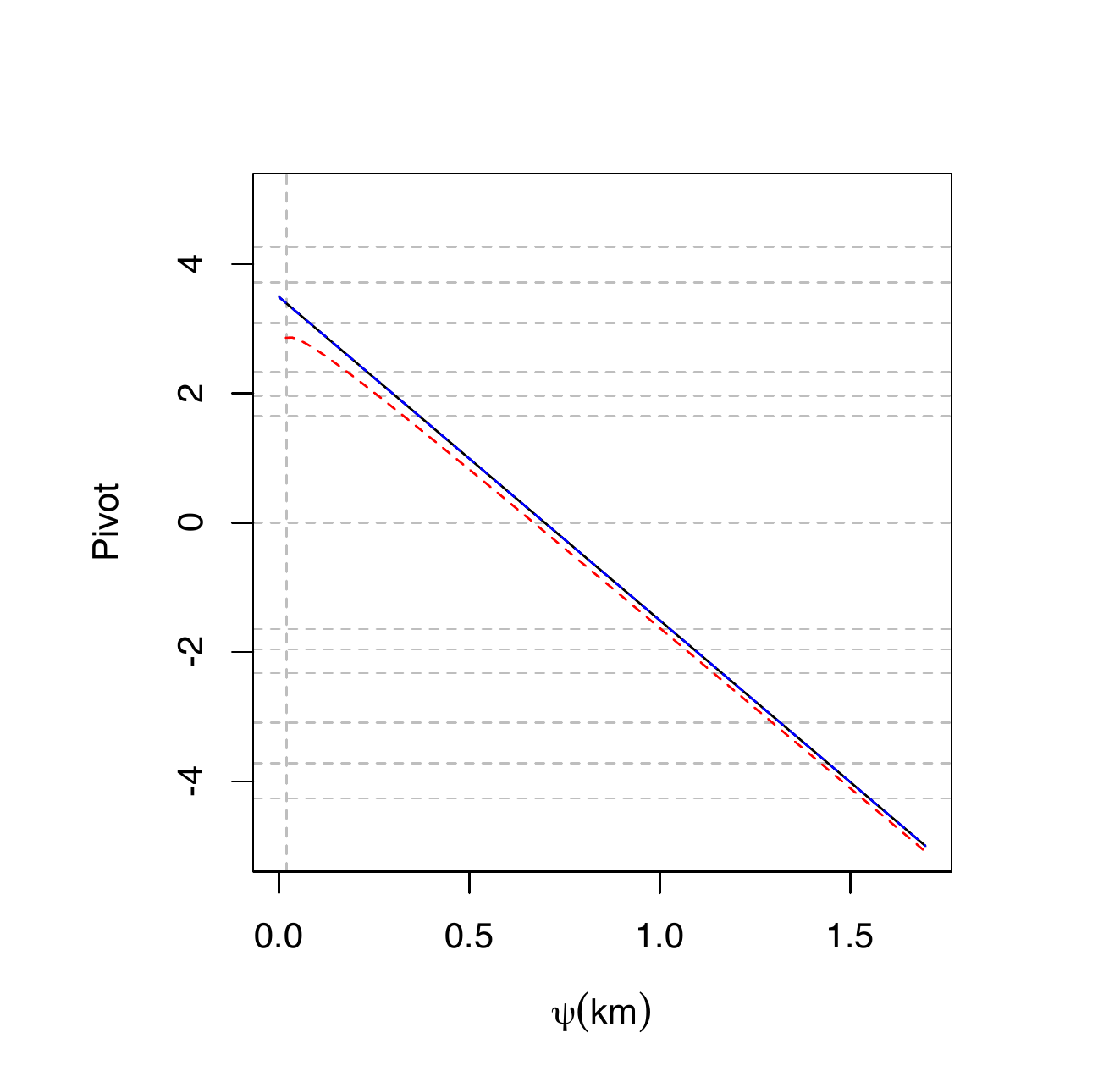}

\caption{Summaries for US and Russian satellite collision event.  Left:  significance functions based on likelihood root $r^\o(\psi)$ (solid black),Wald statistic $w(\psi)$ (dotted blue), and modified likelihood root $r^{*\o}(\psi)$ (red dashes).  Right: the likelihood root $r^\o(\psi)$ (solid black),Wald statistic $w(\psi)$ (dotted blue), and modified likelihood root $r^{*\o}(\psi)$ themselves.}
\label{fig:fig6}
\end{figure}

Our second example is the 10 February 2009 collision of the U.S.~operational communications satellite Iridium 33 and the decommissioned Russian communications satellite Cosmos $2251$, whose relative configuration is given in  Table~\ref{tab:tab2}. Figure~\ref{fig:fig6} shows the significance functions for this conjunction.   \red{With a safety threshold of 20m, corresponding to $\psi_0=0.02$ (the dashed vertical line in both panels), the significance functions all suggest that evasive action is essential except that the probability of collision is $1.14687\times 10^{-5}$ smaller than typically-employed remediation threshold of $10^{-4}$, meaning that a conjunction remediation action is not needed. The Wald statistic and the likelihood root are indistinguishable, but for the modified likelihood root is shifted slightly to the left, reducing the upward bias and slightly weakening any evidence that the true miss distance is greater than $\psi_0$.  The significance probabilities for testing $\psi=\psi_0=20$~m against $H_+: \psi>\psi_0$ are $1.2\times10^{-3}$ for for the Wald statistic and the likelihood root, and $7.2\times10^{-3}$ for the modified likelihood root, so the same conclusions would be drawn for both approaches if the threshold is $\v=10^{-4}$, but clearly there is potential for different conclusions if we consider different uncertainties, and other values of $\v$ and $\psi_0$.  The posterior probability $\pr(\psi\leq \psi_0\mid y^\o)$ would be similar to the significance probability for the modified likelihood root.}

Table~\ref{tab:tab5} gives left and right error rates for different values of $\sigma^2$ and $\tau$ for this conjunction. Its first row corresponds to  the case where the standard deviation of the position error along each axis is $10 \mathrm{m}$ and the standard deviation of the velocity errors is $10 \mathrm{m} / \mathrm{s}$, giving $(\sigma,\tau)=(10^{-2},1)$. In lower rows we first increase uncertainty on the position while keeping uncertainty on the velocity fixed by increasing $\tau$, and then increase both velocity and position errors by increasing $\sigma^2$. For these simulations the true miss distance and relative velocity are taken to have  $\psi=698$ m and $\|\nu\|=11.648\times 10^3$ m/s,  so the errors we consider vary from $10$ to $200$ for both velocity (m/s) and distance (m).

The error rates for the Wald statistic and the likelihood root are almost identical, implying that the corresponding pivots are indistinguishable. For small velocity errors, with $\sigma^2< 10^{-3}$, there is no significant difference in the error rates for the three statistics. For larger velocity variance, although the overall error rates  found by summing the left and the right error rates equal the nominal values, left error always dominates the sum.  In these cases the modified likelihood root  is more symmetric and shows fewer extreme values, especially for large $\tau$, and thus interval estimates based on $r^*$  are more reliable.

\begin{table}[p]
\caption{ Left and right error rates ($\%$) for two-sided nominal $10 \%, 5 \%,$ and $1 \%$ confidence intervals for the parameter $\psi$ of the U.S.~and Russian satellite collision event, Case Study B, based on $10^4$ Monte Carlo replications. The standard errors (SE) appear in the last line.}\label{tab:tab4}
\center
\begin{adjustbox}{width=\textwidth,totalheight=0.8\textheight,keepaspectratio}
\begin{tabular}{crccccccc}
\hline \hline 
&&\multicolumn{3}{c}{Left tail $(\%)$}&&\multicolumn{3}{c}{Right tail $(\%)$} \\
\cline{3-5} \cline{7-9}
Uncertainty & Statistic&5 & 2.5 & 0.5&& 5 & 2.5 & 0.5 \\
\hline
$(\sigma^2,\tau)=(10^{-4},1)$ & $w$ & $5.06$ & $2.54$ & $0.42$ && $5.00$ & $2.53$ & $0.64$ \\ 
& $r$ & $5.06$ & $2.54$ & $0.42$ && $5.00$ & $2.53$ & $0.64$ \\ 
& $r^*$ &$4.98$ & $2.46$ & $0.41$ && $5.05$ & $2.62$ & $0.65$ \\ 
\hline 
$(\sigma^2,\tau)=(10^{-4},2)$ & $w$ &$5.27$ & $2.73$ & $0.60$ && $4.66$ & $2.29$ & $0.52$ \\ 
& $r$ & $5.27$ & $2.73$ & $0.60$ && $4.66$ & $2.290$ & $0.52$ \\ 
& $r^*$ &$5.17$ & $2.67$ & $0.56$ && $4.78$ & $2.33$ & $0.53$ \\ 
\hline
$(\sigma^2,\tau)=(10^{-4},4)$ & $w$ &$4.67$ & $2.29$ & $0.49$ && $4.53$ & $2.34$ & $0.43$ \\ 
& $r$ &$4.67$ & $2.29$ & $0.49$ && $4.53$ & $2.34$ & $0.43$ \\ 
& $r^*$ & $4.58$ & $2.21$ & $0.47$ && $4.67$ & $2.40$ & $0.47$ \\ 
 \hline
$(\sigma^2,\tau)=(10^{-4},10^{2})$ & $w$ & $5.89$ & $2.98$ & $0.74$ && $4.33$ & $2.05$& $0.47$ \\ 
& $r$ &$5.89$ & $2.98$ & $0.74$ && $4.33$ & $2.05$ & $0.47$ \\ 
& $r^*$ &$5.18$ & $2.73$ & $0.65$ && $5.14$ & $2.59$ & $0.61$ \\ 
\hline 
$(\sigma^2,\tau)=(10^{-3},1)$ & $w$ & $5.17$ & $2.66$ & $0.50$ && $4.60$ & $2.16$ & $0.40$ \\ 
& $r$ & $5.17$ & $2.66$ & $0.50$ && $4.60$ & $2.16$ & $0.40$ \\ 
& $r^*$ & $4.96$ & $2.55$ & $0.45$ && $4.88$ & $2.37$ & $0.44$ \\ 
\hline 
$(\sigma^2,\tau)=(10^{-3},2)$ & $w$ &$5.00$ & $2.50$ & $0.48$ && $4.93$ & $2.35$ & $0.37$ \\ 
& $r$ &$5.00$ & $2.50$ & $0.48$ && $4.93$ & $2.35$ & $0.37$ \\ 
& $r^*$ &$4.76$ & $2.34$ & $0.45$ && $5.31$ & $2.54$ & $0.42$ \\ 
\hline 
$(\sigma^2,\tau)=(10^{-3},4)$ & $w$ & $5.25$ & $2.73$ & $0.61$ && $4.54$ & $2.23$ & $0.38$ \\ 
& $r$ & $5.25$ & $2.73$ & $0.61$ && $4.54$ & $2.23$ & $0.38$ \\ 
& $r^*$ & $4.74$ & $2.52$ & $0.52$ && $5.00$ & $2.45$ & $0.43$ \\ 
\hline

$(\sigma^2,\tau)=(10^{-2},1)$& $w$ & $5.75$ & $2.96$ & $0.54$ && $4.53$ & $2.06$ & $0.31$ \\ 
& $r$ & $5.75$ & $2.96$ & $0.54$ && $4.54$ & $2.07$ & $0.32$ \\ 
& $r^*$ &  $5.09$ & $2.56$ & $0.44$ && $5.33$ & $2.70$ & $0.43$ \\ 
\hline
$(\sigma^2,\tau)=(10^{-2},2)$ & $w$ & $5.96$ & $2.94$ & $0.51$ && $3.69$ & $1.92$ & $0.30$ \\ 
& $r$  & $5.96$ & $2.94$ & $0.51$ && $3.69$ & $1.92$ & $0.30$ \\ 
& $r*$ &  $4.89$ & $2.39$ & $0.38$ && $4.87$ & $2.53$ & $0.49$ \\ 
\hline 
$(\sigma^2,\tau)=(10^{-2},4)$ & $w$ &  $6.20$ & $3.26$ & $0.57$ && $2.73$ & $1.14$ & $0$ \\ 
& $r$ & $6.20$ & $3.26$ & $0.57$ && $2.73$ & $1.14$ & $0$ \\ 
& $r*$ & $4.78$ & $2.45$ & $0.41$ && $4.96$ & $2.30$ & $0.31$ \\ 
 \hline
 SE && 0.22 &0.16 & 0.07 &&  0.22 &0.16 & 0.07\\
 \hline \hline 
\end{tabular}
\end{adjustbox}

\end{table}

\newpage
\subsection{Case study C}\label{CARA.sec}

\red{ Our third case study is based on a test case posted by NASA\footnote{
 \url{https://github.com/nasa/CARA_Analysis_Tools/tree/master/two-dimension_Pc/UnitTest/InputFiles}} as one of the limited datasets for their publicly-released Conjunction Assessment Risk Analysis (CARA) tool. Each test case contains data that can be readily converted to the Earth-centered inertial (ECI) reference frame  for both the primary and secondary, $\eta_{s1}$ and $\eta_{s2}$, respectively. The test cases also contain state covariance data, originally expressed in the Cartesian radial, in-track and cross-track (RIC) reference frame (also called the UVW frame). The RIC covariance matrices can be transformed to the ECI state covariances for the primary and secondary, respectively denoted by $\Omega_{s1}^{-1}$ and $\Omega_{s1}^{-1}$, which are $6 \times 6$ matrices. These transformed quantities can then be used to define the relative ECI state at close approach, $\eta=\eta_{s1}-\eta_{s2}$,  and the associated combined covariance, $\Omega^{-1}=\Omega_{s1}^{-1}+\Omega_{s2}^{-1}$.}
 
\begin{figure}[t]
  \includegraphics[width=0.5\textwidth]{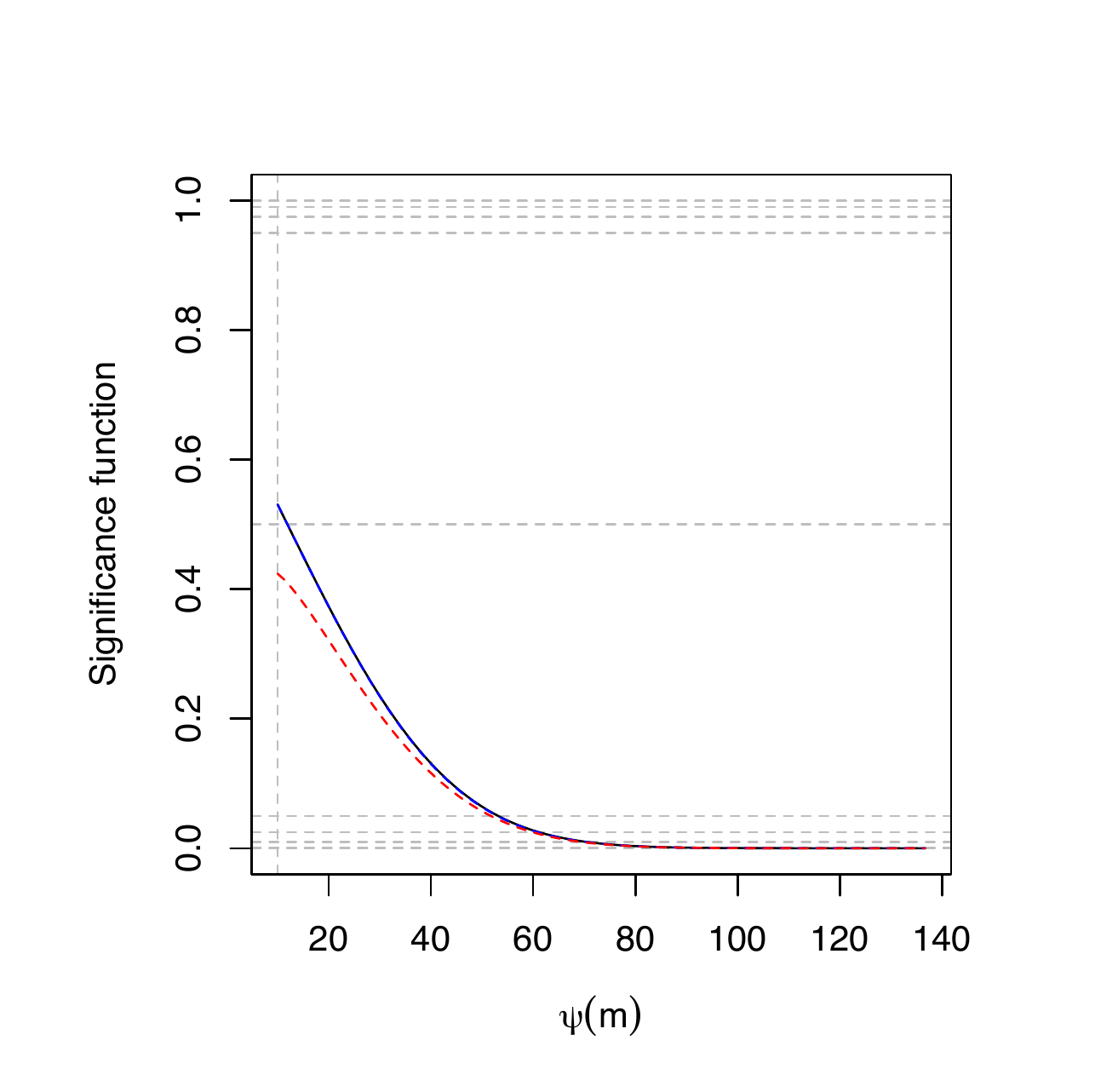}
  \includegraphics[width=0.5\textwidth]{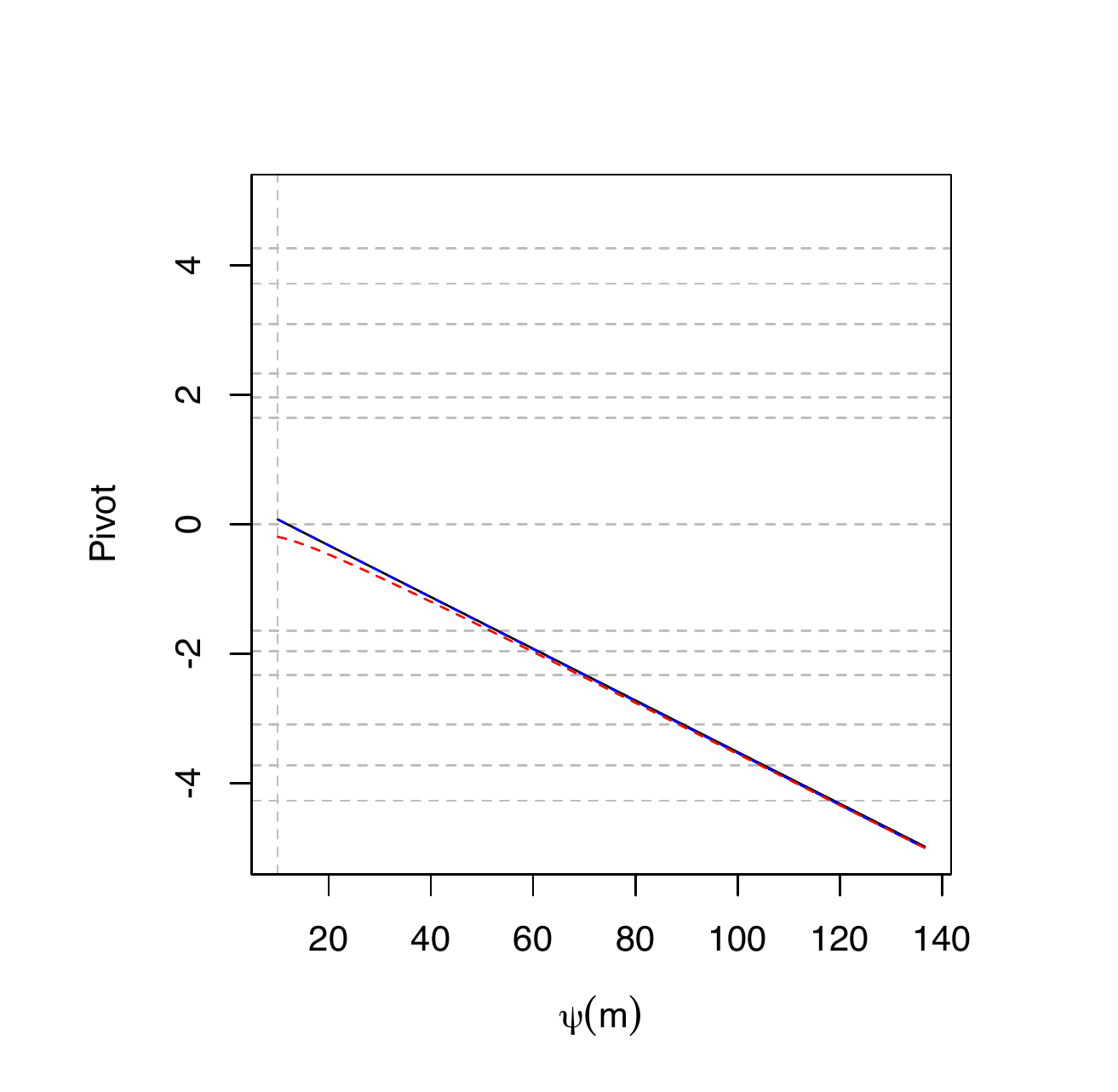}
\caption{Summaries of the CARA event.  Left:  significance functions based on likelihood root $r^\o(\psi)$ (solid black),Wald statistic $w(\psi)$ (dotted blue), and modified likelihood root $r^{*\o}(\psi)$ (red dashes),  Right: the same quantities transformed to the standard normal scale.}
\label{fig:fig7}
\end{figure}

\red{Column C of Table~\ref{tab:tab2} shows the conjunction elements for this example. Since the motion is linear and uncertainty on the velocity can be ignored, we use the reduced two-dimensional model representing conjunction in the encounter plane. The projected quantities in this plane, as descried in Appendix C, are given by 
$$
A=(C V, v /\|v\|)=\begin{bmatrix}
  0.72 & 0.11 & -0.69 \\ 
  0.68 & -0.30 & 0.67 \\ 
  0.13 & 0.95 & 0.28 \\ 
   \end{bmatrix}, \quad 
 x^\o=A^{\mathrm{T}} y^\o=\begin{bmatrix}
  11.84 \\ 
  -1.36 \\ 
   \end{bmatrix}, \quad 
   D=\operatorname{diag}\left(d_{1}^{2}, d_{2}^{2}\right)= \begin{bmatrix}
 25.1^2 & 0.00 \\ 
  0.00 & 11.61^2 \\ 
   \end{bmatrix}, 
$$ 
so $\hat\psi=\|x\|= 11.92$~m.  Not surprisingly, the significance functions plotted in Figure \ref{fig:fig7} show that with $\psi_0=10$~m the significance probabilities $p_\obs$ are all of order 0.5, though that for the modified likelihood root is closer to 0.6; clearly evasive action would be essential in this case.}

\red{Table~\ref{tab:tab5} shows empirical error rates for the three pivots based on $10^6$ values of $x$ simulated from the bivariate normal distribution with  $\xi=x^\o$ and covariance matrix $c^2D$, with $c^2$ varying from $0.005$ to $4$, as in Figure \ref{fig:fig2}. In order to study the error rates for very rare events, we took values of $\alpha$ ranging from 2.5\% down to 0.005\%,  the latter corresponding to two-sided 99.99\% confidence intervals; the one-sided error rates for $\alpha=0.05\%$ and $\alpha=0.005\%$ span the level $\v=0.01\%$ above which evasive action might be considered.  Under the conditions of the CARA event,  $c^2=1$, the left-tail error is systematically high and the right-tail error is exactly zero.  This is unsurprising, because in this setting the pivots can be close to zero, and with increasing uncertainty the upper confidence limit is almost invariably larger than $\psi_0$. As $c^2$ decreases, the uncertainty becomes unrealistically small and the properties of all three pivots improve.  When $c^2$ increases, the error rates for $r$ and $w$ remain poor, and  we see that $r^*$ has better error rate properties, particularly in the left tail, which is of most interest.  Although the errors for $r^*$ are closer to the nominal level, it should be jointly interpreted with the other pivots, especially when the uncertainties are large relative to the miss distance.  In a second experiment we fixed $D$ to the real uncertainty matrix but increased $\xi$ to $c'\xi$, for $c'>1$. A situation in which we expect  $\E(\|x\|)\approx\psi_0$ for large $c'$, hence smaller bias in the probability of collision. Table~\ref{tab:tab6} shows the resulting error rates, again based on $10^6$ simulated values of $x$.  All the error rates approach their nominal values as $c'$ increases, but  $r^*$ again performs best overall, particularly in the left tail.}

\begin{table}[p]
\center
\caption{Empirical left- and right-tail error rates ($\%$) at nominal levels $\alpha= 2.5 \%,0.5 \%, 0.05\%$ and $0.005 \%$ for the CARA event, Case Study C, with variance matrix $c^2D$, based on $10^6$ Monte Carlo samples.  The standard errors (SE) appear in the last line.}
\label{tab:tab5}
\begin{adjustbox}{width=1\textwidth,totalheight=0.8\textheight,keepaspectratio}
$\begin{array}{ll cccc c cccc}
\hline \hline& & \multicolumn{4}{c}{ \text{Left tail } (\%)} && \multicolumn{4}{c}{ \text{Right tail } (\%)}  \\
\cline { 3 - 6 } \cline { 8 - 11 }
 c^2 & \text{Statistic} &   2.5 & 0.5 & 0.05&0.005&& 2.5 & 0.5&0.05&0.005\\
 
  \hline& w & 2.5986 & 0.5229 & 0.0504 & 0.0062 && 2.4616 & 0.4985 & 0.0542 & 0.0062 \\ 
     0.005& r & 2.6115 & 0.5275 & 0.0507 & 0.0062 && 2.4274 & 0.4817 & 0.0494 & 0.0047 \\ 
     & r^{*} & 2.5206 & 0.5059 & 0.0490 & 0.0060 && 2.5256 & 0.5031 & 0.0524 & 0.0048 \\ 
  
 \hline& w & 2.6242 & 0.5278 & 0.0531 & 0.0054 && 2.4530 & 0.5194 & 0.0665 & 0.0117 \\ 
   0.01& r  & 2.6411 & 0.5347 & 0.0548 & 0.0057 && 2.3701 & 0.4644 & 0.0433 & 0.0039 \\ 
   & r^{*} & 2.5064 & 0.5044 & 0.0500 & 0.0053 && 2.5087 & 0.5008 & 0.0476 & 0.0042 \\ 
 
   \hline&w& 2.7484 & 0.5611 & 0.0538 & 0.0051 && 3.1974 & 1.3743 & 0.4281 & 0.0616 \\ 
  0.05& r  &  2.8128 & 0.5804 & 0.0570 & 0.0055 && 0.8226 & 0.0000 & 0.0000 & 0.0000 \\ 
   & r^{*} &2.5127 & 0.5070 & 0.0474 & 0.0047 && 1.3042 & 0.0000 & 0.0000 & 0.0000 \\

\hline& w & 2.7974 & 0.5681 & 0.0556 & 0.0057 && 2.0243 & 0.3167 & 0.0000 & 0.0000 \\ 
  0.1& r  &2.9171 & 0.6048 & 0.0623 & 0.0062 && 0.0000 & 0.0000 & 0.0000 & 0.0000 \\ 
  & r^{*} & 2.4808 & 0.5020 & 0.0481 & 0.0048 && 0.0041 & 0.0004 & 0.0001 & 0.0000 \\

 \hline&w&2.8968 & 0.5855 & 0.0602 & 0.0061 && 0.1192 & 0.0000 & 0.0000 & 0.0000 \\ 
0.2& r  &  3.1142 & 0.6499 & 0.0680 & 0.0071 && 0.0000 & 0.0000 & 0.0000 & 0.0000 \\ 
  & r^{*} & 2.4780 & 0.4964 & 0.0497 & 0.0050 && 0.0414 & 0.0079 & 0.0013 & 0.0004 \\ 

  \hline& w &  3.1192 & 0.6139 & 0.0650 & 0.0082 && 0.0000 & 0.0000 & 0.0000 & 0.0000 \\ 
0.5& r  &   3.6289 & 0.7466 & 0.0807 & 0.0104 && 0.0000 & 0.0000 & 0.0000 & 0.0000 \\ 
  & r^{*} & 2.4578 & 0.4864 & 0.0510 & 0.0070 && 0.3016 & 0.0862 & 0.0212 & 0.0080 \\ 

 \hline&w& 3.3729 & 0.6352 & 0.0666 & 0.0088 && 0.0000 & 0.0000 & 0.0000 & 0.0000 \\ 
 0.8& r  &  4.1558 & 0.8421 & 0.0891 & 0.0110 && 0.0000 & 0.0000 & 0.0000 & 0.0000 \\ 
  & r^{*} & 2.4974 & 0.4877 & 0.0508 & 0.0066 && 0.6050 & 0.2070 & 0.0618 & 0.0264 \\

\hline& w&   3.5192 & 0.6687 & 0.0619 & 0.0050 && 0.0000 & 0.0000 & 0.0000 & 0.0000 \\ 
1& r  &  4.4715 & 0.9130 & 0.0920 & 0.0075 && 0.0000 & 0.0000 & 0.0000 & 0.0000 \\ 
  & r^{*} & 2.5252 & 0.5021 & 0.0475 & 0.0036 && 0.7748 & 0.2829 & 0.0899 & 0.0380 \\

 \hline& w&4.1574 & 0.7724 & 0.0737 & 0.0081 && 0.0000 & 0.0000 & 0.0000 & 0.0000 \\ 
2& r  &   5.7200 & 1.1922 & 0.1211 & 0.0129 && 0.0000 & 0.0000 & 0.0000 & 0.0000 \\ 
  & r^{*} & 2.6649 & 0.5265 & 0.0512 & 0.0052 && 1.5817 & 0.6505 & 0.2498 & 0.1222 \\

\hline & w &   4.6041 & 0.8318 & 0.0761 & 0.0070 && 0.0000 & 0.0000 & 0.0000 & 0.0000 \\ 
3& r  &  6.5660 & 1.3730 & 0.1428 & 0.0131 && 0.0000 & 0.0000 & 0.0000 & 0.0000 \\ 
  & r^{*} & 2.7466 & 0.5273 & 0.0516 & 0.0052 && 2.2104 & 0.9820 & 0.4040 & 0.2065 \\

 \hline& w&  4.9401 & 0.8807 & 0.0772 & 0.0070 && 0.0000 & 0.0000 & 0.0000 & 0.0000 \\ 
4& r  &  7.2084 & 1.5009 & 0.1495 & 0.0147 && 0.0000 & 0.0000 & 0.0000 & 0.0000 \\ 
 & r^{*} &  2.8042 & 0.5438 & 0.0520 & 0.0057 && 2.8894 & 1.3420 & 0.5913 & 0.3171 \\

\hline\hline
\mbox{SE } (\times 10^{-3})& & 15.6&  7.05 &  2.24 &  0.71 && 15.6&   7.05& 2.24   & 0.71 \\
\hline\hline 

\end{array}$
\end{adjustbox}
\end{table}

\begin{table}[t]
\center
\caption{Empirical left- and right-tail error rates ($\%$) at nominal levels $\alpha= 2.5 \%,0.5 \%, 0.05\%$ and $0.005 \%$ for the CARA event, Case Study C, with true position vector $c'\xi$ in the encounter plane, based on $10^6$ Monte Carlo samples.  The standard errors (SE) appear in the last line.}
\label{tab:tab6}
\begin{adjustbox}{width=0.8\textwidth,totalheight=0.5\textheight,keepaspectratio}
$\begin{array}{lc cccc c cccc}
\hline \hline& & \multicolumn{4}{c}{ \text{Left tail } (\%)} && \multicolumn{4}{c}{ \text{Right tail } (\%)}  \\
\cline { 3 - 6 } \cline { 8 - 11 }
 c'& \text{Statistic} &   2.5 & 0.5 & 0.05&0.005&& 2.5 & 0.5&0.05&0.005\\

 \hline& \text{Wald} & 2.9354 & 0.5837 & 0.0570 & 0.0066 && 0.0045 & 0.0000 & 0.0000 & 0.0000 \\ 
 2& r  &  3.2085 & 0.6631 & 0.0677 & 0.0079 && 0.0000 & 0.0000 & 0.0000 & 0.0000 \\ 
  & r^{*} &   2.4625 & 0.4821 & 0.0478 & 0.0055 && 0.0749 & 0.0157 & 0.0027 & 0.0007 \\

\hline& \text{Wald} & 2.8360 & 0.5736 & 0.0620 & 0.0063 && 1.6491 & 0.1615 & 0.0000 & 0.0000 \\ 
3& r  &    2.9632 & 0.6124 & 0.0675 & 0.0070 && 0.0000 & 0.0000 & 0.0000 & 0.0000 \\ 
 & r^{*} &    2.5088 & 0.5057 & 0.0530 & 0.0048 && 0.0052 & 0.0005 & 0.0000 & 0.0000 \\

\hline& \text{Wald} &    2.7575 & 0.5586 & 0.0580 & 0.0053 && 3.0591 & 1.2037 & 0.1975 & 0.0077 \\ 

 4& r  &   2.8315 & 0.5846 & 0.0615 & 0.0058 && 0.0000 & 0.0000 & 0.0000 & 0.0000 \\ 
 & r^{*} &  2.4990 & 0.4999 & 0.0506 & 0.0046 && 0.0052 & 0.0000 & 0.0000 & 0.0000 \\

\hline& \text{Wald} &   2.7127 & 0.5559 & 0.0523 & 0.0048 && 3.1485 & 1.3268 & 0.5772 & 0.2385 \\ 
5& r  &  2.7627 & 0.5703 & 0.0554 & 0.0052 && 2.1960 & 0.4215 & 0.0380 & 0.0027 \\ 
 & r^{*} &   2.5079 & 0.5045 & 0.0475 & 0.0040 && 2.5059 & 0.4962 & 0.0480 & 0.0037 \\ 
 
 \hline& \text{Wald} &  2.6819 & 0.5358 & 0.0523 & 0.0047 && 2.7548 & 0.9488 & 0.4112 & 0.2252 \\ 
6& r  &   2.7249 & 0.5494 & 0.0537 & 0.0049 && 2.2379 & 0.4364 & 0.0428 & 0.0051 \\ 
 & r^{*} &  2.4966 & 0.4969 & 0.0483 & 0.0044 && 2.4794 & 0.4978 & 0.0508 & 0.0060 \\ 
 
\hline\hline
\text{SE }(\times 10^{-3})& & 15.6&  7.05 &  2.24 &  0.71 && 15.6&   7.05 & 2.24   & 0.71 \\
\hline\hline 

\end{array}$

\end{adjustbox}
\end{table}

\section{Further remarks}\label{further.sect}


%
In some cases successive six-dimensional observation vectors $y_1,\ldots, y_n$ and corresponding $6\times 6$ variance matrices $\Omega^{-1}_1,\ldots, \Omega^{-1}_n$ are available, with the variance matrices increasingly concentrated as information accrues on a conjunction.  If the observations can be regarded as independent, then the corresponding log likelihood is 
$$
\ell(\psi,\lambda_1,\ldots,\lambda_n) = -\frac12 \sum_{j=1}^n \{y_j-\eta(\vartheta_j)\}^\T \Omega_j  \{y_j-\eta(\vartheta_j)\}, 
$$  
where $\vartheta_j=(\psi,\lambda_j)$, with $\psi$ representing the miss distance common to all the observations and $\lambda_1,\ldots, \lambda_n$ representing the $n$ $5\times 1$ vectors of nuisance parameters corresponding to $y_1,\ldots, y_n$.  The more precise $y_j$ are automatically given higher weight, since the corresponding dispersion matrices $\Omega_j$ are larger.  In this case the overall parameter vector is $\vartheta = (\psi,\lambda_1,\ldots, \lambda_n)$, and the approach of Section~\ref{likelihood.sec} can be applied with minor changes.   \red{A similar but more complex generalisation should be feasible when the observations are dependent due to batch updates.}

More complicated geometry leading to a different form for $\eta(\vartheta_j)$ would be needed if the relative motions could not be considered to be rectilinear. 

%
\red{Likelihood methods rest on distributional assumptions, and one might query whether it is appropriate to assume normality of the observation vector.  Provided suitable data are available, standard methods could be used to check model adequacy and used to elaborate the model if this was found to be necessary. For example, the multivariate Student $t$ or Laplace distributions might be used, though the numerical details would be more complex.}

\red{The incorporation of reliable prior information would be valuable, and might for example be based on the output of a filtering approach to tracking.  One might then treat conjunction analysis as a prediction problem rather than an estimation problem, and then the Bayesian formalism would be attractive.}

\section{Conclusion}

In this paper we formulated conjunction assessment in statistical terms, discussed approximate likelihood inference on the miss distance, and studied the repeated-sampling properties of likelihood statistics with reference to conjunction assessment. \red{Viewed in our framework the usual conjunction probability estimator has a downward bias that seems not to have been noticed previously, and the so-called probability dilution paradox vanishes, since it refers to the properties of an estimator of the conjunction probability, rather than to the probability itself.}  Examples illustrate inference on the miss distance, suggest that standard likelihood confidence intervals may need improvement when uncertainty on the relative distance and velocity is large, and numerical work show how an improved approximation gives appreciably better inferences.  In our setup the estimated conjunction probability is replaced by a significance probability for testing that the true miss distance is larger than a specified safety threshold.  If the model is correctly specified then this significance probability is calibrated in a repeated-sampling sense and thus provides a statistically well-founded basis for avoidance decisions. 

\section*{\red{Appendix A: Known velocity}}\label{known-vel}

\red{The statistical model simplifies when the velocity $\nu$ of the second object relative to the first can be assumed known.  The random data then consist of a $3\times 1 $ vector $y$ containing the estimated position of the second object relative to the first, and it is assumed that $y\sim \N_3(\mu, \Omega^{-1})$, where $\mu$ and $\Omega$ are of respective dimensions $3 \times 1 $ and $3\times 3$. Now let $A$ be a $3\times 3$ orthogonal matrix whose final column is $\nu/\|\nu\|$ and whose other columns are chosen so that the upper left $2\times 2$ corner of $A^\T\Omega^{-1}A$ is diagonal, say, $D=\diag(d_1^2,d_2^2)$. If so, then $x=A^\T y\sim\N_3(A^\T\mu, A^\T\Omega A)$, and the first two components of $x$ are independent, with distribution $\N_2(\xi, D)$, say. Hence $\xi$ is the projection of the true position of the second object along its velocity vector onto the encounter plane, and thus $\psi=\|\xi\|$ is the miss distance.  In terms of the coordinates in the encounter plane defined by $A$ we can write $ \xi=(\psi\cos\lambda, \psi\sin\lambda)$ and $x=(x_1,x_2)$, where $\psi\geq 0$ is the parameter of interest and $\lambda\in[0,2\pi)$ is the nuisance parameter.  We describe how to obtain $A$ at the end of this appendix.}

\red{The transformation from $y$ to $x$ is invertible and depends only on the known quantities $\nu$ and $\Omega$, so no statistical information is lost in using $x$ instead of $y$.  In satellite conjunction analysis it is customary to ignore the coordinate $x_3$ of $y$ in the direction orthogonal to the encounter plane, which in statistical terms amounts to basing inference on the joint density of $x_1$ and $x_2$, leading to the log likelihood~\eqref{radial.lik}.}

\red{To obtain a suitable $3\times 3$ projection matrix $A$, note that if we define 
$$
B = (b_1,b_2,\nu) = \begin{pmatrix} 0& \nu_2^2+\nu_3^2&\nu_1\\ \nu_3&-\nu_1\nu_2&\nu_2\\ -\nu_2&-\nu_1\nu_3&\nu_3 \end{pmatrix} , 
\quad 
N= \begin{pmatrix} \|b_1\|^{-1}&0&0\\ 0&\| b_2\|^{-1}&0\\ 0&0& \|\nu\|^{-1} \end{pmatrix} , 
$$
then the first two columns of the orthonormal matrix $BN$ span the encounter plane.  If $C$ denotes the $3\times 2$ matrix containing these columns, then $C^\T y$
is the orthogonal projection of $y$ onto the encounter plane.  If $V D V^\T $ is the spectral decomposition of $C^\T \Omega^{-1} C$, then $(CV)^\T\Omega^{-1} (CV) = V^\T C^\T \Omega^{-1} C V = D=\diag(d^2_1,d^2_2)$.  Thus $(CV)^\T y$ is bivariate normal with mean $\xi=(CV)^\T \mu$ and covariance matrix $D$. Hence if we let $A=(CV, \nu/\|\nu\|)$, then the first two elements of $x=A^\T y$ are independent, and satisfy $x_1\sim\N(\xi_1,d_1^2)$ and  $x_2\sim\N(\xi_2,d_2^2)$.}

\section*{Appendix B: Theoretical background}

The key to improving first-order approximations is to account for the presence of nuisance parameters.  One important improvement, the modified likelihood root \citep{
Barndorff-Nielsen.Cox:1994,Fraser.Reid.Wu:1999}, is given by  equation~\eqref{rstar.eq}. If the response distribution is continuous and has true parameter $\psi_0$, then $r^{*}(\psi_0)$ is asymptotically standard normal to order $\mathrm{O}\left(n^{-3 / 2}\right)$. 

The  tangent exponential model proposed by Fraser and co-authors \citep[e.g.,][]{Fraser.Reid.Wu:1999} may be used to construct~\eqref{rstar.eq}.   This is based on a local exponential family approximation to the log-likelihood function for $n$ independent observations $y_1,\ldots, y_n$ that has $d\times 1$ canonical parameter
\begin{equation} \label{phi}
\varphi^\T(\vartheta)=\ell_{;V}(\vartheta;y^\o)=V^\T {\partial \ell(\vartheta)\over\partial y} = \sum_{i= 1}^{n}\dfrac{\partial \ell(\vartheta;y)}{\partial y_i} \Bigr\rvert_{y_i = y_i^\o} V_i ,
\end{equation}
and involves a directional derivative of the log-likelihood function evaluated at the observed data point $y^\o=(y_1^\o,\ldots, y_n^\o)^\T$, as  $\ell_{;V}(\vartheta;y^\o)$ is the derivative of 
$\ell(\vartheta)$ in the directions given by the $n$ rows $V_1,\ldots,V_n$ of the $n\times d$ matrix $V$. This matrix can be constructed using a vector of pivotal quantities $z\equiv z(y;\vartheta)=\{z_1(y_1,\vartheta),\ldots,z_n(y_n,\vartheta)\}^\T$ through
\begin{equation}
\displaystyle V=\left.\dfrac{\partial y}{\partial \vartheta^{\T}}\right|_{y=y^\o,\vartheta=\hat\vartheta^\o} =-\left.\left(\dfrac{\partial z}{\partial y^{\T}}\right)^{-1}\right|_{y=y^\o,\vartheta=\hat\vartheta^\o} \left. \times  \dfrac{\partial z}{\partial \vartheta^{\T}}\right|_{y=y^\o,\vartheta=\hat\vartheta^\o},
\end{equation}
where $\hat{\vartheta}^\o$ is the maximum likelihood estimate at $y^\o$. 
The resulting correction term can be written in the form~\eqref{q.eq}. The numerator of the first term of $q$ is the determinant of a $d \times d $ matrix whose first column is $\varphi(\hat{\vartheta})-\varphi(\hat{\vartheta_{\psi}})$ and whose remaining columns are $\varphi_{\lambda}(\hat{\vartheta_{\psi}})$. Further details can be found in \citet[Section~8.5]{Brazzale.Davison.Reid:2007} and a guide to the literature may be found in \citet{Davison.Reid:2022}.

In order to obtain the matrix $V$ for use when the velocity vector $\nu$ is unknown, we define the vector of pivots as $z(y,\vartheta)=\Omega^{1/2}\{y-\eta(\vartheta)\}$; these are independent and standard normal under the model. Partial differentiation yields 
$$
V={\partial\eta(\vartheta)\over \partial\vartheta^\T}=\eta_\vartheta(\vartheta), 
$$ 
evaluated at the maximum likelihood estimate $\hat\vartheta^\o$ corresponding to $y^\o$. The log-likelihood,  $-\frac{1}{2}(y-\eta)^\T\Omega(y-\eta)$, has derivative $\Omega(\eta-y)$ with respect to $y$, so from~\eqref{phi} the canonical parameter of the tangent exponential model may be written in the form 
$\varphi(\vartheta)=G\eta(\vartheta)+a$,   where $G=-\eta^\T_{\vartheta}(\hat\vartheta^\o)  \Omega$ and $a=\eta^\T_{\vartheta}(\hat\vartheta^\o) \Omega y^\o$ are both constant with respect to $\vartheta$ and $G$ is full-rank. Any canonical parameter that is an affine transformation of $\eta$ gives the same expression for $q(\psi)$, because 
\begin{flalign}
\dfrac{\left|\varphi(\hat{\vartheta})-\varphi\left(\hat{\vartheta}_{\psi}\right) \quad \varphi_{\lambda}\left(\widehat{\vartheta}_{\psi}\right)\right|}{\left|\varphi_{\vartheta}(\hat{\vartheta})\right|}
=& \dfrac{\left| G\left\{\eta(\hat{\vartheta})- \eta(\hat{\vartheta}_{\psi})\right\} \quad G \eta_{\lambda}(\hat{\vartheta}_{\psi})\right|}{\left|G \eta_{\vartheta}(\hat{\vartheta})\right|}= \dfrac{\left| \eta(\hat{\vartheta})- \eta(\hat{\vartheta}_{\psi}) \quad \eta_{\lambda}(\hat{\vartheta}_{\psi}) \right|}{\left|\eta_{\vartheta}(\hat{\vartheta})\right|}, \label{q}
 \end{flalign}
where $\eta_{\lambda}(\vartheta)=\partial{\eta}(\vartheta)/\partial \lambda^{\T}$.  Hence we can take the constructed parameter~\eqref{phi} to be $\varphi(\vartheta)=\eta(\vartheta)$. To compute~\eqref{q}, we need the $6\times 6$ Jacobian  $\eta_{\vartheta}$ and the second derivatives $\eta_{\vartheta\vartheta}$, a $6\times 6 \times 6$ tensor containing the second derivatives of $\eta$, which is needed to compute $\jmath_{\lambda\lambda}(\hat\vartheta_\psi)$.
The score equation and the observed Fisher information can respectively be given as
\begin{equation}\label{obsinfo.eq}
{\partial\eta^\T(\vartheta)\over\partial\vartheta} \Omega\{y-\eta(\vartheta)\} = 0, \quad 
{\partial\eta^\T(\vartheta)\over\partial\vartheta} \Omega {\partial\eta(\vartheta)\over\partial\vartheta_r}  - {\partial^2\eta^\T(\vartheta)\over \partial\vartheta\partial\vartheta_r} \Omega\{y-\eta(\vartheta)\}, \quad r=1,\ldots, 6.
%
\end{equation} 
The observed information matrix evaluated at the maximum likelihood estimate for the full model equals $\jmath(\hat\vartheta) = \eta^\T_\vartheta(\hat\vartheta)\Omega \eta_\vartheta(\hat\vartheta)$, because the score equation for the full model implies that $\eta(\hat\vartheta)=y$. 

A fuller version of a similar computation is in~\citet{Fraser.Wong.Wu:1999}. 

\red{In the simplified problem with known $\nu$ there are just two parameters $\psi\geq 0$ and $\lambda\in[0,2\pi)$,  the log likelihood reduces to~\eqref{radial.lik} and $\varphi(\psi,\lambda)=\xi=(\psi\cos\lambda, \psi\sin\lambda)^\T$. The overall maximum likelihood estimates are 
\begin{equation}
\hat\psi= (x_1^2+x_2^2)^{1/2},\quad \hat\lambda = \arctan(x_2/x_1).\label{MLE_2D}
\end{equation} 
 If $\psi$ is fixed, then $\hat\lambda_\psi$ is readily found as the unique minimum of the sum of squares in~\eqref{radial.lik}; a good starting value should be $\hat\lambda$.  Then~  (\ref{r.eq},~\ref{q.eq}) reduce to
\begin{flalign*}
r(\psi)&=\text{sign}(\hat{\psi}-\psi)\dfrac{1}{d_1d_2}\left\{d_2^2\left(x_1-\psi\cos\lambda\right)^2 +d_1^2 \left(x_2-\psi\sin\lambda\right)^2\right\}^{1/2},\\
q(\psi) &= \psi^{1/2}{ x_1\cos\hat\lambda_\psi + x_2\sin\hat\lambda_\psi - \psi\over 
\left\{d_2^2(x_1\cos\hat\lambda_\psi -\psi\cos 2\hat\lambda_\psi) + d_1^2 (x_2\sin\hat\lambda_\psi +\psi\cos 2\hat\lambda_\psi) \right\}^{1/2}}.
\end{flalign*}
}

\red{
When $d_1=d_2=d$, say,  then $\hat\lambda_\psi\equiv \hat\lambda$ does not depend on $\psi$ and after simplification we obtain 
$$
r(\psi)= w(\psi) = (\hat\psi-\psi)/d,\quad q(\psi) = r(\psi)(\psi/\hat\psi)^{1/2}, \quad r^*(\psi) = {\hat\psi-\psi\over d} + {d\over2( \hat\psi-\psi)}\log(\psi/\hat\psi), \quad \psi>0;
$$
note that $r^*(\psi)\to -d/(2\hat\psi)$ when $\psi\to \hat\psi$. The fact that $r^*(\psi)<r(\psi)$ for all $\psi$ implies that confidence intervals for $\psi$ based on $r^*(\psi)$ will be closer to the origin than those based on $r(\psi)$, and significance levels for fixed $\psi$ will be higher, leading to more conservative inferences.  We should consider evasive action based on $r(\psi)$ when $1-\Phi\{r^\o(\psi_0)\}<\v$, i.e.,  when $\hat\psi^\o - dz_\epsilon$ is smaller than the hard-body radius $\psi_0$; if $\epsilon=10^{-4}$ for example, then $z_\epsilon=-3.72$.  Notice that this rule relates the observed distance of the second object from the origin, $\hat\psi^\o$,  to the measurement uncertainty, $d$.  Use of $r^{*\o}(\psi)$ will lead to very similar conclusions in most cases.}

\section*{Appendix C: Bayesian approximation}\label{AppBayes}

The derivation of equation~\eqref{qB.eq} is outlined in Section~8.7 of \citet{Brazzale.Davison.Reid:2007}.  As $\hat\vartheta^\T = (\psi, \hat\lambda_\psi^\T)$, 
$$
 {\D{\ell(\hat\vartheta_\psi)}\over \D{ \psi}}= \ell_\psi(\hat\vartheta_\psi) + {\partial \hat\lambda^\T_\psi\over \partial \psi} \ell_\lambda(\hat\vartheta_\psi)= \ell_\psi(\hat\vartheta_\psi) -  \ell_{\psi\lambda}(\hat\vartheta_\psi)  \ell^{-1}_{\lambda\lambda}(\hat\vartheta_\psi)\ell_\lambda(\hat\vartheta_\psi),
$$
where subscripts on $\ell$ denote partial differentiation and the second equality follows from noting that differentiating the equation $ \ell_\lambda(\hat\vartheta_\psi) =0$ that defines $\hat\lambda_\psi$ yields  
$$
\ell_{\psi \lambda}(\hat\vartheta_\psi) +{\partial \hat\lambda^\T_\psi\over \partial \psi}\, \ell_{\lambda \lambda}(\hat\vartheta_\psi) =0,
$$
with the matrix $\ell_{\lambda \lambda}(\hat\vartheta_\psi)$ invertible because it is the Hessian corresponding to the maximum of $\ell$ in the $\lambda$ direction for fixed $\psi$.  A standard identity for the determinant of a partitioned matrix gives
\begin{equation}\label{nasty.eq}
\left| \begin{matrix}\ell_\vartheta(\hat\vartheta_\psi)&  \ell_{\vartheta\lambda}(\hat\vartheta_\psi)\end{matrix}\right| = 
\left|\begin{matrix} \ell_\psi(\hat\vartheta_\psi)&   \ell_{\psi\lambda}(\hat\vartheta_\psi)\\ \ell_\lambda(\hat\vartheta_\psi)&   \ell_{\lambda\lambda}(\hat\vartheta_\psi) \end{matrix}\right| =
\left\{\ell_\psi(\hat\vartheta_\psi) -  \ell_{\psi\lambda}(\hat\vartheta_\psi)  \ell^{-1}_{\lambda\lambda}(\hat\vartheta_\psi)\ell_\lambda(\hat\vartheta_\psi)\right\}
\times \left| \ell_{\lambda\lambda}(\hat\vartheta_\psi)\right|,
\end{equation}
and the expression for the observed information in~\eqref{obsinfo.eq} implies that we can write
$$
-\ell_{\vartheta\lambda_r}(\vartheta) = \eta^\T_\vartheta(\vartheta)\Omega\eta_{\lambda_r}(\vartheta) + A_r(\vartheta)\{y-\eta(\vartheta)\},\quad r=1,\ldots, d-1, 
$$
where $A(\vartheta)$ involves second derivatives of $\eta$.  Now $y=\eta(\hat\vartheta)$, so if $\hat\psi-\psi$ is of order $n^{-1/2}$, then $y-\eta(\hat\vartheta_\psi)=\eta(\hat\vartheta)-\eta(\hat\vartheta_\psi)$ is also $O(n^{-1/2})$, and hence so too is the second term of $-\ell_{\vartheta \lambda}(\vartheta)$.  Thus~\eqref{nasty.eq} equals 
\begin{eqnarray}
\nonumber
 \left| \begin{matrix}\ell_\vartheta(\hat\vartheta_\psi)&  \ell_{\vartheta\lambda}(\hat\vartheta_\psi)\end{matrix}\right| &=& (-1)^{d-1} \left| \begin{matrix}\ell_\vartheta(\hat\vartheta_\psi)& - \ell_{\vartheta\lambda}(\hat\vartheta_\psi)\end{matrix}\right|, \\ \nonumber
&=&  (-1)^{d-1} \left| \begin{matrix}\eta_\vartheta(\hat\vartheta_\psi)\Omega\{y-\eta(\hat\vartheta_\psi)\}& \eta^\T_\vartheta(\hat\vartheta_\psi)\Omega\eta_\lambda(\hat\vartheta_\psi)\end{matrix}\right| +O(n^{-1/2}),\\  \nonumber
&=& (-1)^{d-1}\left|\eta_\vartheta(\hat\vartheta_\psi)\right|\left|\Omega\right| 
\left| \begin{matrix}\eta(\hat\vartheta)-\eta(\hat\vartheta_\psi)& \eta_\lambda(\hat\vartheta_\psi)\end{matrix}\right| +O(n^{-1/2}).
\end{eqnarray}
As  $\vartheta$ is of dimension $d$ and $|\jmath(\hat\vartheta)| = \left| \eta^\T_\vartheta(\hat\vartheta)\right|^2|\Omega|$,  equation~\eqref{qB.eq} equals 
\begin{eqnarray}
\nonumber
q_B(\psi)&= &{\left| \begin{matrix}\ell_\vartheta(\hat\vartheta_\psi)&  \ell_{\vartheta\lambda}(\hat\vartheta_\psi)\end{matrix}\right|  \over 
{|\ell_{\lambda\lambda}(\hat\vartheta_{\psi})| }}  \times
{|\jmath_{\lambda\lambda}(\hat\vartheta_{\psi})|^{1/2}  \over |\jmath(\hat\vartheta)|^{1/2}}
{f(\hat\vartheta)\over f(\hat\vartheta_\psi)}, \\ \nonumber
 &= &{\left| \begin{matrix}\ell_\vartheta(\hat\vartheta_\psi)&  -\ell_{\vartheta\lambda}(\hat\vartheta_\psi)\end{matrix}\right|  \over 
{|\jmath_{\lambda\lambda}(\hat\vartheta_{\psi})|^{1/2} |\jmath(\hat\vartheta)|^{1/2}}} \times
{f(\hat\vartheta)\over f(\hat\vartheta_\psi)},\\ \nonumber
&=& {\left|\eta_\vartheta(\hat\vartheta_\psi)\right|\left|\Omega\right| 
\left| \begin{matrix}\eta(\hat\vartheta)-\eta(\hat\vartheta_\psi)& \eta_\lambda(\hat\vartheta_\psi)\end{matrix}\right|  \over 
|\jmath_{\lambda\lambda}(\hat\vartheta_{\psi})|^{1/2} 
\left| \eta^\T_\vartheta(\hat\vartheta)\right||\Omega|^{1/2}}  \times
{f(\hat\vartheta)\over f(\hat\vartheta_\psi)} + O(n^{-1/2}), \\ 
&=& q(\psi) \times {\left| \eta^\T_\vartheta(\hat\vartheta_\psi)\right| \over  {\left| \eta^\T_\vartheta(\hat\vartheta)\right|}} \times
{f(\hat\vartheta)\over f(\hat\vartheta_\psi)} + O(n^{-1/2}). \label{nastier.eq}
\end{eqnarray}
where $q(\psi)$ is given in equation~\eqref{q.eq}.

The Jeffreys prior is the root of the determinant of the Fisher information matrix, 
$$
f(\vartheta) \propto \left| \eta^\T_\vartheta(\vartheta)\Omega \eta_\vartheta(\vartheta)\right|^{1/2} \propto \left| \eta_\vartheta(\vartheta)\right|,
$$
as the constant $|\Omega|$ can be ignored.  If this prior is used then~\eqref{nastier.eq} simplifies to  equation~\eqref{useful.eq}, plus a term of $O(n^{-1/2})$.  Hence $r_B(\psi)= r^*(\psi) + O(n^{-1})$, and inferences from~\eqref{r-approx.eq} and~\eqref{B-approx.eq} will be the same to this order of error.

\section*{Appendix D: Implementation details}\label{numdetails}
\red{The main steps for the numerical implementation of higher-order quantities are given in table \ref{tab:tab8}. Numerical experiments have been obtained with Intel® Xeon(R) CPU E5-1650 v3 @3.50GHz × 12 processor with 64GB ram.  The \textsf{R} code can be designed to use parallel loops for speed-up purposes. In terms computational cost (CPU time), it takes roughly 8 seconds to run $10^3$ simulations in the two-dimensional settings. }
\begin{table}[H]
\stepcounter{table}
\renewcommand{\arraystretch}{1.2}
\begin{tabular}{*{1}{@{}L{16cm}}}
    \toprule\toprule
    \multicolumn{1}{@{}l}{\bfseries Numerical details} \tabularnewline
    \bottomrule
    \textbf{Input}: $6\times 1$ state vector $y$ containing the observed $3\times 1$ relative position and velocity vectors $\hat\mu$ and $\hat\nu$, $6\times 6$ covariance matrix $\Omega^{-1}$ for $y$, and a nominal confidence level $\alpha$.
    \tabularnewline
        {\small
    \begin{enumerate}[topsep=0pt]     
    \item Compute the overall maximum likelihood estimates $\hat{\theta}$: (i) In the six-dimensional setting, $\hat{\psi}$ and $\hat{\lambda}=( \hat\theta_1,\hat\phi_1,\hat \|\nu\|,\hat\theta_2,\hat\phi_2)$ are obtained from $(\hat\mu,\hat\nu)$ using equations~\eqref{mu.eq}--\eqref{cosalpha}. (ii) In the simplified two-dimensional setting, we first need to project relative quantities into the encounter plane using the matrix $A$ (see Appendix A), then  $\hat{\psi}$ and $\hat{\lambda}$ are given in equation \eqref{MLE_2D}.
 \item Compute the observed information matrix $\jmath(\hat{\vartheta}) $ given by~\eqref{jhat.eq} (this is a $6\times 6$ or $2\times$ 2 matrix), the estimated variance of each $\hat{\vartheta}_r$ is the $(r,r)$ element of $\jmath(\hat{\vartheta})^{-1}$.  Let $\text{se}(\hat \psi)$ denote the square root of the estimated variance for $\hat\psi$. 
 
   \item Define a grid $\calG=\{\psi'_1, \ldots, \psi'_{n_\psi}\}$ of values of $\psi$ that includes the maximum likelihood estimate $\widehat{\psi}$.  We use a non-uniform grid in the interval 
   $\left[ \max\left(\hat{\psi}-z_{1-\alpha} \text{se}(\hat{\psi}),0\right), \hat{\psi}+z_{\alpha} \text{se}(\hat{\psi})\right]$, such that the mesh is finer for small $\psi$ and coarser for larger values.
 
    \item For each $\psi\in\calG$, obtain the constrained estimates $\hat{\lambda}_{\psi}$ subject to $0 \leq \theta_1,\theta_2 \leq \pi$, $-\pi \leq \phi_1,\phi_2 \leq\pi$ and $0 < ||\nu||$, by minimising the sum of squares in~\eqref{logL.eq}, and store the corresponding values of $\hat\vartheta_\psi=(\psi,\hat\lambda_\psi)$. We have found that 
    \begin{itemize}
    \item  it may help to transform the components of $\vartheta$ to take values in the real line, for example removing the restrictions by maximizing in terms of $\log||\nu||, \log\left\{\theta_i/(\theta_i-\pi)\right\}$, and $\tan (\phi_i/2)$, for $i=1,2$, and 
    \item constrained optimisation using the `Rvmmin' or `nlminb' solvers in the optimx() function of the  \textsf{R} package optimx leads to an algorithm that is overall robust, fast, and generally insensitive to perturbations in initial values. 
 
      \end{itemize}
    \item  Use partial derivatives of the log-likelihood and the mean vector $\eta$ to  evaluate 
    $\ell(\hat{\vartheta}_{\psi})$, $\jmath_{\lambda\lambda}(\hat{\vartheta}_{\psi})$, $\eta(\hat\vartheta_\psi)$ and $\eta_\lambda(\hat\vartheta)$ and then use expressions~\eqref{r.eq},~\eqref{wald.eq} and~\eqref{q.eq} to obtain the values of $r(\psi)$, $w(\psi)$, $q(\psi)$ and $r^*(\psi)$ on $\calG$.
    
    \item  Interpolate $r(\psi)$ and $r^*(\psi)$ on $\calG$  by (for example) a cubic smoothing spline in which the values of $\psi$ are treated as functions of those of $r(\psi)$ and $r^*(\psi)$. Very large values of $r^{*}$ arising in a few cases due to numerical instabilities when $|r|<0.1$ are excluded. 
   
    \item  If required, obtain the point estimate $\widehat{\psi}^*$ of $\psi$ by using the interpolating spline for $r^*(\psi)=0$. This is not needed for the Wald or the likelihood root, as $\hat\psi$ is already known.
    \item Obtain a $(1-2 \alpha)$ confidence interval $(\psi_\alpha, \psi_{1-\alpha})$ based on $r(\psi)$  as the solutions of $r(\psi)=\pm z_\alpha$. If the equation $r(\psi)=-z_\alpha$ cannot be solved, then the lower limit of the confidence interval is $\psi_\alpha=0$.  Confidence intervals based on $r^*(\psi)$ are obtained likewise.
    \end{enumerate}
    }
 \tabularnewline
\textbf{Output}: Point estimates $\hat\psi$ and $\hat\psi^*$ of the miss distance and corresponding two-sided $(1-2\alpha)$ confidence intervals.\\
 \bottomrule \bottomrule
    \end{tabular}
    \caption{Miss distance interval estimation} \label{tab:tab8}
\end{table}

\section*{Acknowledgement} 
We thank Alessandra Brazzale, Nancy Reid and very constructive reviewers for helpful remarks that led to major improvements in our work.  The work was supported by the Swiss National Science Foundation.

\bibliography{ACD}

\end{document}